\documentclass{article}

\usepackage{microtype}
\usepackage{graphicx}
\usepackage{subcaption}
\usepackage{booktabs} 
\usepackage{subcaption}

\usepackage{tikz}

\usepackage{amsmath,amsthm,bm}

\usepackage{hyperref}

\usepackage{algorithm}
\usepackage{algorithmic}

\usepackage[accepted]{icml2025}

\usepackage{amsmath}
\usepackage{amssymb}
\usepackage{mathtools}
\usepackage{amsthm}

\usepackage[capitalize,noabbrev]{cleveref}

\theoremstyle{plain}

\theoremstyle{definition}

\theoremstyle{remark}

\graphicspath{{./img-png/}}

\usepackage[textsize=tiny]{todonotes}
\usepackage{color}
\definecolor{myred}{RGB}{189, 52, 67}
\definecolor{mygreen}{RGB}{19, 136, 8}
\definecolor{myblue}{RGB}{16, 52, 166}
\definecolor{myorange}{RGB}{255, 128, 0}

\newcommand{\karima}[1]{{\color{black} #1}\normalcolor}

\icmltitlerunning{Toward Efficient and Scalable Design of In-Memory Graph-Based Vector Search}

\begin{document}

\twocolumn[
\icmltitle{Toward Efficient and Scalable Design of In-Memory Graph-Based Vector Search}

\icmlsetsymbol{equal}{*}
\begin{icmlauthorlist}
\icmlauthor{Ilias Azizi}{etis}
\icmlauthor{Karima Echihabi}{um6p}
\icmlauthor{Themis Palpanas}{lipade}
\icmlauthor{Vassilis Christophides}{etis}
\end{icmlauthorlist}

\icmlaffiliation{etis}{ETIS UMR-8051 Laboratory, CY Cergy Paris Université, ENSEA, CNRS, France}
\icmlaffiliation{um6p}{College of Computer Science, University Mohammed VI Polytechnic, Morocco}
\icmlaffiliation{lipade}{LIPADE, Université Paris Cité - IUF, France}

\icmlcorrespondingauthor{Ilias Azizi}{ilias.azizi@ensea.fr}
\icmlkeywords{Vector Databases, Approximate KNN search}

\vskip 0.3in
]



\printAffiliationsAndNotice{} 

\begin{abstract}
Vector data is prevalent across business and scientific applications, and its popularity is growing with the proliferation of learned embeddings. Vector data collections often reach billions of vectors with thousands of dimensions, thus, increasing the complexity of their analysis. Vector search is the backbone of many critical analytical tasks, and graph-based methods have become the best choice for analytical tasks that do not require guarantees on the quality of the answers. Although several paradigms (seed selection, incremental insertion, neighborhood propagation, neighborhood diversification, and divide-and-conquer) have been employed to design in-memory graph-based vector search algorithms, a systematic comparison of the key algorithmic advances is still missing. We conduct an exhaustive experimental evaluation of twelve state-of-the-art methods on seven real data collections, with sizes up to 1 billion vectors. We share key insights about the strengths and limitations of these methods; e.g., the best approaches are typically based on incremental insertion and neighborhood diversification, and the choice of the base graph can hurt scalability. Finally, we discuss open research directions, such as the importance of devising more sophisticated data adaptive seed selection and diversification strategies. This work was presented at the ICML 2025 VecDB Workshop. An extended version appeared in ACM SIGMOD 2025.
\end{abstract}

\section{Introduction}
\label{sec:introduction}

Vector data is common in various scientific and business domains, and its prevalence is expected to grow in the future with the proliferation of learned embeddings. The volume and dimensionality of this data, which can exceed multiple terabytes and thousands of dimensions, make its analysis very challenging. A critical component of these data analysis tasks is vector search ~\cite{conf/icde/echihabi2021,palpanas2015data}. It supports recommendation~\cite{conf/kdd/wang2018}, information retrieval~\cite{conf/williams2014}, clustering~\cite{journal/JMLR/bubeck2009}, classification~\cite{DBLP:conf/icdm/PetitjeanFWNCK14}, entity resolution~\cite{christophides2020overview}, outlier detection~\cite{series2graph} and AI explainability~\cite{kaneko2023local} in many fields including bioinformatics, computer vision, finance. More recently, vector search has been playing a crucial role in improving the performance and interpretability of Large Language Models and reducing their hallucinations~\cite{retrieval-diffusion-models,dense-passage-retrieval}. 

As these applications scale to ever-larger datasets, often terabytes in size and thousands of dimensions, vector search becomes not only a computational bottleneck, but also a significant contributor to system latency and energy consumption \cite{Huang24}. In many AI pipelines, especially those deployed at inference time, similarity search dominates runtime cost~\cite{ai-inference-energy}. Optimizing vector search is therefore essential to building applications that are not only faster and more responsive, but also more efficient in terms of operational resources and environmental impact~\cite{chiang2025efficient}.

A vector search algorithm over a dataset ${S}$ of $n$ $d$-dimensional vectors returns answers in ${S}$ that are similar to a given input vector $V_Q$. The brute-force approach (a.k.a. a sequential or serial scan) compares $V_Q$ to every single element in ${S}$, with time complexity $O\left(nd\right)$—which becomes impractical and energy-intensive for large, high-dimensional datasets. To address this, state-of-the-art methods reduce the dimensionality $d$ via compact representations and/or minimize the number of comparisons $n$ through efficient indexing and pruning strategies. These improvements not only accelerate search but also reduce memory usage and computational overhead, contributing to more energy-efficient systems. Some approaches compute exact answers, while others $\epsilon$- and $\delta$-$\epsilon$-approximate answers, with deterministic and probabilistic guarantees on the accuracy of the answers, or ng-approximate answers without any theoretical guarantees, but high accuracy in practice ~\cite{lernaeanhydra2}. 

A large body of work has been dedicated to approximate vector search, which trades off accuracy for efficiency. These approaches are based on scans~\cite{vafile,va+file}, trees~\cite{isax2+,leafi}, graphs~\cite{hnsw,vamana,hcnng}, inverted indexes~\cite{gist, journal/pami/babenko15}, hashing~\cite{conf/vldb/sun14,srs}, or a combination of these data structures~\cite{SPTAG4,elpis}. Over the last decade, graph-based techniques have emerged as the method of choice for vector search in many real applications that can relax theoretical guarantees to achieve a query latency of a few milliseconds on terabyte-scale collections~\cite{alibabaknngml,faiss}. 

Graph-based methods typically represent a dataset $S$ as a proximity graph $G(V,E)$, where $V$ is the set of vertices corresponding to the $n$ data points, and $E$ is the set of edges connecting similar points. Query answering for a given query node $V_Q$ usually involves running an ng-approximate beam search, starting with an initial set of seed nodes to warm up the candidate priority queue, followed by a greedy graph traversal expanding promising candidates. State-of-the-art (SotA) graph-based methods share this beam-search approach but differ mainly in their graph construction strategies and seed selection mechanisms.

Although several paradigms (seed selection, incremental insertion, neighborhood propagation, neighborhood diversification, and divide-and-conquer) have been employed to design in-memory graph-based vector search algorithms, a systematic comparison of the key algorithmic advances is still missing. The only empirical study dedicated entirely to this family of techniques is ~\cite{graph-survey-vldb} without providing conclusive results due to the small scale of datasets used in experiments, not exceeding 1M vectors. As we will show in Section~\ref{sec:experiments} that trends change as dataset size increases. Moreover, existing benchmarks~\cite{aumuller2017ann,neurips-2021-ann-competition} for evaluating vector search methods do not focus on graph-based approaches, and thus, do not shed light on why the best methods have superior query performance. To the best of our knowledge, our work is the first proposing a taxonomy of algorithmic advances grounded in the actual indexing and search design principles, rather than on abstract graph categories. This enables new insights into how key design choices impact the indexing and search performance across varying dataset sizes, addressing limitations in prior studies.

In this paper, we make the following contributions\footnote{Full paper appeared as~\cite{azizi2025graph}.}:

\noindent$\bullet$ We identify five core paradigms underlying graph-based vector search: Seed Selection (SS), Neighborhood Propagation (NP), Incremental Insertion (II), Neighborhood Diversification (ND), and Divide-and-Conquer (DC). We briefly overview these paradigms and focus deeply on SS and ND, due to their significant impact on performance.

\noindent$\bullet$ We propose a taxonomy categorizing existing methods according to these paradigms, highlighting their chronological evolution and influence.

\noindent$\bullet$ We briefly survey state-of-the-art methods, discussing their design principles, strengths, and limitations.

\noindent$\bullet$ We conduct an extensive evaluation of twelve state-of-the-art methods on synthetic and real-world datasets (up to 1 billion vectors). Our experiments validate conventional wisdom, such as the superiority of incremental insertion methods in query performance and scalability~\cite{elpis,graph-survey-vldb}. However, we also find differences from recent studies~\cite{graph-survey-vldb}, notably demonstrating better-than-previously-reported efficiency of SPTAG-BKT~\cite{SPTAG4} on small datasets and superior indexing efficiency of HNSW~\cite{hnsw}. Contrary to prior findings~\cite{elpis}, we also show that Vamana~\cite{vamana} is competitive on large scale datasets.

\noindent$\bullet$ We provide novel insights, such as (i) identifying the most effective ND technique for large datasets, and (ii) analyzing the impact of different SS methods on query and indexing performance on various dataset scales.

\noindent$\bullet$ Finally, we highlight promising research directions, including developing scalable NP and ND graph structures, data-adaptive seed selection techniques, theoretical analysis of ND methods, and tailored strategies (e.g., clustering, diversification) for DC-based approaches.

\section{Graph-Based Vector Search}
\label{sec:survey}
We now present an overview of the main SotA graph-based $ng$-approximate vector search methods. 
We outline the base data structures and algorithms in this field, and identify five main paradigms exploited by the SotA approaches. 
We propose a new taxonomy that categorizes these approaches along the five paradigms, highlighting also their chronological development and influence map. 
\subsection{A Primer}

A proximity graph is a graph $G\left({V},{E}\right)$ in which two vertices $V_i$ and $V_j$ are connected by an edge if and only if they satisfy particular geometric requirements, namely the \textit{neighborhood criterion}~\cite{shamos1975closest}.
A proximity graph can be constructed using different distances such as the dot product~\cite{mipsg}, nevertheless the Euclidean distance remains the most popular one~\cite{edelsbrunner87}. One of the earliest proximity graphs in the literature is the Delaunay Graph (\textit{DG}). It is a planar dual graph for the Voronoi Diagram~\cite{vd95}, where each vertex is the center of its own voronoi cell, and two vertices are linked if and only if their corresponding voronoi cells share at least one edge. A DG 
satisfies the Delaunay Triangulation :$\forall q,p,r \in {V}, \left(q,p\right), \left(q,r\right), \left(r,p\right) \in {E}$
if the circumcircle of the triangle $q, p, r$ is empty~\cite{aurenhammer2013voronoi}.
A beam search~\cite{reddy77bm} (Algorithm~\ref{alg:beamsearch}) on a \textit{DG} can find the exact nearest neighbors~\cite{dobkin1990delaunay} when the dataset has a high dimensionality or the beam search uses a large beam width.
However, using a DG in high dimensions is impractical, because the graph becomes almost fully connected as the dimensionality $d$ grows~\cite{dobkin1990delaunay}. 
Thus, SotA methods build alternative graph structures and use beam search to support efficient query answering
~\cite{gabriel69,matula80,toussaint02}.

\begin{algorithm}[tb]
   \caption{Beam Search (G, $q$, \textit{s}, \textit{k}, \textit{l})}
   \label{alg:beamsearch}
\begin{algorithmic}[1]
   \STATE {\bfseries Input:} Graph $G$, query vector $q$, initial seeds $s$, result size $k$, beam width $l \geq k$
   \STATE {\bfseries Output:} $k$ approximate nearest neighbors to $q$
   
   \STATE Initialize candidate set $C \gets s$
   \STATE Initialize visited list $V \gets \emptyset$
   
   \WHILE{$C \setminus V \neq \emptyset$}
      \STATE $p^{*} \gets \operatorname*{argmin}_{x_i \in C \setminus V} dist(q, x_i)$
      \STATE $C \gets C \cup N_{out}(p^*)$
      \STATE $V \gets V \cup \{p^*\}$
      
      \IF{$|C| > l$}
         \STATE Retain only the closest $L$ points in $C$ to $q$
      \ENDIF
   \ENDWHILE
   
   \STATE \textbf{Return} the closest $k$ candidates in $C$ to $q$
\end{algorithmic}
\end{algorithm}

\subsection{Main Paradigms}

We provide a brief overview of the five main paradigms exploited by SotA methods.
Then, we describe in more detail the two  paradigms that have the greatest impact on query performance (as will be demonstrated in Section~\ref{sec:experiments}).

\noindent{\bf{Seed Selection (SS)}} chooses initial nodes to visit during search. It is also used during index building by approaches that exploit a beam search during the construction of the index to decide which edges to build. Some methods simply select one or more seed(s) randomly, while others use special data structures, e.g., a K-D Tree.

\noindent{\bf{Neighborhood Propagation (NP)}} refines a pre-existing graph following a user-defined number of iterations, a.k.a. NNDescent~\cite{nndescent}. 
During each iteration, the potential neighbors of a given node are sourced both from its immediate neighbors and the neighbors of its neighbors. Then, the node only keeps the $m$ closest neighbors, where $m$ is a user-parameter. 
The pre-existing graph could be a random graph  or some other type of graph.

\noindent{\bf{Incremental Insertion (II)}} refers to building a graph by inserting one vertex at a time. 
Each vertex is connected using bi-directional edges to its nearest neighbors and some distant vertices. The neighbors are selected using a beam search on the already inserted portion of the graph. At the end of graph construction, some vertices retain early connections which act as long-range links. 
This approach was first proposed in the object-based peer-to-peer overlay network VoroNet~\cite{voronet}, with the idea of adding long-range links being inspired from Kleinberg's small-world model~\cite{kleinberg2000, kleinberg2002}, with the difference that the latter selects the long-range links randomly. 

\noindent{\bf{Neighborhood Diversification (ND)}} was first introduced by the Relative Neighborhood Graph (RNG)~\cite{rng}. It aims to sparsify the graph by pruning unnecessary edges while preserving connectivity. For each node, ND exploits the geometrical properties of the graph to remove edges to neighbors that lead to redundant regions or directions. This indirectly causes the creation of long-range links allowing nodes to maintain diversified neighborhood lists, which reduces the number of comparisons during search.

\noindent{\bf{Divide-and-Conquer (DC)}} is a strategy that splits a dataset into multiple, possibly overlapping, partitions, then builds a separate graph on each partition. 
Some approaches such as SPTAG~\cite{SPTAG4} and HCNNG~\cite{hcnng} combine the individual graphs into one large graph, on which a beam search is performed, while ELPIS~\cite{elpis} maintains the graphs separate and searches them in parallel.

\subsection{Seed Selection}
\label{sec:ss}
While state-of-the-art graph-based vector search methods vary in graph construction strategies, nearly all rely on beam search (Algorithm~\ref{alg:beamsearch}) for query answering, as it efficiently retrieves good answers in well-connected graphs. However, the choice of initial nodes—referred to as \emph{seeds}—significantly impacts search efficiency: better seeds lead to fewer visited nodes and faster searches. Typically, an entry node or multiple candidate entry nodes are used to warm the beam search's priority queue; in the case of multiple seeds, the search picks the one closest to the query as the initial node, retaining others in the priority queue. While several methods propose additional indexes built on data samples to facilitate seed selection, none provide comprehensive evidence supporting their effectiveness. In this paper, we systematically examine the seed-selection techniques proposed in the literature:

\noindent (1) \textbf{Stacked-NSW (SN)}: HNSW~\cite{hnsw}, inspired by skip lists~\cite{skiplist}, builds hierarchical NSW graphs by sampling nodes from lower layers. Nodes are assigned a top layer \(L=-\ln(\xi)/\ln(M/2)\), with \(\xi\sim U(0,1)\) and out-degree \(M\). Queries descend greedily from a fixed top-layer entry point, returning the closest sampled node to the query as entry point.

\noindent (2) \textbf{K-D Trees (KD)}: Constructs a single or multiple K-D Tree(s)~\cite{kdtree} on a dataset sample. During search,retrieves the set of seed points is retrieved through depth-first search traversal on the K-D Tree structure(s).

\noindent (3) \textbf{LSH}: Constructs an LSH index on a sample of the dataset to retrieve the seeds during search. 

\noindent (4) \textbf{Medoid (MD)}: 
Uses medoid node as seed and entry point during query answering.

\noindent (5) \textbf{Single Fixed Random Entry Point (SF)}: A random node is selected and fixed as the entry point for all searches.

\noindent (6) \textbf{K-Sampled Random Seeds (KS)}: 
For each query, \( k \) random nodes are selected as seed points. 

\noindent (7) \textbf{Balanced K-means Trees (KM)}: Constructs Balanced K-means Trees (BKT) ~\cite{bkmtree} on a dataset sample. During search, seed points are retrieved via depth-first search on the BKT structure(s).

\subsection{Neighborhood Diversification}
\label{sec:nd}
\begin{figure}[t]    
  \centering
  \begin{subfigure}[b]{0.3\linewidth}
      \includegraphics[width=\linewidth]{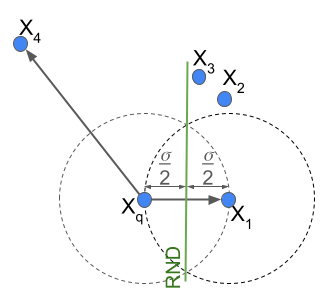}
      \caption{RND}
      \label{fig:ND:RND}
  \end{subfigure}\hfill
  \begin{subfigure}[b]{0.3\linewidth}
      \includegraphics[width=\linewidth]{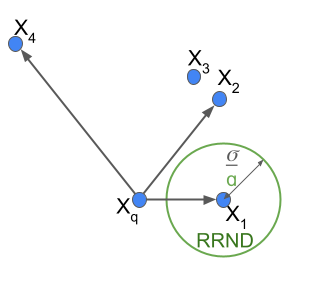}
      \caption{RRND}
      \label{fig:ND:RRND}
  \end{subfigure}\hfill
  \begin{subfigure}[b]{0.3\linewidth}
      \includegraphics[width=\linewidth]{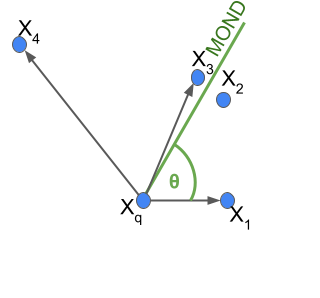}
      \caption{MOND}
      \label{fig:ND:MOND}
  \end{subfigure}

  \caption{Neighborhood diversification approaches}
  \label{fig:ND:example}
\end{figure}

Neighborhood Diversification (ND) builds sparse graphs by selectively pruning edges to balance short and long-range connections, thus reducing redundant comparisons. Initially proposed by RNG~\citep{rng,toussaint02}, ND was adapted for approximate vector search by several graph-based methods. Three main ND strategies exist: Relative Neighborhood Diversification (RND) used by HNSW~\cite{hnsw}, NSG~\cite{nsg}, SPTAG~\cite{SPTAG4} and ELPIS~\cite{elpis}; Relaxed RND (RRND) introduced by Vamana~\cite{vamana}; and Maximum-Oriented ND (MOND), proposed by DPG~\cite{dpg} and NSSG~\cite{nssg}. RND connects node \(X_q\) to candidate \(X_j\) only if no current neighbor \(X_i\) is closer to \(X_j\). RRND relaxes this condition by factor \(\alpha \geq 1\), reducing pruning and increasing edges. MOND selects edges based on an angle threshold \(\theta \geq 60^\circ\), favoring diverse edge orientations. RRND and MOND prune fewer edges than RND, thus resulting in denser graphs (cf. Figure~\ref{fig:ND:example}). Note that any nodes pruned by RRND and MOND will eventually be pruned by RND, but not vice versa. 
Refer to~\cite{url/GASS} for a detailed proof.

\subsection{A Taxonomy}
\begin{figure}[t]        
  \centering   
  \includegraphics[width=\linewidth]{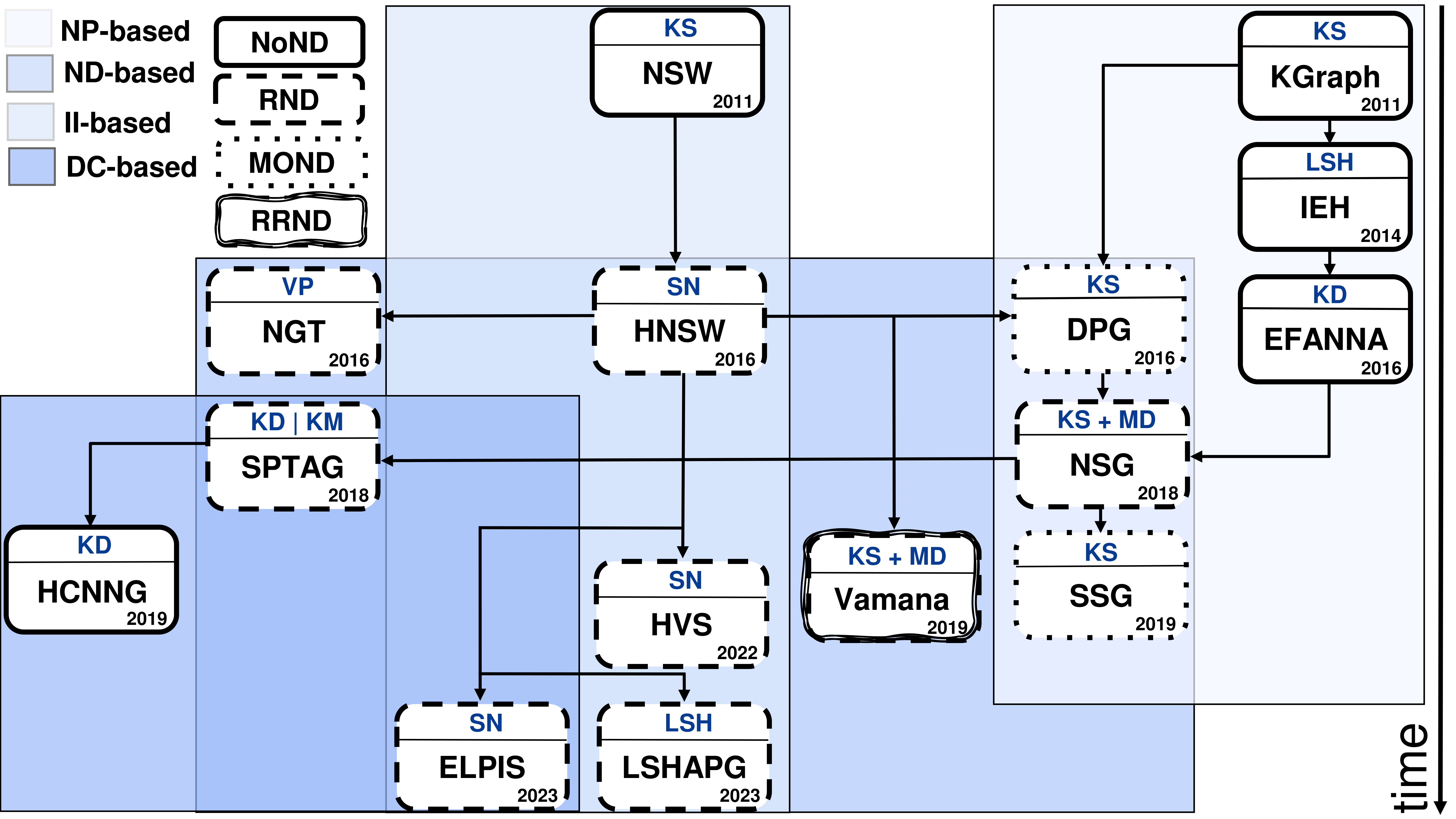}
  \caption{Graph-based ANN indexing paradigms}
  \label{fig:roadmap}
\vspace{-0.8cm}
\end{figure}

Figure~\ref{fig:roadmap} depicts the SotA graph-based approaches, classified based on the five design paradigms: SS, NP, II, ND, and DC. The taxonomy also reflects the chronological development of the methods. 
Directed arrows indicate the influence of one method on another. Within the ND category, distinctions are made between different strategies, i.e., No Neighborhood Diversification (NoND), RND, RRND, and MOND (cf. Section~\ref{sec:nd}). 
We identify the SS strategy of each method: KS, KD, SN, MD, LSH, and KM (SF is not used by any SotA method, but we consider it as an alternative strategy). 

Additionally, some methods use more than one strategy (e.g. NSG and VAMANA use KS and MD), or offer the flexibility to use different strategies (e.g., SPTAG can use either KD or KM). 
Note that a method can exploit one or more paradigms; e.g., HNSW uses incremental node insertion and prunes each node's neighbors using the RND approach, thereby being classified as both II and ND. 
KGraph~\cite{kgraph} was the first to use NP to approximate the exact k-NN graph (k-NNG) (with quadratic complexity), and influenced numerous subsequent methods, including IEH~\cite{ieh} and EFANNA~\cite{efanna}. In parallel, NSW~\cite{nsw11} introduced the II strategy for graph construction.

HNSW~\cite{hnsw} and DPG~\cite{dpg} leveraged ND to enhance NSW and KGraph, respectively. 
The good performance of HNSW and DPG encouraged more methods to adopt the ND paradigm, including NGT~\cite{ngt_library}, NSG~\cite{nsg} and SSG~\cite{nssg}, which apply ND on the NP-based graph EFANNA. SPTAG~\cite{SPTAG4} combined DC with ND. 

Vamana~\cite{vamana} adopts NSG's idea of constructing the graph through beam search and ND. However, Vamana constructs its graph by refining an initial base random graph in two rounds of pruning, using RRND and RND. Inspired by HNSW, Vamana and NGT proposed variants that support incremental graph building~\cite{diskanncode, ngt_library}, but we classify them as ND-based per the ideas proposed in the original papers. HCNNG~\cite{hcnng} was influenced by SPTAG and adopted a DC approach for constructing the graph without adopting ND. ELPIS~\cite{elpis} also adopted a DC strategy but leveraged both II and ND. 
HVS~\cite{hvs} and LSHAPG~\cite{lshapg} both propose new seed selection structures for HNSW, with the latter additionally adopting a new probabilistic rooting approach. Note that earlier approaches, except from NSW, were mainly NP-based; however, recent studies have focused on devising methods that leverage the ND, II, and DC paradigms because they lead to superior performance 
(cf. Section~\ref{sec:experiments}).

\section{Experimental Evaluation}
\label{sec:experiments}

We experimentally evaluate twelve state-of-the-art graph-based vector search methods, based on the two key paradigms, namely seed selection and neighborhood diversification. To single out the effect of each strategy, we first implement a basic II-based method, where nodes are inserted incrementally and each node $i$ acquires its list of candidate neighbors through a beam search on the current partial graph of already inserted nodes.  Then, we implement each strategy independently on the resulting graph. Finally, we assess the indexing and query-answering performance of these methods on a variety of real and synthetic datasets. All artifacts are available in~\cite{url/GASS}.

\subsection{Framework}

\noindent{\bf Setup.} 
All methods were compiled with GCC 8.2 on Ubuntu 20.04 and run on a 4-socket Intel Xeon Platinum 8276 server (28 cores/socket, 1 thread/core) with 1.5TB RAM.

\noindent{\bf Algorithms.} We cover the following methods: HNSW~\cite{hnsw}, NSG~\cite{nsg}, Vamana~\cite{vamana}, DPG~\cite{dpg}, EFANNA~\cite{efanna}, HCNNG~\cite{hcnng}, KGraph~\cite{kgraph}, NGT~\cite{ngt_library}, DPG~\cite{dpg}, and two versions of SPTAG~\cite{SPTAG4} (SPTAG-BKT and SPTAG-KDT, using BKT and K-D Trees, respectively). We also include ELPIS~\cite{elpis} and LSHAPG~\cite{lshapg}. IEH and FANNG are excluded due to suboptimal performance~\cite{graph-survey-vldb, nsg}, and HVS due to difficulties running the official implementation~\cite{hvsgithub}. Euclidean distance is employed as the sole similarity metric throughout all methods and experiments. We use the most efficient publicly available C/C++ implementations for each algorithm, leveraging multithreading and SIMD vectorization to optimize performance. We also disabled the optimizations that would lead to an unfair evaluation such as cache pre-warming and L2-normalized Euclidean distance. Since all methods except ELPIS and HNSW use a single priority queue, we modified their original two-queue implementations~\cite{url/Elpis,url/hnsw} to use a single queue. The modified versions are documented in~\cite{url/GASS}.

\noindent{\bf Datasets.} 
We use seven real-world datasets from various domains, including deep network embeddings, computer vision, neuroscience, and seismology:  
(i) \emph{Deep}~\cite{url/data/deep1b} contains 1 billion 96-dimensional vectors extracted from the final layers of a convolutional neural network;  
(ii) \emph{Sift}~\cite{url/data/sift} consists of 1 billion 128-dimensional SIFT vectors representing image feature descriptors;  
(iii) \emph{Sald}~\cite{url/data/eeg} provides neuroscience MRI data with 200 million 128-dimensional data series;  
(iv) \emph{Seismic}~\cite{url/data/seismic} contains 100 million 256-dimensional time series representing earthquake recordings from seismic stations worldwide;  
(v) \emph{Text-to-Image}~\cite{url/data/text2image} offers 1 billion 200-dimensional image embeddings from Se-ResNext-101 along with 50 million DSSM-embedded text queries for cross-modal retrieval under domain shifts;  
(vi) \emph{Gist}~\cite{gist} contains 1 million 960-dimensional vectors, using GIST descriptors~\cite{gistdesc} to capture spatial structure and color layout of images;  and
(vii) \emph{ImageNet1M}, a new dataset that we generated from the original ImageNet~\cite{imagenet}, producing embeddings of 1 million original vectors using the ResNet50 model~\cite{resnet}, with PCA applied to reduce dimensionality to 256. 
We select subsets of different sizes from the Sift, Deep, SALD and Seismic datasets, and we refer to each subset with the name of the dataset followed by the subset size in GBs (e.g., Deep25GB). 
We refer to the 1-million and 1-billion datasets with the 1M and 1B prefixes, respectively. 
To test robustness across distributions, we generate three 25GB synthetic datasets (RandPow0, RandPow5, RandPow50; 256 dimensions) using power law exponents 0 (uniform), 5, and 50 (high skew). Power law follows $Y = kX^a$, with skewness increasing with exponent $a$.

\begin{figure}[th]                           
  \centering
  \captionsetup[subfigure]{justification=centering}

  \begin{subfigure}[b]{0.48\linewidth}      
    \includegraphics[width=\linewidth]{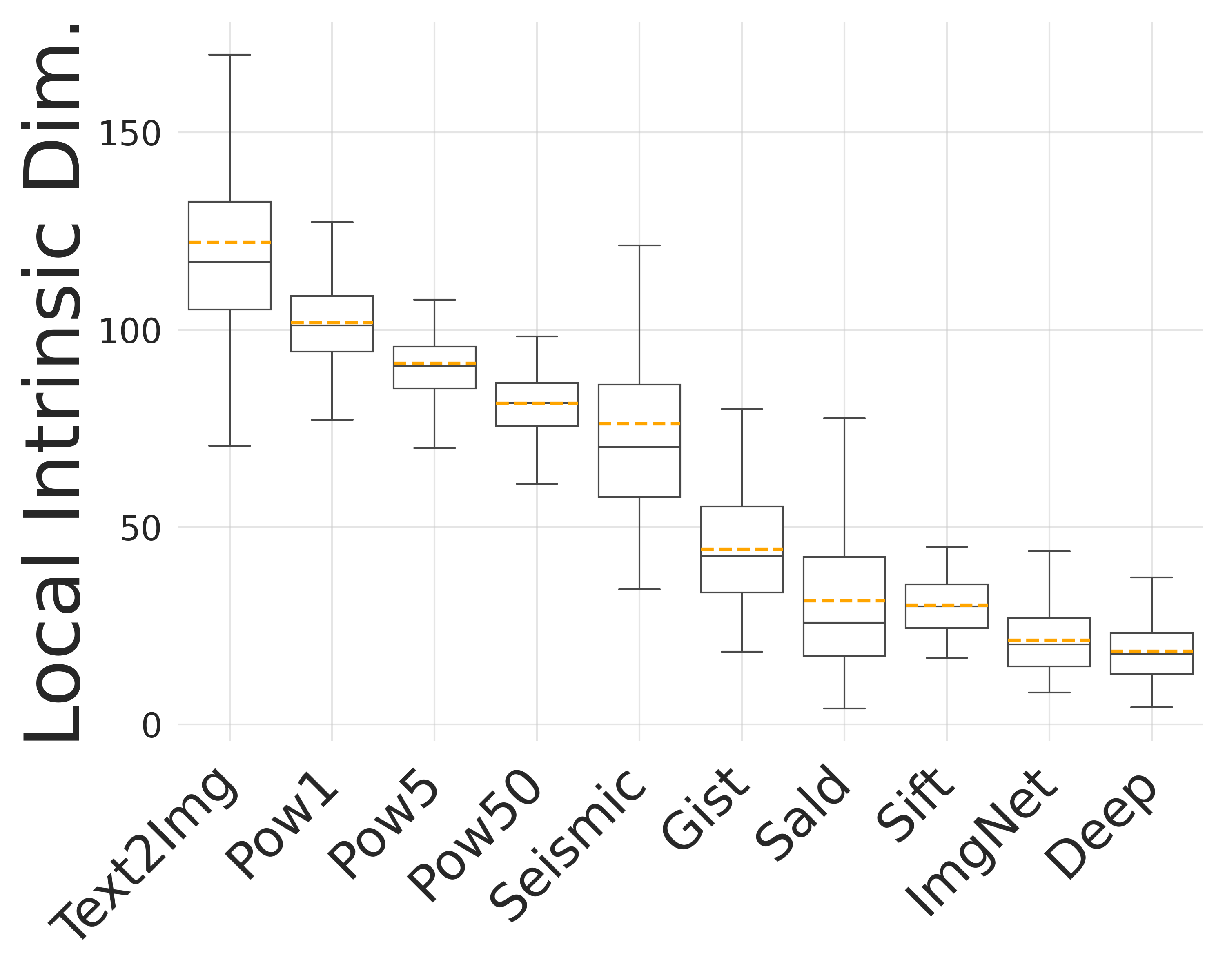}
    \caption{Local intrinsic dim.}
    \label{fig:datacomp:lid}
  \end{subfigure}\hfill                     
  \begin{subfigure}[b]{0.48\linewidth}
    \includegraphics[width=\linewidth]{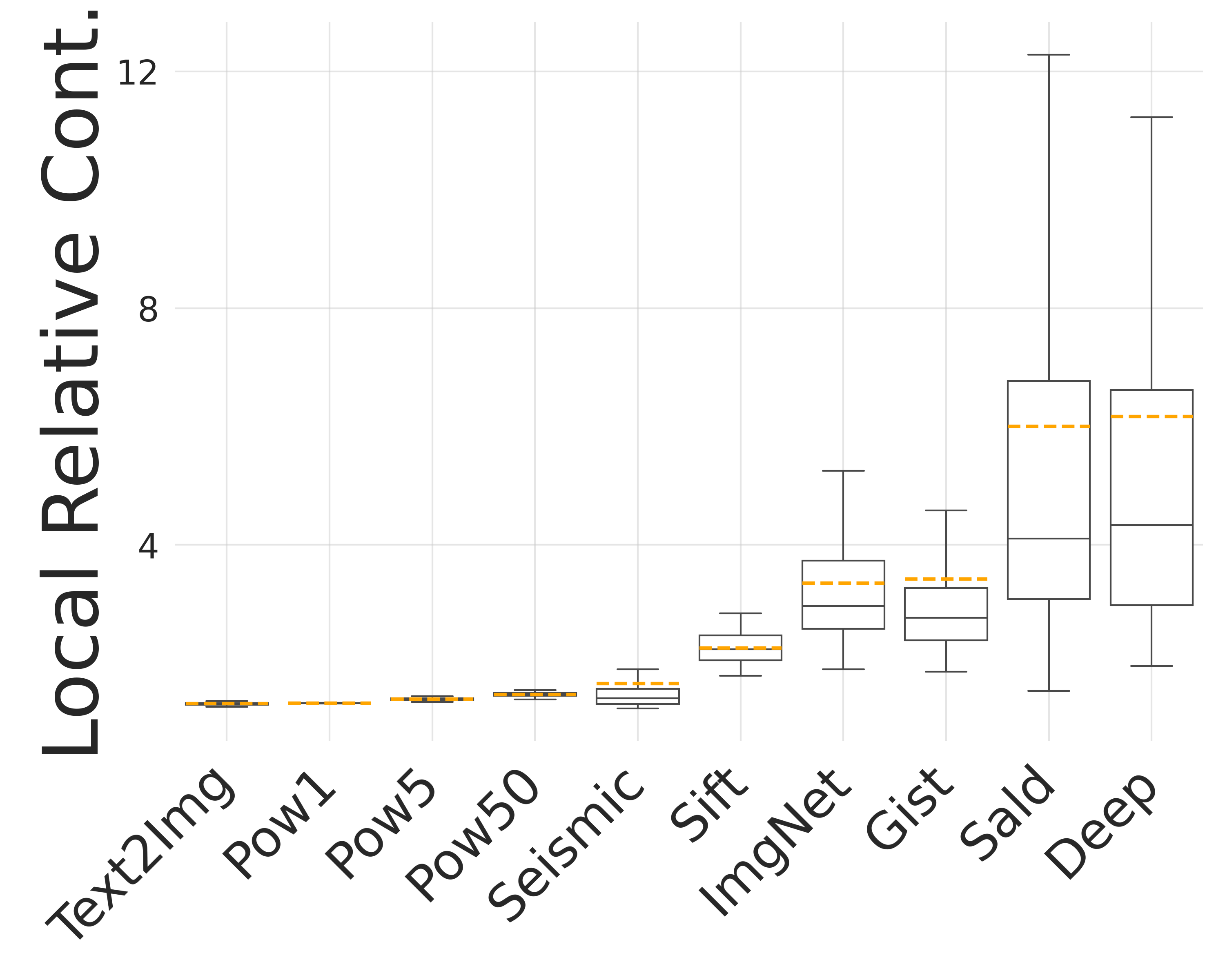}
    \caption{Local relative contrast}
    \label{fig:datacomp:rc}
  \end{subfigure}

  \caption{\karima{Dataset complexity}}
  \label{fig:datacomp}
\end{figure}

\noindent{\bf Dataset Complexity.} 
We measure dataset complexity using Local Intrinsic Dimensionality (LID) and Local Relative Contrast (LRC)~\cite{rc,DBLP:journals/is/AumullerC21}, defined respectively as:

\begin{equation}
\text{LID}(x) = -\left(\frac{1}{k} \sum_{i=1}^{k} \log \frac{\text{dist}_i(x)}{\text{dist}_{k}(x)}\right)^{-1} \hfill 
\end{equation}
\begin{equation}
\text{LRC}(x) = \frac{\text{dist}_{\text{mean}}(x)}{\text{dist}_k(x)}
\hfill
\end{equation}

where $\text{dist}_i(x)$ is the Euclidean distance from $x$ to its $i$-th nearest neighbor, $\text{dist}_{mean}(x)$ is the average distance from $x$ to all dataset points, and $k=100$. Lower LID and higher LRC values indicate easier search tasks. Figure~\ref{fig:datacomp} (computed on 1M-point dataset samples) confirms consistency between these metrics, with orange lines indicating dataset mean values. Pow0, Pow5, Pow50, Seismic, and Text2Img have the highest LID and lowest LRC (hardest datasets), whereas Sift, Deep, and ImageNet have the lowest LID and highest LRC (easiest datasets).

\noindent{\bf Queries.} 
Each query workload consists of 100 vectors processed sequentially to simulate realistic, unpredictable query streams~\cite{itsawreport,conf/sigmod/gogolou20}; results for 1M queries are extrapolated from these workloads. For Deep, Sift, Gist, and Text-to-Image(Text2Img), queries are randomly sampled from their provided workloads; for Sald, ImageNet, and Seismic, queries are randomly selected from datasets and excluded from index building. Query hardness experiments use Deep dataset vectors perturbed by Gaussian noise (\(\mu=0\), variance \(\sigma^2\) from 0.01 to 0.1), labeled as 1\%–10\% noise based on the variance~\cite{johannesjoural2018}. Queries for power-law datasets follow the original distribution with new random seeds. Unless stated otherwise, experiments use standard 10-NN queries~\cite{neurips-2021-ann-competition,aumuller2017ann}, except for dataset complexity and seed-selection analyses, where 100-NN queries are used for higher precision and to reflect increased overhead.


\noindent\textbf{Measures.} We measure the {\it wall clock time} and {\it distance calculations} for both indexing and query answering. We also measure the accuracy of each k-NN query using {\it Recall} which quantifies the fraction of the true nearest neighbors that the query $S_Q$ successfully returns.  

\subsection{Neighborhood Diversification}
We compare existing ND methods with the baseline NOND on II-graph (HNSW) (Fig.\ref{fig:nd_choice:deep1b}). RND consistently ranks first, slightly ahead of MOND, followed by RRND; NOND performs worst across all scales (see Appendix for full results).

\subsection{Seed Selection} We evaluate seed Selection methods on II-graph with RND pruning across different datasets and scales. Both SN and KS lead in query performance. KS edges out SN on small scale (Fig.\ref{fig:ss_choice:deep25gb}), while SN outperforms at large scale (Fig.\ref{fig:ss_choice:deep1b}) (see full results in appendix).

\newcommand{\figsize}{0.495\linewidth} 

\begin{figure}[t]
  \begin{minipage}{0.345\linewidth}
    \centering
    \captionsetup{justification=centering}
        \includegraphics[width=1.1\linewidth]{Experiments/RNG/legend.png}
    \includegraphics[width=\linewidth]{Experiments/RNG/DC_DEEP1B.png}
    \caption{Impact of ND choice on query performance (Deep1B)}
    \label{fig:nd_choice:deep1b}
  \end{minipage}%
  \hfill
  \begin{minipage}{0.65\linewidth}
    \centering
    \captionsetup{justification=centering}
    \begin{subfigure}[b]{0.03\columnwidth}
			\includegraphics[width=\textwidth]{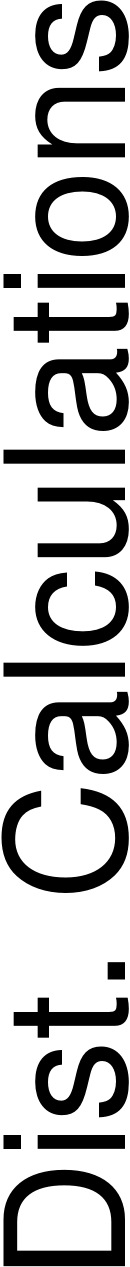}
            \vspace{0.2in}
	\end{subfigure}%
    \begin{subfigure}[b]{\figsize}
      \centering
      \includegraphics[width=\linewidth]{Experiments/EP/DEEP_25GB_100.png}
          \caption{Deep25GB}
          \label{fig:ss_choice:deep25gb}
    \end{subfigure}%
    \begin{subfigure}[b]{\figsize}
      \centering
      \includegraphics[width=\linewidth]{Experiments/EP/DEEP_1B_100.png}
        \caption{Deep1B}
        \label{fig:ss_choice:deep1b}
    \end{subfigure}%
    \captionsetup{justification=centering}
    \caption{Impact of SS choice on query performance}
    \label{fig:ss_choice}
  \end{minipage}
\end{figure}

\subsection{Indexing Efficiency}
Figure~\ref{fig:idx:time} shows II-based methods offer the fastest indexing. ELPIS leads, being 2.7× faster than HNSW and 4× faster than NSG on 1M and 25GB. NSG is slowed by its EFANNA base, while SPTAG variants are ~24× slower than ELPIS due to costly multi-tree construction. Most graph-based methods fail to scale to large collections; only HNSW, ELPIS, and Vamana index the 1B scale efficiently. ELPIS remains fastest, outperforming HNSW and Vamana by 2× and 2.7×, respectively. We also assess the memory footprint of SOTA methods; see Appendix for details.

\subsection{Query Performance}
We evaluate query performance across varying dataset scale and difficulties (Figure~\ref{fig:combined_experiments}). On small, low-LID datasets, NSG, SSG, and HNSW perform best (Figure~\ref{fig:query_deep_1m}). As dataset complexity increases—e.g., Seismic1M (Figure~\ref{fig:query_seismic_1m}) and Deep25GB with 10\% noise (Figure~\ref{fig:query_10noise})—DC-based methods like SPTAG and ELPIS outperform others. Several methods, including NSG, fail to scale due to memory or indexing constraints. At billion-scale, ELPIS achieves the best performance (Figure~\ref{fig:query_deep_1b}), leveraging its multithreaded design. More results across various datasets and scales are available in the Appendix.

\newcommand{\subfigsize}{0.243\textwidth}
\newcommand{\ylabelsize}{0.017\textwidth} 
\begin{figure*}[t]
  \centering
  \includegraphics[width=\textwidth]{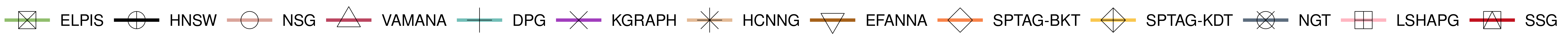}

  \begin{minipage}[t]{0.18\textwidth}
    \centering
  
    \includegraphics[width=\linewidth]{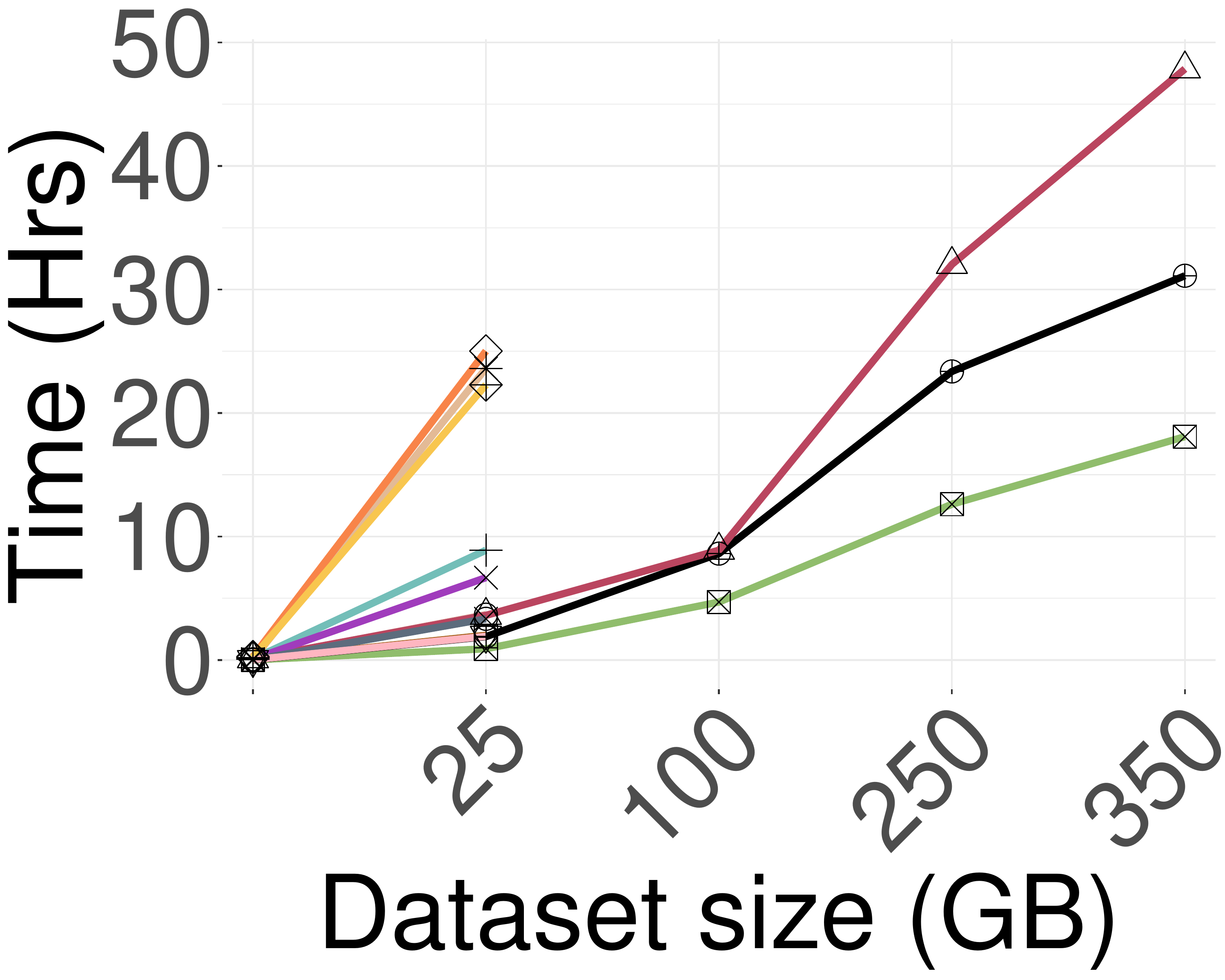}
      \captionsetup{justification=centering}
    \caption{Indexing time}
    \label{fig:indexing_time}
  \end{minipage}%
  \hfill
  \begin{minipage}[t]{0.81\textwidth}
    \centering
    \begin{subfigure}{\ylabelsize}
      \raisebox{0.8cm}{\includegraphics[width=\linewidth]{Experiments/time}}
    \end{subfigure}
       \hfill
    \begin{subfigure}[t]{\subfigsize}
      \centering
    \includegraphics[width=\linewidth]{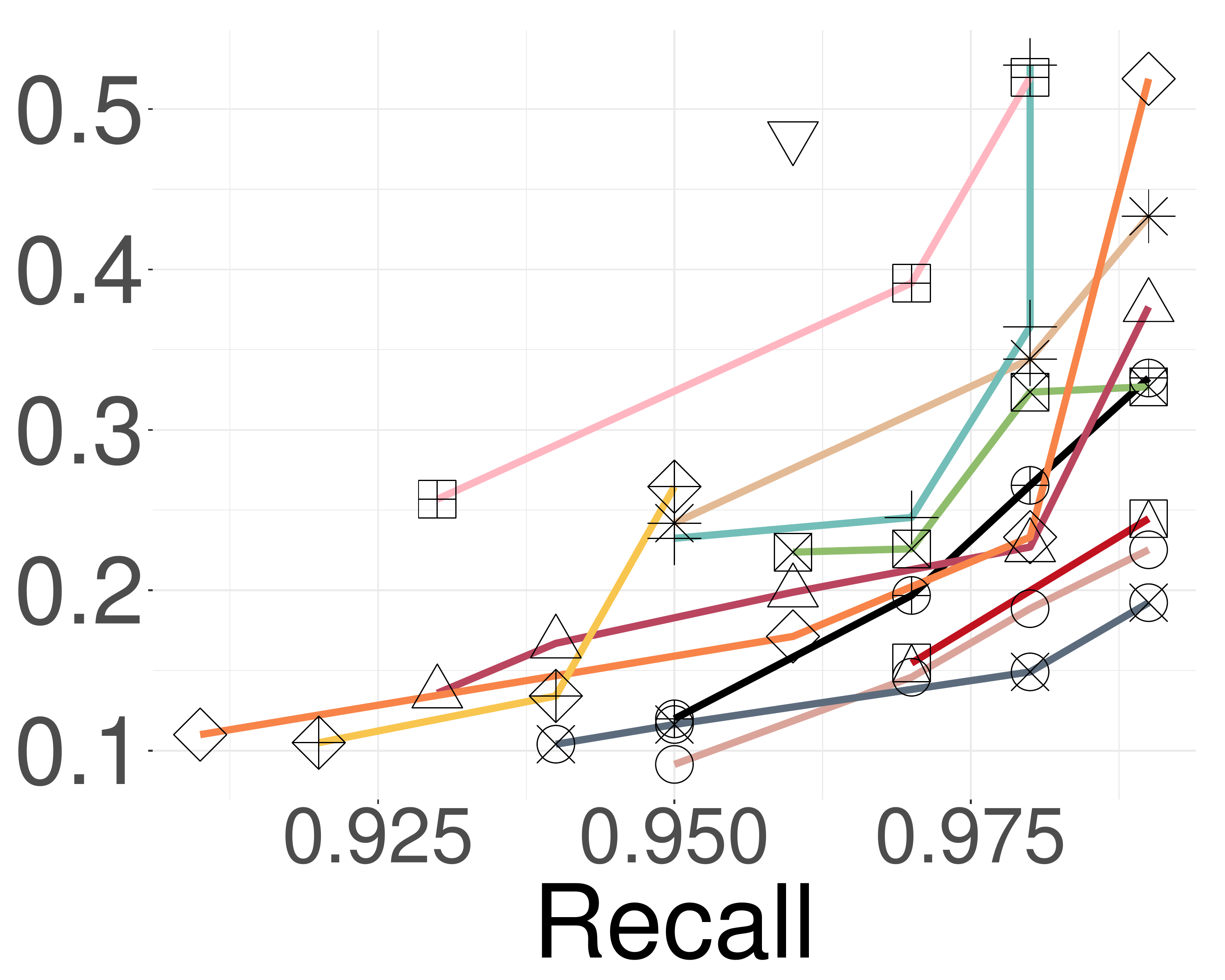}
      \caption{Deep1M}
      \label{fig:query_deep_1m}
    \end{subfigure}%
    \hfill
    \begin{subfigure}[t]{\subfigsize}
      \centering
      \includegraphics[width=\linewidth]{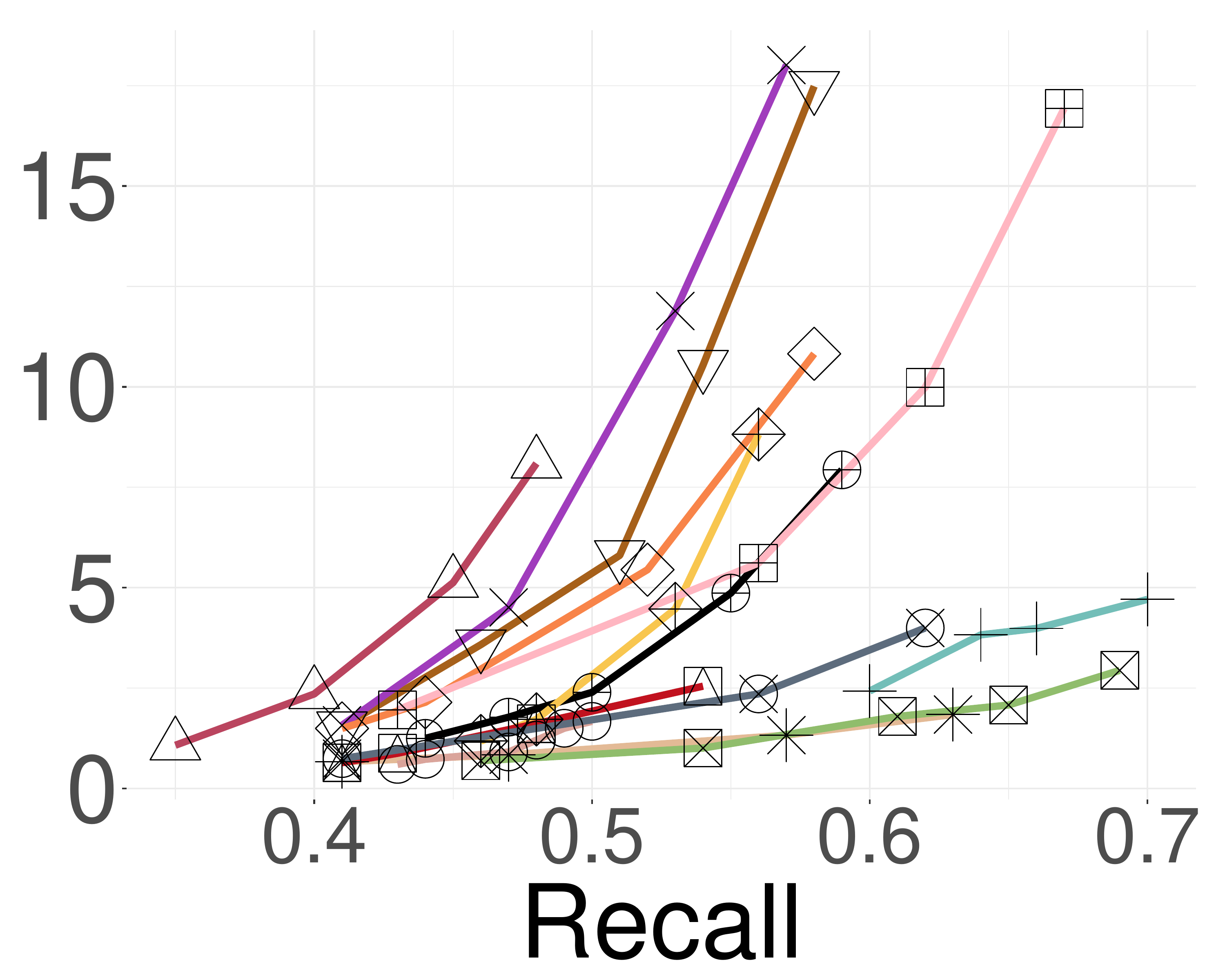}
      \caption{Seismic1M}
      \label{fig:query_seismic_1m}
    \end{subfigure}%
    \hfill
    \begin{subfigure}[t]{\subfigsize}
      \centering
      \includegraphics[width=\linewidth]{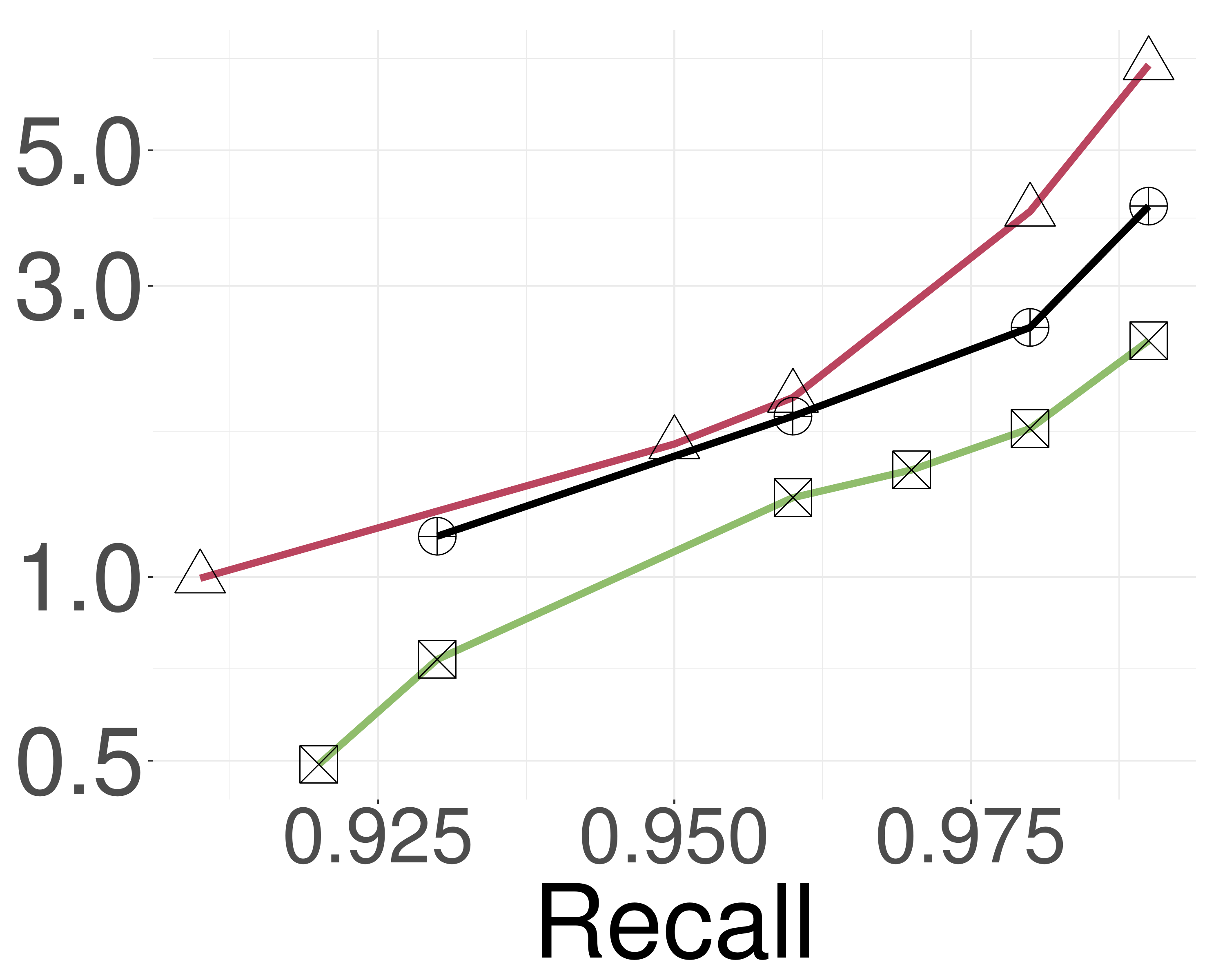}
      \caption{Deep1B}
      \label{fig:query_deep_1b}
    \end{subfigure}%
    \hfill
    \begin{subfigure}[t]{\subfigsize}
      \centering
    \includegraphics[width=\linewidth]{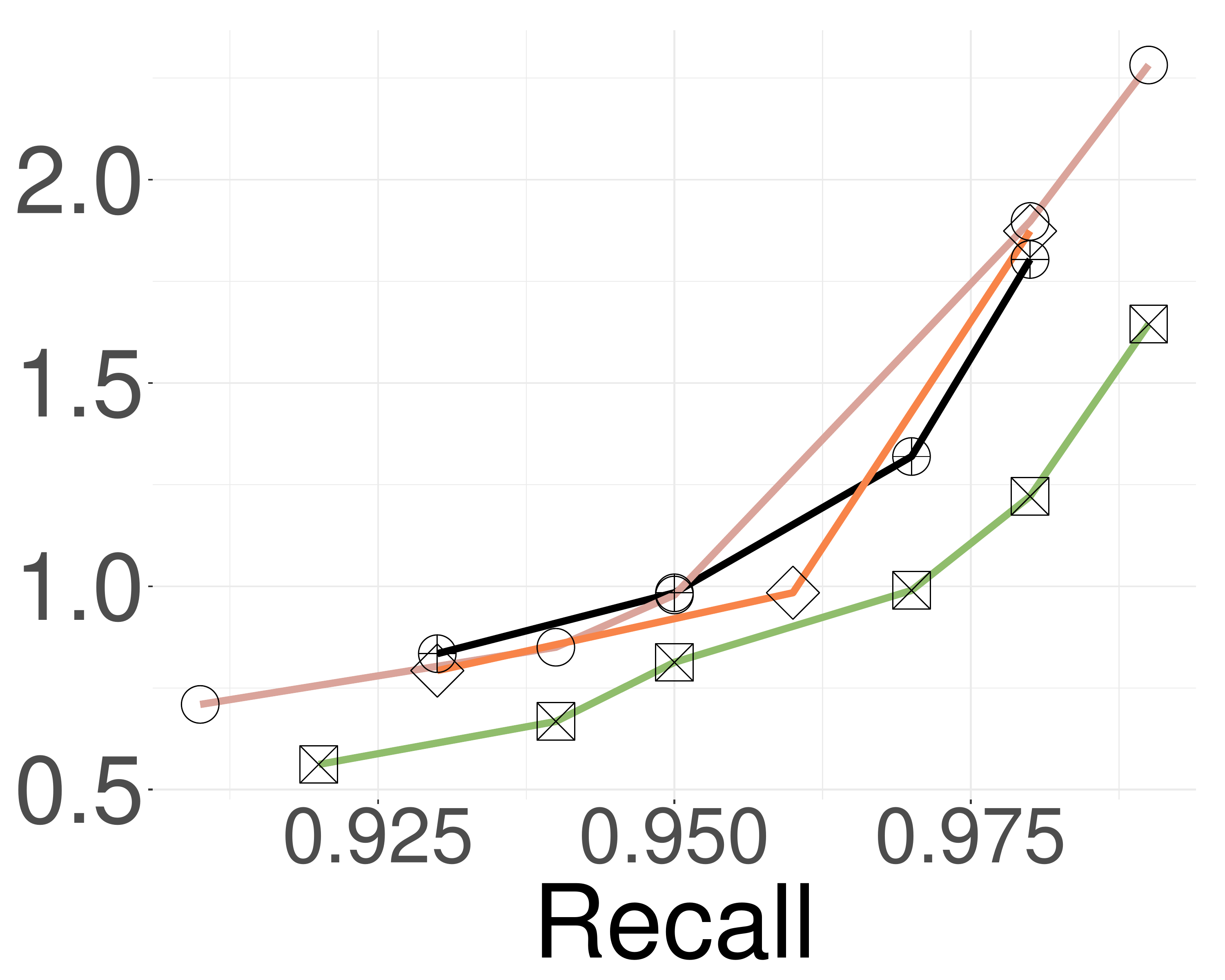}
      \caption{10\% noise}
      \label{fig:query_10noise}
    \end{subfigure}%
  \end{minipage}

  \caption{Query performance}
  \label{fig:combined_experiments}
\vspace{-.5cm}
\end{figure*}
\section{Discussion}
\label{sec:discussion}
In the previous section, we presented the results of an extensive evaluation of twelve state-of-the-art graph-based vector search methods. Table~\ref{tab:comp} summarizes the evaluation across key criteria: 
for search, we assess efficiency, accuracy, and the number of tunable parameters; 
for indexing, we evaluate efficiency at high recall, memory footprint, and parameter tuning complexity. 

The best-performing methods, HNSW, VAMANA, and ELPIS, have the best search performance and index efficiency. However, ELPIS requires an extra parameter during both indexing (leaf size) and search (nprobes), whereas VAMANA requires an extra parameter to tune during indexing (alpha). NSG and SSG exhibit efficient query performance, but their indexing capability is hindered because of their base graph EFANNA, which similarly to KGraph, is tedious to tune and suffers from high indexing time and footprint. Both SPTAG-BKT and NGT show satisfactory performance during search; however, they do not scale well to large datasets and require more tuning compared to the best methods. The assessment of HCNNG is based on the optimized parlayANN implementation which has shown competitive performance on large-scale datasets. 

We now summarize the key insights and pinpoint promising research directions.

\newcommand{\tablecolwidth}{\columnwidth}   %
\begin{table}[t]
  \centering
  \captionsetup{justification=centering}
  \small             
  
  \makebox[\tablecolwidth][c]{%
    $\checkmark$ Good \quad
    $\sim$ Medium \quad
    $\times$ Bad
  }\vskip 1pt   
  \resizebox{\tablecolwidth}{!}{%
  \begin{tabular}{lccc|ccc}
    \toprule
    \textbf{Method} &
      \multicolumn{3}{c|}{\textbf{Query answering}} &
      \multicolumn{3}{c}{\textbf{Index building}} \\[2pt]
      & \textbf{Efficiency} & \textbf{Accuracy} & \textbf{Tuning}
      & \textbf{Efficiency} & \textbf{Footprint} & \textbf{Tuning} \\
    \midrule
    HNSW   & $\checkmark$ & $\checkmark$ & $\checkmark$ & $\checkmark$ & $\checkmark$ & $\checkmark$ \\
    VAMANA & $\checkmark$ & $\checkmark$ & $\checkmark$ & $\checkmark$ & $\checkmark$ & $\sim$       \\
    NSG    & $\checkmark$ & $\checkmark$ & $\checkmark$ & $\sim$       & $\sim$       & $\sim$       \\
    SSG    & $\checkmark$ & $\checkmark$ & $\checkmark$ & $\sim$       & $\sim$       & $\sim$       \\
        ELPIS  & $\checkmark$ & $\checkmark$ & $\sim$       & $\checkmark$ & $\checkmark$ & $\sim$       \\
    EFANNA & $\times$     & $\sim$       & $\times$     & $\times$     & $\times$     & $\times$     \\
    KGRAPH & $\times$     & $\times$     & $\times$     & $\times$     & $\times$     & $\times$     \\
    DPG    & $\times$     & $\sim$       & $\sim$       & $\sim$       & $\sim$       & $\sim$       \\
    SPTAG-BKT  & $\sim$       & $\checkmark$ & $\times$     & $\times$     & $\checkmark$ & $\times$     \\
    HCNNG  & $\checkmark$ & $\checkmark$ & $\checkmark$ & $\checkmark$ & $\checkmark$ & $\sim$       \\
    LSHAPG & $\times$     & $\sim$       & $\times$     & $\sim$       & $\checkmark$ & $\checkmark$ \\
    NGT    & $\sim$       & $\sim$       & $\times$     & $\times$     & $\checkmark$ & $\times$     \\
    SPTAG-KDT & $\sim$       & $\sim$       & $\sim$       & $\times$     & $\checkmark$ & $\times$     \\
    \bottomrule
  \end{tabular}} 

  \caption{Comparative analysis of graph-based ANN methods.}
  \label{tab:comp}
\vspace{-0.8cm}
\end{table}

\noindent{\bf Unexpected Results.} Our results lead to interesting observations that warrant further study. 

\noindent
\textit{(1) Stacked NSW:} while hierarchical layers of NSW graphs have shown promise in improving search performance on billion-scale datasets (Figure~\ref{fig:ss:search}), our experiments demonstrate that a simpler approach like K-random sampling can achieve better results on smaller and medium-sized datasets. 

\noindent
\textit{(2) Scalability of Graph Approaches:} while all graph-based vector search methods can efficiently build indexes on small datasets, most approaches face significant scalability challenges. 
Some methods (SPTAG, NGT, NSG, and SSG) demonstrate
impressive search performance on 1M and 25GB datasets (Figs. ~\ref{fig:query:performance:1M:sift:10NN}, ~\ref{fig:query:performance:1M:deep:10NN}, ~\ref{fig:query:performance:25GB:seismic:10NN},~\ref{fig:query:performance:25GB:deep:10NN}, and ~\ref{fig:query:performance:25GB:sald:10NN}) but their index construction could not scale to 100GB and billion-scale datasets. An important research direction is to improve the indexing scalability for these methods either by adopting summarization techniques during index construction or by using a scalable data structure to construct the base graph (i.e IVFPQ~\cite{faiss} to find the neighbors of nodes during insertion). 

\noindent
\textit{(3) DC-based approach for hard datasets and workloads:} an interesting finding was the superior performance of DC-based approaches compared to other methods like HNSW, NSG, and Vamana on challenging datasets/workloads such as Seismic, RandPow0, RandPow50 and Deep hard query workload for 1M and 25GB dataset sizes. We believe the DC strategy helps in this context because the graphs are built on clustered subsets of data, which facilitates beam search in retrieving more accurate answers, as opposed to running the search on the entire dataset, resulting in lower accuracy (Figures~\ref{fig:query:performance:1M:seismic:10NN}, \ref{fig:query:performance:25GB:seismic:10NN}, \ref{fig:query:performance:25GB:rand:pow1:10NN}, and~\ref{fig:query:performance:25GB:rand:pow50:10NN}).


\noindent{\bf Neighborhood Diversification.} Adopting an ND strategy to sparsify the graph \textit{always} leads to better search performance, especially as dataset size grows (Figures \ref{fig:query:performance:1M:seismic:10NN}, \ref{fig:query:performance:25GB:seismic:10NN}). Our experiments show that RND and MOND achieve the best performance overall (Figure \ref{fig:ND:search:real}). While RRND can mimic RND by setting \(\alpha=1\), relaxing pruning via \(\alpha > 1\) allows control over edge density, which benefits disk-based and in-memory searches differently. For example, DiskANN \cite{vamana} uses denser graphs to reduce disk I/O, trading off additional distance computations. We see promising improvements from increased density on hard datasets, but further theoretical work is needed to balance proximity and sparsity for efficient, well-connected graphs.

\noindent{\bf Seed Selection:} Our experiments demonstrate that the SS strategy plays a crucial role in enhancing not only search performance (Figure~\ref{fig:ss:search}) but also indexing efficiency (Table \ref{tab:ss:idx}). An important research direction is to develop novel, lightweight SS strategies. Such strategies could significantly improve the overall performance of graph-based vector search, both in terms of indexing and query-answering. Additionally, they could enhance the ability to handle out-of-distribution queries, particularly for large datasets where efficient seed selection becomes even more critical (Figures \ref{fig:ss:deep1b}, \ref{fig:ss:sift1b}).


\noindent{\bf Data-Adaptive Techniques.} Our experiments evaluate the performance of various graph-building paradigms within our taxonomy (SS, NP, II, ND, and DC). While NP-based methods perform the worst overall and are the least scalable, there is no clear winner across all dataset sizes and query workloads. \textit{(1) Scalability:} II-based approaches have superior efficiency during indexing and higher scalability in both querying and indexing.
\textit{(2) Query Answering:} ND-based methods have the best query performance overall. 
Meanwhile, DC-based approaches are superior on challenging datasets (High LID\&Low RC, Fig. \ref{fig:datacomp}) and hard query workloads \karima{(Fig. \ref{fig:query:performance:1B:t2i:10NN},
\ref{fig:query:performance:1M:seismic:10NN}, 
\ref{fig:query:performance:25GB:seismic:10NN}, 
\ref{fig:search:query:performance:25GB:hard},
\ref{fig:query:performance:25GB:rand:pow50:10NN},
\ref{fig:query:performance:25GB:rand:pow1:10NN})}. 
A promising research direction would be to develop techniques that adapt to dataset characteristics such as dataset size, dimensionality, RC and LID to excel both in indexing and query answering across a variety of query workloads and dataset sizes.

\noindent{\bf Hybrid Design.}  Most recent methods use a mix of paradigms. HNSW leverages II to scale index construction to large datasets and ND to support efficient query answering. ELPIS incorporates a DC-based strategy during both index building and search to further enhance the scalability of HNSW across varying dataset difficulty levels. Interestingly, Vamana, relying only on the ND paradigm, achieves good search performance and scalability, however its indexing time is prohibitive. 
A promising research direction is building hybrid approaches that combine the key strengths of different techniques, particularly II, ND and DC. Besides, devising novel base graphs, clustering and summarization techniques tailored for DC-based methods can further improve their performance.

\noindent{\bf Optimized Libraries.}  
Our experiments with ParlayANN~\cite{parlayann} (Fig.~\ref{fig:optimized_impl}) highlight the value of optimized implementations. For example, HCNNG\_Opt scaled to 1B datasets, while the non-optimized version failed beyond 25GB. Other methods could similarly benefit, underscoring the need for broader community support of libraries like ParlayANN.

\noindent{\bf Green Vector Search.} Our findings highlight that choices like Neighborhood Diversification (ND) and Seed Selection (SS) reduce not only search time and memory, but also the number of distance computations and overall floating-point operations, which are key contributors to energy usage. This makes them simple yet effective levers for aligning vector search with Green AI goals. We advocate for incorporating such computation-aware metrics into future evaluations to better guide the development of efficient and sustainable search methods.

\begin{figure}[tb]
  \captionsetup{justification=centering}
  \centering
    \includegraphics[width=0.9\columnwidth]{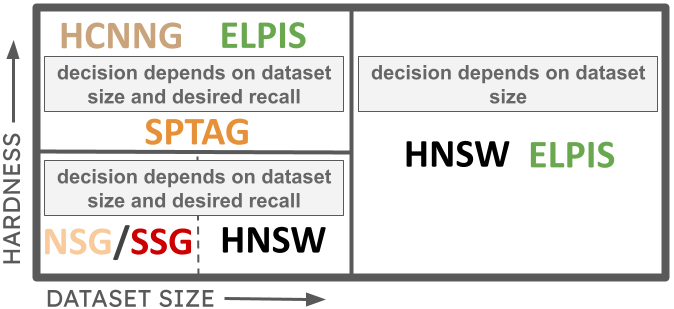}
    \vspace*{-0.4cm}
    \caption{{Recommendations (Indexing + 10K queries)} }
   \vspace*{-0.2cm}
    \label{fig:recgann}
\vspace{-0.4cm}
\end{figure}

\karima{\noindent{\bf Recommendations.}
Our study demonstrates varying performance trends across datasets of different sizes, query workloads of different hardness and desired recall values. Figure~\ref{fig:recgann} provides recommendations for methods based on these criteria.  For small to medium-sized datasets (25GB and below), II-ND and ND based methods consistently demonstrate excellent performance on easier datasets (Fig.\ref{fig:query:performance:1M:deep:10NN}, \ref{fig:query:performance:1M:sift:10NN}, \ref{fig:query:performance:1M:imagenet:10NN}, \ref{fig:query:performance:1M:gist:10NN}). On harder datasets, DC-based methods prove more efficient (Figs. \ref{fig:query:performance:1M:seismic:10NN} \ref{fig:query:performance:25GB:seismic:10NN}, \ref{fig:query:performance:1M:sald:10NN}, \ref{fig:query:performance:25GB:sald:10NN}, \ref{fig:search:query:performance:25GB:hard:1p}, \ref{fig:search:query:performance:25GB:hard:10p}). 
On large datasets (100GB and above), II-ND based methods consistently rank as top choices (Figs.\ref{fig:query:performance:100GB}, \ref{fig:query:performance:1B}). 
}

\section{Conclusions}
\label{sec:conclusions}
In this paper, we conduct a survey of the SotA graph-based methods for in-memory $ng$-approximate vector search, proposing a new taxonomy based on five key design paradigms.
%
Through extensive experimentation on datasets with up to 1B vectors, we highlight the scalability challenges faced by most methods, with incremental insertion methods showing the best scalability on datasets exceeding 100GB. 
We observe that light-weight hierarchical structures help select better seeds to start the search on billion-scale datasets, and that neighborhood diversification is a key contributor in improving the query answering performance, with RND and MOND being the best techniques. 
We also propose promising research directions.

\section*{Acknowledgements}

Supported by EU Horizon projects AI4Europe ($101070000$), TwinODIS ($101160009$), ARMADA ($101168951$), DataGEMS ($101188416$), RECITALS ($101168490$), $Y \Pi AI \Theta A$ \& NextGenerationEU project HARSH ($Y\Pi 3TA-0560901$), and CNRS-UM6P program for postdoctoral fellowships.

\section*{Impact Statement}

Paper presents work that advances field of Machine Learning; many potential societal consequences, none of which must be specifically highlighted here.

\bibliographystyle{icml2025}
\bibliography{CR/ref,CR/icde-tutorial,CR/pargis,CR/parisinmemory}

\begin{thebibliography}{96}
\providecommand{\natexlab}[1]{#1}
\providecommand{\url}[1]{\texttt{#1}}
\expandafter\ifx\csname urlstyle\endcsname\relax
  \providecommand{\doi}[1]{doi: #1}\else
  \providecommand{\doi}{doi: \begingroup \urlstyle{rm}\Url}\fi

\bibitem[url(2019)]{url/hnsw}
{Hnswlib - fast approximate nearest neighbor search}.
\newblock \url{https://github.com/nmslib/hnswlib}, 2019.

\bibitem[url(2022)]{url/Elpis}
{ELPIS Archive}.
\newblock \url{https://github.com/scalablesimilaritysearch/ELPIS}, 2022.

\bibitem[url(2025)]{url/GASS}
Graph-based vector search: An experimental evaluation of the state-of-the-art
  (archive).
\newblock \url{https://github.com/iliasazizi/GVS}, 2025.

\bibitem[Aum{\"{u}}ller \& Ceccarello(2021)Aum{\"{u}}ller and
  Ceccarello]{DBLP:journals/is/AumullerC21}
Aum{\"{u}}ller, M. and Ceccarello, M.
\newblock The role of local dimensionality measures in benchmarking nearest
  neighbor search.
\newblock \emph{Inf. Syst.}, 101:\penalty0 101807, 2021.
\newblock \doi{10.1016/J.IS.2021.101807}.
\newblock URL \url{https://doi.org/10.1016/j.is.2021.101807}.

\bibitem[Aum{\"u}ller et~al.(2017)Aum{\"u}ller, Bernhardsson, and
  Faithfull]{aumuller2017ann}
Aum{\"u}ller, M., Bernhardsson, E., and Faithfull, A.
\newblock Ann-benchmarks: A benchmarking tool for approximate nearest neighbor
  algorithms.
\newblock In \emph{International Conference on Similarity Search and
  Applications}, pp.\  34--49. Springer, 2017.

\bibitem[Aurenhammer et~al.(2013)Aurenhammer, Klein, and
  Lee]{aurenhammer2013voronoi}
Aurenhammer, F., Klein, R., and Lee, D.-T.
\newblock \emph{Voronoi diagrams and Delaunay triangulations}.
\newblock World Scientific Publishing Company, 2013.

\bibitem[Azizi et~al.(2023)Azizi, Echihabi, and Palpanas]{elpis}
Azizi, I., Echihabi, K., and Palpanas, T.
\newblock Elpis: Graph-based similarity search for scalable data science.
\newblock \emph{{PVLDB}}, 16\penalty0 (6), 2023.

\bibitem[Azizi et~al.(2025)Azizi, Echihabi, and Palpanas]{azizi2025graph}
Azizi, I., Echihabi, K., and Palpanas, T.
\newblock Graph-based vector search: An experimental evaluation of the
  state-of-the-art.
\newblock \emph{Proceedings of the ACM on Management of Data}, 3\penalty0
  (1):\penalty0 1--31, 2025.

\bibitem[Babenko \& Lempitsky(2015)Babenko and
  Lempitsky]{journal/pami/babenko15}
Babenko, A. and Lempitsky, V.
\newblock {The Inverted Multi-Index}.
\newblock \emph{TPAMI}, 37\penalty0 (6), 2015.

\bibitem[Baranchuk \& Babenko(2021)Baranchuk and Babenko]{url/data/text2image}
Baranchuk, D. and Babenko, A.
\newblock Text-to-image dataset for billion-scale similarity search.
\newblock
  \url{https://research.yandex.com/datasets/text-to-image-dataset-for-billion-scale-similarity-search},
  2021.

\bibitem[Beaumont et~al.(2007{\natexlab{a}})Beaumont, Kermarrec, Marchal, and
  Rivi{\`e}re]{voronet}
Beaumont, O., Kermarrec, A.-M., Marchal, L., and Rivi{\`e}re, {\'E}.
\newblock Voronet: A scalable object network based on voronoi tessellations.
\newblock In \emph{2007 IEEE International Parallel and Distributed Processing
  Symposium}, pp.\  1--10. IEEE, 2007{\natexlab{a}}.

\bibitem[Beaumont et~al.(2007{\natexlab{b}})Beaumont, Kermarrec, and
  Rivi{\`e}re]{beaumont07}
Beaumont, O., Kermarrec, A.-M., and Rivi{\`e}re, {\'E}.
\newblock Peer to peer multidimensional overlays: Approximating complex
  structures.
\newblock In \emph{International Conference On Principles Of Distributed
  Systems}, pp.\  315--328. Springer, 2007{\natexlab{b}}.

\bibitem[Beis \& Lowe(1997)Beis and Lowe]{kdtree}
Beis, J.~S. and Lowe, D.~G.
\newblock Shape indexing using approximate nearest-neighbour search in
  high-dimensional spaces.
\newblock In \emph{Proceedings of IEEE computer society conference on computer
  vision and pattern recognition}, pp.\  1000--1006. IEEE, 1997.

\bibitem[Blattmann et~al.(2022)Blattmann, Rombach, Oktay, M{\"u}ller, and
  Ommer]{retrieval-diffusion-models}
Blattmann, A., Rombach, R., Oktay, K., M{\"u}ller, J., and Ommer, B.
\newblock Retrieval-augmented diffusion models.
\newblock \emph{Advances in Neural Information Processing Systems}, 35, 2022.

\bibitem[Boniol \& Palpanas(2020)Boniol and Palpanas]{series2graph}
Boniol, P. and Palpanas, T.
\newblock {Series2Graph: Graph-based Subsequence Anomaly Detection for Time
  Series}.
\newblock \emph{{PVLDB}}, 2020.

\bibitem[Bubeck \& von Luxburg(2009)Bubeck and von
  Luxburg]{journal/JMLR/bubeck2009}
Bubeck, S. and von Luxburg, U.
\newblock Nearest neighbor clustering: {A} baseline method for consistent
  clustering with arbitrary objective functions.
\newblock \emph{JMLR}, 10, 2009.

\bibitem[Camerra et~al.(2014)Camerra, Shieh, Palpanas, Rakthanmanon, and
  Keogh]{isax2+}
Camerra, A., Shieh, J., Palpanas, T., Rakthanmanon, T., and Keogh, E.
\newblock {Beyond One Billion Time Series: Indexing and Mining Very Large Time
  Series Collections With $i$SAX2+}.
\newblock \emph{Knowledge and information systems}, 39\penalty0 (1):\penalty0
  123--151, 2014.

\bibitem[Chen et~al.(2018)Chen, Wang, Li, Ren, Li, Zhu, Li, Liu, Zhang, and
  Wang]{SPTAG4}
Chen, Q., Wang, H., Li, M., Ren, G., Li, S., Zhu, J., Li, J., Liu, C., Zhang,
  L., and Wang, J.
\newblock \emph{SPTAG: A library for fast approximate nearest neighbor search},
  2018.
\newblock URL \url{https://github.com/Microsoft/SPTAG}.

\bibitem[Chiang et~al.(2025)Chiang, Huang, Cheng, Tseng, Lee, and
  Wu]{chiang2025efficient}
Chiang, H.-W., Huang, C.-T., Cheng, H.-Y., Tseng, P.-H., Lee, M.-H., and Wu,
  A.-Y.
\newblock Efficient and reliable vector similarity search using asymmetric
  encoding with nand-flash for many-class few-shot learning.
\newblock In \emph{Proceedings of the 30th Asia and South Pacific Design
  Automation Conference}, pp.\  93--99, 2025.

\bibitem[Christophides et~al.(2020)Christophides, Efthymiou, Palpanas,
  Papadakis, and Stefanidis]{christophides2020overview}
Christophides, V., Efthymiou, V., Palpanas, T., Papadakis, G., and Stefanidis,
  K.
\newblock An overview of end-to-end entity resolution for big data.
\newblock \emph{ACM Computing Surveys (CSUR)}, 53\penalty0 (6):\penalty0 1--42,
  2020.

\bibitem[Croft et~al.(2021)Croft, Gupta, Josifovski, Narayanan, Wu, and
  Xue]{diskanncode}
Croft, D., Gupta, M., Josifovski, V., Narayanan, P., Wu, W., and Xue, Y.
\newblock Diskann: Fast approximate nearest neighbor search on disk.
\newblock GitHub repository, 2021.
\newblock URL \url{https://github.com/microsoft/DiskANN}.
\newblock Accessed: 2024-10-25.

\bibitem[Dasgupta \& Freund(2008)Dasgupta and Freund]{dasgupta2008random}
Dasgupta, S. and Freund, Y.
\newblock Random projection trees and low dimensional manifolds.
\newblock In \emph{Proceedings of the fortieth annual ACM symposium on Theory
  of computing}, pp.\  537--546, 2008.

\bibitem[DBAIWangGroup(2023)]{dpgrepo}
DBAIWangGroup.
\newblock Nns benchmark: Evaluating approximate nearest neighbor search
  algorithms in high dimensional euclidean space - dpg algorithm.
\newblock
  \url{https://github.com/DBAIWangGroup/nns_benchmark/tree/master/algorithms/DPG},
  2023.
\newblock GitHub repository.

\bibitem[Desislavov et~al.(2023)]{ai-inference-energy}
Desislavov, R. et~al.
\newblock Trends in ai inference energy consumption: Beyond the
  performance-vs-parameter laws of deep learning.
\newblock \emph{Sustainable Computing: Informatics and Systems}, 38, 2023.

\bibitem[Dobkin et~al.(1990)Dobkin, Friedman, and Supowit]{dobkin1990delaunay}
Dobkin, D.~P., Friedman, S.~J., and Supowit, K.~J.
\newblock Delaunay graphs are almost as good as complete graphs.
\newblock \emph{Discrete \& Computational Geometry}, 5\penalty0 (4):\penalty0
  399--407, 1990.

\bibitem[Dong(2022)]{kgraph}
Dong, W.
\newblock {Kgraph, an open source library for k-nn graph construction and
  nearest neighbor search}.
\newblock \url{www.kgraph.org}, 2022.

\bibitem[Dong et~al.(2011)Dong, Moses, and Li]{nndescent}
Dong, W., Moses, C., and Li, K.
\newblock Efficient k-nearest neighbor graph construction for generic
  similarity measures.
\newblock In \emph{Proceedings of the 20th international conference on World
  wide web}, pp.\  577--586, 2011.

\bibitem[Echihabi et~al.(2019)Echihabi, Zoumpatianos, Palpanas, and
  Benbrahim]{lernaeanhydra2}
Echihabi, K., Zoumpatianos, K., Palpanas, T., and Benbrahim, H.
\newblock {Return of the Lernaean Hydra: Experimental Evaluation of Data Series
  Approximate Similarity Search}.
\newblock \emph{{PVLDB}}, 2019.

\bibitem[Echihabi et~al.(2021)Echihabi, Zoumpatianos, and
  Palpanas]{conf/icde/echihabi2021}
Echihabi, K., Zoumpatianos, K., and Palpanas, T.
\newblock High-dimensional similarity search for scalable data science.
\newblock ICDE, 2021.

\bibitem[Echihabi et~al.(2022)Echihabi, Fatourou, Zoumpatianos, Palpanas, and
  Benbrahim]{hercules}
Echihabi, K., Fatourou, P., Zoumpatianos, K., Palpanas, T., and Benbrahim, H.
\newblock {Hercules Against Data Series Similarity Search}.
\newblock \emph{{PVLDB}}, 15\penalty0 (10), 2022.

\bibitem[Edelsbrunner(2012)]{edelsbrunner87}
Edelsbrunner, H.
\newblock \emph{Algorithms in Combinatorial Geometry}.
\newblock Springer Publishing Company, Incorporated, 1st edition, 2012.
\newblock ISBN 3642648738.

\bibitem[Engel et~al.(2004)Engel, Monasson, and Hartmann]{erconnect}
Engel, A., Monasson, R., and Hartmann, A.~K.
\newblock On large deviation properties of erd{\"o}s--r{\'e}nyi random graphs.
\newblock \emph{Journal of Statistical Physics}, 117\penalty0 (3):\penalty0
  387--426, 2004.

\bibitem[Ferhatosmanoglu et~al.(2000)Ferhatosmanoglu, Tuncel, Agrawal, and
  El~Abbadi]{va+file}
Ferhatosmanoglu, H., Tuncel, E., Agrawal, D., and El~Abbadi, A.
\newblock Vector approximation based indexing for non-uniform high dimensional
  data sets.
\newblock In \emph{Proceedings of the ninth international conference on
  Information and knowledge management}, pp.\  202--209, 2000.

\bibitem[for Seismology~with Artificial~Intelligence(2018)]{url/data/seismic}
for Seismology~with Artificial~Intelligence, I. R.~I.
\newblock {Seismic Data Access}.
\newblock \url{http://ds.iris.edu/data/access/}, 2018.

\bibitem[Fortune(1995)]{vd95}
Fortune, S.
\newblock Voronoi diagrams and delaunay triangulations.
\newblock \emph{Computing in Euclidean geometry}, pp.\  225--265, 1995.

\bibitem[Fu \& Cai(2016)Fu and Cai]{efanna}
Fu, C. and Cai, D.
\newblock Efanna: An extremely fast approximate nearest neighbor search
  algorithm based on knn graph.
\newblock \emph{arXiv preprint arXiv:1609.07228}, 2016.

\bibitem[Fu et~al.(2019)Fu, Xiang, Wang, and Cai]{nsg}
Fu, C., Xiang, C., Wang, C., and Cai, D.
\newblock Fast approximate nearest neighbor search with the navigating
  spreading-out graph.
\newblock \emph{Proc. {VLDB} Endow.}, 12\penalty0 (5):\penalty0 461--474, 2019.

\bibitem[Fu et~al.(2021)Fu, Wang, and Cai]{nssg}
Fu, C., Wang, C., and Cai, D.
\newblock High dimensional similarity search with satellite system graph:
  Efficiency, scalability, and unindexed query compatibility.
\newblock \emph{IEEE Transactions on Pattern Analysis and Machine
  Intelligence}, 2021.

\bibitem[Gabriel \& Sokal(1969)Gabriel and Sokal]{gabriel69}
Gabriel, K.~R. and Sokal, R.~R.
\newblock A new statistical approach to geographic variation analysis.
\newblock \emph{Systematic zoology}, 18\penalty0 (3):\penalty0 259--278, 1969.

\bibitem[Gionis et~al.(1999)Gionis, Indyk, Motwani, et~al.]{iehlsh}
Gionis, A., Indyk, P., Motwani, R., et~al.
\newblock Similarity search in high dimensions via hashing.
\newblock In \emph{Vldb}, volume~99, pp.\  518--529, 1999.

\bibitem[Gogolou et~al.(2020)Gogolou, Tsandilas, Echihabi, Palpanas, and
  Bezerianos]{conf/sigmod/gogolou20}
Gogolou, A., Tsandilas, T., Echihabi, K., Palpanas, T., and Bezerianos, A.
\newblock {Data Series Progressive Similarity Search with Probabilistic Quality
  Guarantees}.
\newblock In \emph{SIGMOD}, 2020.

\bibitem[He et~al.(2012)He, Kumar, and Chang]{rc}
He, J., Kumar, S., and Chang, S.-F.
\newblock On the difficulty of nearest neighbor search.
\newblock \emph{arXiv preprint arXiv:1206.6411}, 2012.

\bibitem[He et~al.(2016)He, Zhang, Ren, and Sun]{resnet}
He, K., Zhang, X., Ren, S., and Sun, J.
\newblock Deep residual learning for image recognition.
\newblock In \emph{Proceedings of the IEEE conference on computer vision and
  pattern recognition}, pp.\  770--778, 2016.

\bibitem[Huang et~al.(2024)Huang, Chang, Cheng, and Wu]{Huang24}
Huang, C.-T., Chang, C.-Y., Cheng, H.-Y., and Wu, A.-Y.
\newblock Bore: Energy-efficient banded vector similarity search with optimized
  range encoding for memory-augmented neural network.
\newblock In \emph{2024 Design, Automation \& Test in Europe Conference \&
  Exhibition (DATE)}, pp.\  1--6, 2024.

\bibitem[Iwasaki(2016)]{ngtpanng1}
Iwasaki, M.
\newblock Pruned bi-directed k-nearest neighbor graph for proximity search.
\newblock In \emph{International Conference on Similarity Search and
  Applications}, pp.\  20--33. Springer, 2016.

\bibitem[J{\'e}gou et~al.(2011)J{\'e}gou, Douze, and Schmid]{gist}
J{\'e}gou, H., Douze, M., and Schmid, C.
\newblock Product quantization for nearest neighbor search.
\newblock In \emph{IEEE Transactions on Pattern Analysis and Machine
  Intelligence}, volume~33, pp.\  117--128. IEEE, 2011.

\bibitem[Jin et~al.(2014)Jin, Zhang, Hu, Lin, Cai, and He]{ieh}
Jin, Z., Zhang, D., Hu, Y., Lin, S., Cai, D., and He, X.
\newblock Fast and accurate hashing via iterative nearest neighbors expansion.
\newblock \emph{IEEE transactions on cybernetics}, 44\penalty0 (11):\penalty0
  2167--2177, 2014.

\bibitem[Johnson et~al.(2019)Johnson, Douze, and J{\'e}gou]{faiss}
Johnson, J., Douze, M., and J{\'e}gou, H.
\newblock Billion-scale similarity search with {GPUs}.
\newblock \emph{IEEE Transactions on Big Data}, 7\penalty0 (3):\penalty0
  535--547, 2019.

\bibitem[Kaneko(2023)]{kaneko2023local}
Kaneko, H.
\newblock Local interpretation of nonlinear regression model with k-nearest
  neighbors.
\newblock \emph{Digital Chemical Engineering}, 6:\penalty0 100078, 2023.

\bibitem[Karpukhin et~al.(2020)Karpukhin, O{\u{g}}uz, Min, Lewis, Wu, Edunov,
  Chen, and Yih]{dense-passage-retrieval}
Karpukhin, V., O{\u{g}}uz, B., Min, S., Lewis, P., Wu, L., Edunov, S., Chen,
  D., and Yih, W.-t.
\newblock Dense passage retrieval for open-domain question answering.
\newblock \emph{arXiv preprint arXiv:2004.04906}, 2020.

\bibitem[Kleinberg et~al.(2002)]{kleinberg2002}
Kleinberg, J. et~al.
\newblock Small-world phenomena and the dynamics of information.
\newblock \emph{Advances in neural information processing systems}, 1:\penalty0
  431--438, 2002.

\bibitem[Kleinberg(2000)]{kleinberg2000}
Kleinberg, J.~M.
\newblock Navigation in a small world.
\newblock \emph{Nature}, 406\penalty0 (6798):\penalty0 845--845, 2000.

\bibitem[Li et~al.(2019)Li, Zhang, Sun, Wang, Li, Zhang, and Lin]{dpg}
Li, W., Zhang, Y., Sun, Y., Wang, W., Li, M., Zhang, W., and Lin, X.
\newblock {Approximate nearest neighbor search on high dimensional data:
  experiments, analyses, and improvement}.
\newblock \emph{IEEE Transactions on Knowledge and Data Engineering},
  32\penalty0 (8):\penalty0 1475--1488, 2019.

\bibitem[Lu(2023)]{hvsgithub}
Lu, K.
\newblock Hvs: Hierarchical graph structure based on voronoi diagrams for
  solving approximate nearest neighbor search, 2023.
\newblock URL \url{https://github.com/Kejing-Lu/hvs}.

\bibitem[Lu et~al.(2021)Lu, Kudo, Xiao, and Ishikawa]{hvs}
Lu, K., Kudo, M., Xiao, C., and Ishikawa, Y.
\newblock Hvs: hierarchical graph structure based on voronoi diagrams for
  solving approximate nearest neighbor search.
\newblock \emph{Proceedings of the VLDB Endowment}, 15\penalty0 (2):\penalty0
  246--258, 2021.

\bibitem[Malinen \& Fr{\"a}nti(2014)Malinen and Fr{\"a}nti]{bkmtree}
Malinen, M.~I. and Fr{\"a}nti, P.
\newblock Balanced k-means for clustering.
\newblock In \emph{Structural, Syntactic, and Statistical Pattern Recognition:
  Joint IAPR International Workshop, S+ SSPR 2014, Joensuu, Finland, August
  20-22, 2014. Proceedings}, pp.\  32--41. Springer, 2014.

\bibitem[Malkov et~al.(2014)Malkov, Ponomarenko, Logvinov, and Krylov]{nsw14}
Malkov, Y., Ponomarenko, A., Logvinov, A., and Krylov, V.
\newblock Approximate nearest neighbor algorithm based on navigable small world
  graphs.
\newblock \emph{Information Systems}, 45:\penalty0 61--68, 2014.

\bibitem[Malkov \& Yashunin(2020)Malkov and Yashunin]{hnsw}
Malkov, Y.~A. and Yashunin, D.~A.
\newblock Efficient and robust approximate nearest neighbor search using
  hierarchical navigable small world graphs.
\newblock \emph{{IEEE} Trans. Pattern Anal. Mach. Intell.}, 42\penalty0
  (4):\penalty0 824--836, 2020.

\bibitem[Manohar et~al.(2024)Manohar, Shen, Blelloch, Dhulipala, Gu, Simhadri,
  and Sun]{parlayann}
Manohar, M.~D., Shen, Z., Blelloch, G., Dhulipala, L., Gu, Y., Simhadri, H.~V.,
  and Sun, Y.
\newblock Parlayann: Scalable and deterministic parallel graph-based
  approximate nearest neighbor search algorithms.
\newblock In \emph{Proceedings of the 29th ACM SIGPLAN Annual Symposium on
  Principles and Practice of Parallel Programming}, pp.\  270--285, 2024.

\bibitem[Matula \& Sokal(1980)Matula and Sokal]{matula80}
Matula, D.~W. and Sokal, R.~R.
\newblock Properties of gabriel graphs relevant to geographic variation
  research and the clustering of points in the plane.
\newblock \emph{Geographical analysis}, 12\penalty0 (3):\penalty0 205--222,
  1980.

\bibitem[Morozov \& Babenko(2018)Morozov and Babenko]{mipsg}
Morozov, S. and Babenko, A.
\newblock Non-metric similarity graphs for maximum inner product search.
\newblock \emph{Advances in Neural Information Processing Systems}, 31, 2018.

\bibitem[Munoz et~al.(2019)Munoz, Gon{\c{c}}alves, Dias, and Torres]{hcnng}
Munoz, J.~V., Gon{\c{c}}alves, M.~A., Dias, Z., and Torres, R. d.~S.
\newblock Hierarchical clustering-based graphs for large scale approximate
  nearest neighbor search.
\newblock \emph{Pattern Recognition}, 96:\penalty0 106970, 2019.

\bibitem[Oliva \& Torralba(2001)Oliva and Torralba]{gistdesc}
Oliva, A. and Torralba, A.
\newblock Modeling the shape of the scene: A holistic representation of the
  spatial envelope.
\newblock \emph{International journal of computer vision}, 42:\penalty0
  145--175, 2001.

\bibitem[Palpanas(2015)]{palpanas2015data}
Palpanas, T.
\newblock Data series management: The road to big sequence analytics.
\newblock \emph{ACM SIGMOD Record}, 44\penalty0 (2):\penalty0 47--52, 2015.

\bibitem[Palpanas \& Beckmann(2019)Palpanas and Beckmann]{itsawreport}
Palpanas, T. and Beckmann, V.
\newblock {Report on the First and Second Interdisciplinary Time Series
  Analysis Workshop (ITISA)}.
\newblock \emph{{ACM SIGMOD Record}}, 48\penalty0 (3), 2019.

\bibitem[Petitjean et~al.(2014)Petitjean, Forestier, Webb, Nicholson, Chen, and
  Keogh]{DBLP:conf/icdm/PetitjeanFWNCK14}
Petitjean, F., Forestier, G., Webb, G.~I., Nicholson, A.~E., Chen, Y., and
  Keogh, E.~J.
\newblock Dynamic time warping averaging of time series allows faster and more
  accurate classification.
\newblock In \emph{{ICDM}}, 2014.

\bibitem[Ponomarenko et~al.(2011)Ponomarenko, Malkov, Logvinov, and
  Krylov]{nsw11}
Ponomarenko, A., Malkov, Y., Logvinov, A., and Krylov, V.
\newblock Approximate nearest neighbor search small world approach.
\newblock In \emph{International Conference on Information and Communication
  Technologies \& Applications}, volume~17, 2011.

\bibitem[Pugh(1990)]{skiplist}
Pugh, W.
\newblock Skip lists: a probabilistic alternative to balanced trees.
\newblock \emph{Communications of the ACM}, 33\penalty0 (6):\penalty0 668--676,
  1990.

\bibitem[Reddy et~al.(1977)]{reddy77bm}
Reddy, D.~R. et~al.
\newblock Speech understanding systems: A summary of results of the five-year
  research effort.
\newblock \emph{Department of Computer Science. Camegie-Mell University,
  Pittsburgh, PA}, 17:\penalty0 138, 1977.

\bibitem[Russakovsky et~al.(2015)Russakovsky, Deng, Su, Krause, Satheesh, Ma,
  Huang, Karpathy, Khosla, Bernstein, et~al.]{imagenet}
Russakovsky, O., Deng, J., Su, H., Krause, J., Satheesh, S., Ma, S., Huang, Z.,
  Karpathy, A., Khosla, A., Bernstein, M., et~al.
\newblock Imagenet large scale visual recognition challenge.
\newblock \emph{International journal of computer vision}, 115:\penalty0
  211--252, 2015.

\bibitem[Shamos \& Hoey(1975)Shamos and Hoey]{shamos1975closest}
Shamos, M.~I. and Hoey, D.
\newblock Closest-point problems.
\newblock In \emph{16th Annual Symposium on Foundations of Computer Science
  (sfcs 1975)}, pp.\  151--162. IEEE, 1975.

\bibitem[Simhadri et~al.(2022)Simhadri, Williams, Aum{\"{u}}ller, Douze,
  Babenko, Baranchuk, Chen, Hosseini, Krishnaswamy, Srinivasa, Subramanya, and
  Wang]{neurips-2021-ann-competition}
Simhadri, H.~V., Williams, G., Aum{\"{u}}ller, M., Douze, M., Babenko, A.,
  Baranchuk, D., Chen, Q., Hosseini, L., Krishnaswamy, R., Srinivasa, G.,
  Subramanya, S.~J., and Wang, J.
\newblock Results of the neurips'21 challenge on billion-scale approximate
  nearest neighbor search.
\newblock \emph{CoRR}, abs/2205.03763, 2022.
\newblock \doi{10.48550/arXiv.2205.03763}.
\newblock URL \url{https://doi.org/10.48550/arXiv.2205.03763}.

\bibitem[{Skoltech Computer Vision}(2018)]{url/data/deep1b}
{Skoltech Computer Vision}.
\newblock {Deep billion-scale indexing}.
\newblock \url{http://sites.skoltech.ru/compvision/noimi}, 2018.

\bibitem[Song et~al.(2020)Song, Pan, Zhao, Yang, Chen, Zhang, Xu, and
  Jin]{alibabaknngml}
Song, L., Pan, P., Zhao, K., Yang, H., Chen, Y., Zhang, Y., Xu, Y., and Jin, R.
\newblock Large-scale training system for 100-million classification at
  alibaba.
\newblock In \emph{Proceedings of the 26th ACM SIGKDD International Conference
  on Knowledge Discovery \& Data Mining}, pp.\  2909--2930, 2020.

\bibitem[Subramanya et~al.(2019)Subramanya, Kadekodi, Krishaswamy, and
  Simhadri]{vamana}
Subramanya, S.~J., Kadekodi, R., Krishaswamy, R., and Simhadri, H.~V.
\newblock Diskann: Fast accurate billion-point nearest neighbor search on a
  single node.
\newblock In \emph{Proceedings of the 33rd International Conference on Neural
  Information Processing Systems}, pp.\  13766--13776, 2019.

\bibitem[Sugawara et~al.(2016)Sugawara, Kobayashi, and Iwasaki]{ngtpanng2}
Sugawara, K., Kobayashi, H., and Iwasaki, M.
\newblock On approximately searching for similar word embeddings.
\newblock In \emph{Proceedings of the 54th Annual Meeting of the Association
  for Computational Linguistics (Volume 1: Long Papers)}, pp.\  2265--2275,
  2016.

\bibitem[Sun et~al.(2014{\natexlab{a}})Sun, Wang, Qin, Zhang, and
  Lin]{conf/vldb/sun14}
Sun, Y., Wang, W., Qin, J., Zhang, Y., and Lin, X.
\newblock {SRS: Solving c-approximate Nearest Neighbor Queries in High
  Dimensional Euclidean Space with a Tiny Index}.
\newblock \emph{PVLDB}, 8\penalty0 (1), 2014{\natexlab{a}}.

\bibitem[Sun et~al.(2014{\natexlab{b}})Sun, Wang, Qin, Zhang, and Lin]{srs}
Sun, Y., Wang, W., Qin, J., Zhang, Y., and Lin, X.
\newblock {SRS: solving c-approximate nearest neighbor queries in high
  dimensional euclidean space with a tiny index}.
\newblock \emph{Proceedings of the VLDB Endowment}, 2014{\natexlab{b}}.

\bibitem[Tao et~al.(2009)Tao, Yi, Sheng, and Kalnis]{lsb}
Tao, Y., Yi, K., Sheng, C., and Kalnis, P.
\newblock Quality and efficiency in high dimensional nearest neighbor search.
\newblock In \emph{Proceedings of the 2009 ACM SIGMOD International Conference
  on Management of data}, pp.\  563--576, 2009.

\bibitem[Team(2023)]{parlayanncode}
Team, T.~P.
\newblock Parlayann: A deep learning library for parallel computation.
\newblock \url{https://github.com/parlayann/parlayann}, 2023.
\newblock Accessed: 2024-10-25.

\bibitem[{TEXMEX Research Team}(2018)]{url/data/sift}
{TEXMEX Research Team}.
\newblock {Datasets for approximate nearest neighbor search}.
\newblock \url{http://corpus-texmex.irisa.fr/}, 2018.

\bibitem[Toussaint(1980)]{rng}
Toussaint, G.~T.
\newblock The relative neighbourhood graph of a finite planar set.
\newblock \emph{Pattern recognition}, 12\penalty0 (4):\penalty0 261--268, 1980.

\bibitem[Toussaint(2002)]{toussaint02}
Toussaint, G.~T.
\newblock Proximity graphs for nearest neighbor decision rules: recent
  progress.
\newblock \emph{Interface}, 34, 2002.

\bibitem[University(2018)]{url/data/eeg}
University, S.
\newblock {Southwest University Adult Lifespan Dataset (SALD)}.
\newblock
  \url{http://fcon_1000.projects.nitrc.org/indi/retro/sald.html?utm_source=newsletter&utm_medium=email&utm_content=See20Data&utm_campaign=indi-1},
  2018.

\bibitem[Wang et~al.(2013)Wang, Wang, Jia, Li, Zeng, Zha, and Hua]{tptree}
Wang, J., Wang, N., Jia, Y., Li, J., Zeng, G., Zha, H., and Hua, X.-S.
\newblock Trinary-projection trees for approximate nearest neighbor search.
\newblock \emph{IEEE transactions on pattern analysis and machine
  intelligence}, 36\penalty0 (2):\penalty0 388--403, 2013.

\bibitem[Wang et~al.(2018)Wang, Huang, Zhao, Zhang, Zhao, and
  Lee]{conf/kdd/wang2018}
Wang, J., Huang, P., Zhao, H., Zhang, Z., Zhao, B., and Lee, D.~L.
\newblock Billion-scale commodity embedding for e-commerce recommendation in
  alibaba.
\newblock In \emph{KDD}, 2018.

\bibitem[Wang et~al.(2021)Wang, Xu, Yue, and Wang]{graph-survey-vldb}
Wang, M., Xu, X., Yue, Q., and Wang, Y.
\newblock A comprehensive survey and experimental comparison of graph-based
  approximate nearest neighbor search.
\newblock \emph{Proc. VLDB Endow.}, 14\penalty0 (11):\penalty0 1964–1978, jul
  2021.
\newblock ISSN 2150-8097.
\newblock \doi{10.14778/3476249.3476255}.
\newblock URL \url{https://doi.org/10.14778/3476249.3476255}.

\bibitem[Wang et~al.(2025)Wang, Ileana, and Palpanas]{leafi}
Wang, Q., Ileana, I., and Palpanas, T.
\newblock {LeaFi: Data Series Indexes on Steroids with Learned Filters}.
\newblock \emph{Proc. {ACM} Manag. Data}, 2025.

\bibitem[Watts \& Strogatz(1998)Watts and Strogatz]{watts98}
Watts, D.~J. and Strogatz, S.~H.
\newblock Collective dynamics of ‘small-world’networks.
\newblock \emph{nature}, 393\penalty0 (6684):\penalty0 440--442, 1998.

\bibitem[Weber et~al.(1998)Weber, Schek, and Blott]{vafile}
Weber, R., Schek, H.-J., and Blott, S.
\newblock {A Quantitative Analysis and Performance Study for Similarity-Search
  Methods in High-Dimensional Spaces}.
\newblock In \emph{Proc. VLDB}, pp.\  194--205, 1998.

\bibitem[{Williams} et~al.(2014){Williams}, {Li}, {Khabsa}, {Wu}, {Shih}, and
  {Giles}]{conf/williams2014}
{Williams}, K., {Li}, L., {Khabsa}, M., {Wu}, J., {Shih}, P.~C., and {Giles},
  C.~L.
\newblock A web service for scholarly big data information extraction.
\newblock In \emph{ICWS}, 2014.

\bibitem[{Yahoo Japan Corporation}(2022)]{ngt_library}
{Yahoo Japan Corporation}.
\newblock Ngt: Neighborhood graph and tree for high-dimensional data.
\newblock \url{https://github.com/yahoojapan/NGT}, 2022.
\newblock Accessed: 2024-10-20.

\bibitem[Yianilos(1993)]{vptree}
Yianilos, P.~N.
\newblock Data structures and algorithms for nearest neighbor search in general
  metric spaces.
\newblock In \emph{Soda}, volume~93, pp.\  311--21, 1993.

\bibitem[Zhang et~al.(2013)Zhang, Huang, Geng, and Liu]{iehitq}
Zhang, Y.-M., Huang, K., Geng, G., and Liu, C.-L.
\newblock Fast knn graph construction with locality sensitive hashing.
\newblock In \emph{Joint European Conference on Machine Learning and Knowledge
  Discovery in Databases}, pp.\  660--674. Springer, 2013.

\bibitem[Zhao et~al.(2023)Zhao, Tian, Huang, Zheng, and Zhou]{lshapg}
Zhao, X., Tian, Y., Huang, K., Zheng, B., and Zhou, X.
\newblock Towards efficient index construction and approximate nearest neighbor
  search in high-dimensional spaces.
\newblock \emph{Proceedings of the VLDB Endowment}, 16\penalty0 (8):\penalty0
  1979--1991, 2023.

\bibitem[Zoumpatianos et~al.(2018)Zoumpatianos, Lou, Ileana, Palpanas, and
  Gehrke]{johannesjoural2018}
Zoumpatianos, K., Lou, Y., Ileana, I., Palpanas, T., and Gehrke, J.
\newblock Generating data series query workloads.
\newblock \emph{The VLDB Journal}, 27\penalty0 (6):\penalty0 823--846, December
  2018.
\newblock ISSN 1066-8888.
\newblock \doi{10.1007/s00778-018-0513-x}.
\newblock URL \url{https://doi.org/10.1007/s00778-018-0513-x}.

\end{thebibliography}

\appendix
\section{State-of-the-Art Approaches}
This section presents the graph-based approximate nearest neighbor (ANN) methods studied in our experiments, highlighting their core principles:

\noindent{\bf KGraph}~\cite{kgraph} reduces the construction cost of an exact k-NNG, which has a quadratic worst-case complexity. It constructs an approximate k-NNG by refining a random initial graph with an empirical cost of \( O\left(n^{1.14}\right) \)~\cite{nndescent}. 
This refinement process, also known as NNDescent~\cite{nndescent} (Neighborhood Propagation), aims at improving the approximation of the \( k \)-NN graph by assuming that the neighbors of a vertex \( u \) are more likely to be neighbors of each other. 
The process iterates over all graph vertices \( u \in {V} \): 
for each vertex \( u \) and pair \( (x,y) \) of its neighbors, it adds \( x \) to the neighbors of \( y \) and vice-versa, keeping the closest \( k \) neighbors of \( u \).

 \noindent{\bf Navigable Small World (NSW)}~\cite{nsw11,nsw14} is an approximation of a Delaunay graph which guarantees the small world property~\cite{watts98}, i.e. the number of hops $L$ between two randomly chosen vertices grows to the logarithm of graph size $n$ such that $L \propto Log\left(n\right)$.
An {\it NSW} graph is based on the VoroNet graph~\cite{voronet}, an extention of Kleinberg's variant of Watts-Strogatz's small world model graph~\cite{kleinberg2000,kleinberg2002},  The VoroNet graph is built incrementally by inserting a randomly picked vertex to the graph and connecting it to 2d+1 
neighbors selected using a beam search on the existing vertices in the graph.
Once this process completes, the first built edges would serve as long-range edges to quickly converge toward nearest neighbors~\cite{voronet}. The resulting graph was proved to guarantee the small world network property~\cite{voronet,beaumont07}. 

 \noindent{\bf Iterative Expanding Hashing (IEH)}~\cite{ieh} follows the same process as KGraph to construct an approximate k-NNG; however, it refines an initial graph where the candidates for each node are generated using a hashing function.
Two extensions of IEH have been proposed to better leverage advanced hashing methods for generating initial candidates: IEH-LSH~\cite{iehlsh} and IEH-ITQ~\cite{iehitq}. All these methods use NNDescent to finalize the graph connections.

 \noindent{\bf EFANNA}~\cite{efanna} selects seeds similarly to KGraph~\cite{kgraph} and IEH~\cite{ieh} and refines candidates using NNdescent. It builds an approximate $k$-NNG by selecting initial neighbors of each node using randomized truncated K-D Trees \cite{dasgupta2008random} and refining the graph using NNDescent~\cite{nndescent}. 
During search, EFANNA uses the pre-built trees to select seeds, then
runs a beam search on the graph index.


\noindent{\bf Hierarchical Navigable Small World (HNSW)}~\cite{hnsw} improves the scalability of NSW~\cite{nsw11,nsw14}  by proposing RND to sparsify the graph and a hierarchical seed selection strategy (SN) 
to shorten the search path during index building and query answering. Each hierarchical layer includes all nodes in the layer above it, with the bottom (a.k.a. {\it base}) layer containing all points of the dataset \( {S} \), 
HNSW builds an NSW graph incrementally. However, HNSW diverges from NSW in that it refines the candidate nearest neighbors, identified through beam search on the nodes already in that layer using RND. 
During query answering, HNSW utilizes SN to quickly find an entry point in the base layer to start the beam search.


\noindent{\bf Diversified Proximity Graph (DPG)}~\cite{dpg} 
extends KGraph~\cite{kgraph} by diversifying the neighborhoods of its nodes through edge orientation, a technique we refer to as Maximum-Oriented Neighborhood Diversification (MOND) in Section 3.4.
MOND’s main objective is to maximize the angles between neighboring nodes, contributing to a sparsed graph structure. This process is iteratively applied to all nodes. After that, the directed graph is transformed into an undirected one, enhancing its connectivity. Nevertheless, note that DPG's publicly available implementation~\cite{dpgrepo} utilizes RND rather than MOND for neighborhood diversification.

\noindent{\bf NGT}~\cite{ngt_library} is an approximate nearest neighbor (ANN) search library developed by Yahoo Japan. It offers two construction methods: one extends KNN graphs with reverse edges, forming bi-directed KNN graphs~\cite{ngtpanng1}, while the other incrementally builds graphs similar to HNSW with a range-based search strategy~\cite{ngtpanng2}.  In this study, we consider the former~\cite{ngtpanng1}. Additionally, the library includes methods that employ quantization for highly efficient search.
NGT maintains efficiency by pruning neighbors via RND and using Vantage-Point Trees~\cite{vptree} to select seed nodes for accurate query results.

\noindent{\bf Navigating Spreading-out Graph (NSG)}~\cite{nsg}, similarly to DPG, builds an approximate k-NNG first. But, unlike DPG, it builds an EFANNA graph rather than a KGraph. It then diversifies the graph using RND. 
At the end, NSG creates a depth-first search tree to verify the connectivity of the graph. If there is a vertex that is disconnected from the tree, NSG connects it to the nearest node in the tree to ensure graph connectivity.

\noindent{\bf SPTAG}~\cite{SPTAG4} is a library for approximate vector search proposed by Microsoft. 
SPTAG follows a DC approach and is based on multiple existing works.  
It selects small dataset samples on which it builds either K-D Trees~\cite{kdtree} or Balanced K-means Trees~\cite{bkmtree}. These strutures will be used for seed selection during query answering. Then it clusters the full dataset using multiple hierarchical random divisions of TP Trees~\cite{tptree}, builds an exact k-NN graph on each cluster (i.e., leaf) and refines each graph using ND. The graphs are merged into one large graph index for query processing. 

\noindent{\bf Vamana}~\cite{vamana} is similar to NSG in considering the set of visited nodes when building long-range edges within the graph. However, instead of using EFANNA~\cite{efanna}, Vamana uses a randomly generated graph with node degree~$\geq~log\left(n\right)$ to ensure the initial graph connectivity~\cite{erconnect}. 
Then, for each node, Vamana runs a beam search on the graph structure to get the visited node list $R$, which will be refined in the first round using RRND. After adding bi-directional edges to selected neighbors, the neighbors that exceed the maximum allowed out-degree will refine their neighborhood list following an RND process. Then, Vamana repeats the same refinement process a second time to improve the graph quality, this time using RRND with $\alpha \geq 1$ to increase the connectivity within the graph.

\noindent{\bf SSG}~\cite{nssg} integrates the MOND approach from DPG~\cite{dpg} and closely follows the steps of NSG~\cite{nsg} and DPG~\cite{dpg} in index building from a foundational graph. Instead of performing a search for each node to acquire candidates, SSG~\cite{nssg} employs a breadth-first search on each node to assemble candidate neighbors through local expansion on a base graph (EFANNA). When the maximum size for the candidate neighbors is achieved, SSG reduces the neighbors in the list by enforcing the MOND diversification strategy, pruning the candidate nodes forming an angle smaller than a user-defined parameter $\theta$ with the already existing neighbors of the concerned node. After iteratively applying this method to all nodes, SSG~\cite{nssg} enhances connectivity by constructing multiple DFS trees from various random points, in contrast to NSG's~\cite{nsg} singular DFS approach.

\noindent{\bf Hierarchical Clustering-based Nearest Neighbor Graph (HCNNG)}~\cite{hcnng} was inspired by SPTAG. It employs hierarchical clustering to randomly divide the dataset into multiple subsets. This subdivision process is executed several times, resulting in a collection of intersecting subsets. On each subset, HCNNG constructs a Minimum Spanning Tree (MST) graph. 
Following this, the vertices and edges from all the MSTs are merged to form a single, connected graph. 
To facilitate the search process, HCNNG constructs multiple K-D Trees~\cite{kdtree}, to identifying entry points during query search. 

\noindent{\bf HVS}~\cite{hvs} extends HNSW's base layer by refining the construction of hierarchical layers. Instead of random selection, nodes are assigned to layers based on local density to better capture data distribution. Each layer forms a Voronoi diagram 
and uses multi-level quantization, increasing dimensionality by a factor of 2 in each lower layer.
Search at the base layer is similar to that of HNSW.

\noindent{\bf LSHAPG}~\cite{lshapg} combines HNSW graphs with multiple hash tables based on the LSB-Tree structure~\cite{lsb} to enhance search efficiency. It leverages $L$ hash tables to retrieve seeds for beam search on the base layer, unlike HNSW, which selects a single seed through SN. LSHAPG also utilizes these hash tables for probabilistic rooting during search, pruning neighbors based on the projected distance 
before evaluating and pruning the raw vectors.
\noindent{\bf ELPIS}~\cite{elpis} is a DC-based approach that splits the dataset into subsets using the Hercules EAPCA tree~\cite{hercules}, where each leaf corresponds to a different subset, then builds in parallel a graph-based index for each leaf using HNSW~\cite{hnsw}. During search, ELPIS first selects heuristically an initial leaf 
and executes a beam search on its respective graph. 
The retrieved set of answers feed the search priority queues for the other leaves. 
Only a subset of leaves is selected based on the answers and the lower-bounding distances of the query to the EAPCA summarization of each leaf.  Then, ELPIS initiates multiple concurrent beam searches on the graph structures of the candidate leaves. 
Finally, ELPIS aggregates all results from candidate clusters and returns the top-k answers. 

\section{Experimental Results}
This section presents detailed experimental results across various datasets and sizes. We analyze key design choices in Neighborhood Diversification and Seed Selection, assessing their impact on query and indexing performance. We also evaluate twelve state-of-the-art methods on real and synthetic datasets of varying scale, dimensionality, and complexity. The importance of optimized implementation is highlighted. Additional details are available in the supplementary material~\cite{url/GASS}.

\noindent {\bf Procedure.} We tune each method to achieve the best trade-offs in accuracy/efficiency.  
Then, we carry experiments in two steps: indexing building and query answering, with caches cleared before each step and kept warm during the same query workload. Methods were allowed at most 48 hours to build a single index. During timed experiments, the server was used exclusively to ensure accurate measurements. 
 For each query workload, we ran the experiment six times; we excluded the two best and worst, and reported the mean of the remaining performances. For reproducibility, all parametrization details are provided in~\cite{url/GASS}.
 
\noindent \textbf{Implementations.} The implementations of various methods compared in this study were carefully examined and sourced from their official open-source repositories. Some approaches, such as VAMANA~\cite{vamana} and NGT~\cite{ngt_library}, are complex and continuously evolving; for instance, VAMANA offers an improved version with incremental insertion building~\cite{diskanncode}. However, we utilized the earlier version as described in the original paper, employing the two pruning steps and starting from a random graph. Ideally, all methods would be re-implemented from scratch to ensure uniformity, but this is highly time-consuming. Instead, by carefully adjusting each implementation to adhere as closely as possible to the original algorithms and applying comparable levels of optimization across methods, we provide a fair basis for comparing different algorithmic approaches for construction and search. This also enables a thorough analysis of the impact of various strategies on indexing and search performance. For comparisons involving seed selection and neighborhood diversification components, we built upon the official HNSW codebase~\cite{url/hnsw}, implementing additional components on top of it. All code is publicly available~\cite{url/GASS}. We believe our implementation will facilitate future research by enabling reproducibility of our findings across different scenarios, as well as serving as a foundation for exploring novel approaches to neighborhood diversification and seed selection.

\noindent \textbf{Comparison.} We compare and evaluate different methods based on indexing time, memory footprint, and disk size, while search comparisons focus on memory usage and query time. Search time is reported for evaluating state-of-the-art (SOTA) methods. Following reviewers comments, we believe that distance calculations provide a more neutral metric for comparing different SOTA approaches, as they typically employ similar search procedures. However, methods like ELPIS leverage lower bounding distances across multi-resolution dimensionalities, which complicates the straightforward summation of distance calculations. Nonetheless, in future extensions of this work, we plan to report the number of distance calculations alongside query time for a more comprehensive evaluation of SOTA methods.

It is also important to note that this study focuses on in-memory ANN search. While our insights and results apply to scenarios where the entire dataset fits within a single machine’s memory, we do not claim they extend to settings where data or graph indices must be maintained on disk or in distributed environments. We believe that techniques such as relaxed RND and \(K\)-random sampling may be preferable choices in such settings, including disk-based graph ANN. Future work could extend this study by experimenting with various graph ANN components and techniques across different settings and ANN search variants.

\label{subsec:experiments-ND}

\subsection{Neighborhood Diversification}
We now evaluate the ND strategies covered in Section~\ref{sec:survey}, i.e., RND, RRND, and MOND against a baseline without ND (NoND). 
We apply each strategy individually to an II-based graph, where each node is inserted sequentially and linked with a pruned list of neighbors, determined via a beam search with maximum out-degree $R=60$ and beam width $L=800$. Bi-directional edges are added to neighbors, and the neighborhood list is pruned to size $R$ using the same ND strategy.
Graphs are built on Deep and Sift (25GB, 100GB and 1B). 
For RRND and MOND, we run experiments with different values of $\alpha$ ($1.2-2$) and $\theta$ ($50^\circ-80^\circ$), respectively, and selected the best values $\alpha = 1.2$ and $\theta = 60^\circ$, which align with recommendation in \cite{vamana, nssg}. 
Then, we execute workloads with 100 queries against each dataset, and measure the accuracy/efficiency tradeoff using the recall and the number of distance calculations incurred during the search. The results in Figure~\ref{fig:ND:search:real} indicate that both RND and MOND consistently outperform, followed by RRND. NoND is the worst performer overall. As the dataset size increases, the performance gap between NoND and ND methods widens, particularly at high Recall (Figures~\ref{fig:ND:sift1b}, \ref{fig:ND:deep1b}). 
This is due to the higher number of hops needed to find the answers and the density of the neighborhoods in the NoND nodes since no pruning was applied. 
These results indicate the key role played by the ND paradigm in improving query-answering performance and the superiority of the RND and MOND strategies.
\newcommand{\sffive}{0.27\columnwidth}
\begin{figure}[tb]
	\captionsetup{justification=centering}
	\centering	
		\begin{subfigure}{\columnwidth}
			\centering
			\captionsetup{justification=centering}	
			\includegraphics[width=0.35\columnwidth]{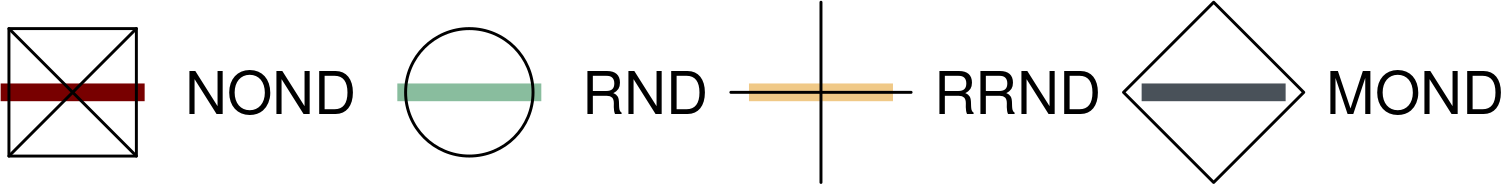}
			\label{fig:ND:legend}
		\end{subfigure}\\
		\begin{subfigure}{\sffive}
			\centering
			\captionsetup{justification=centering}	
			\includegraphics[width=\textwidth]{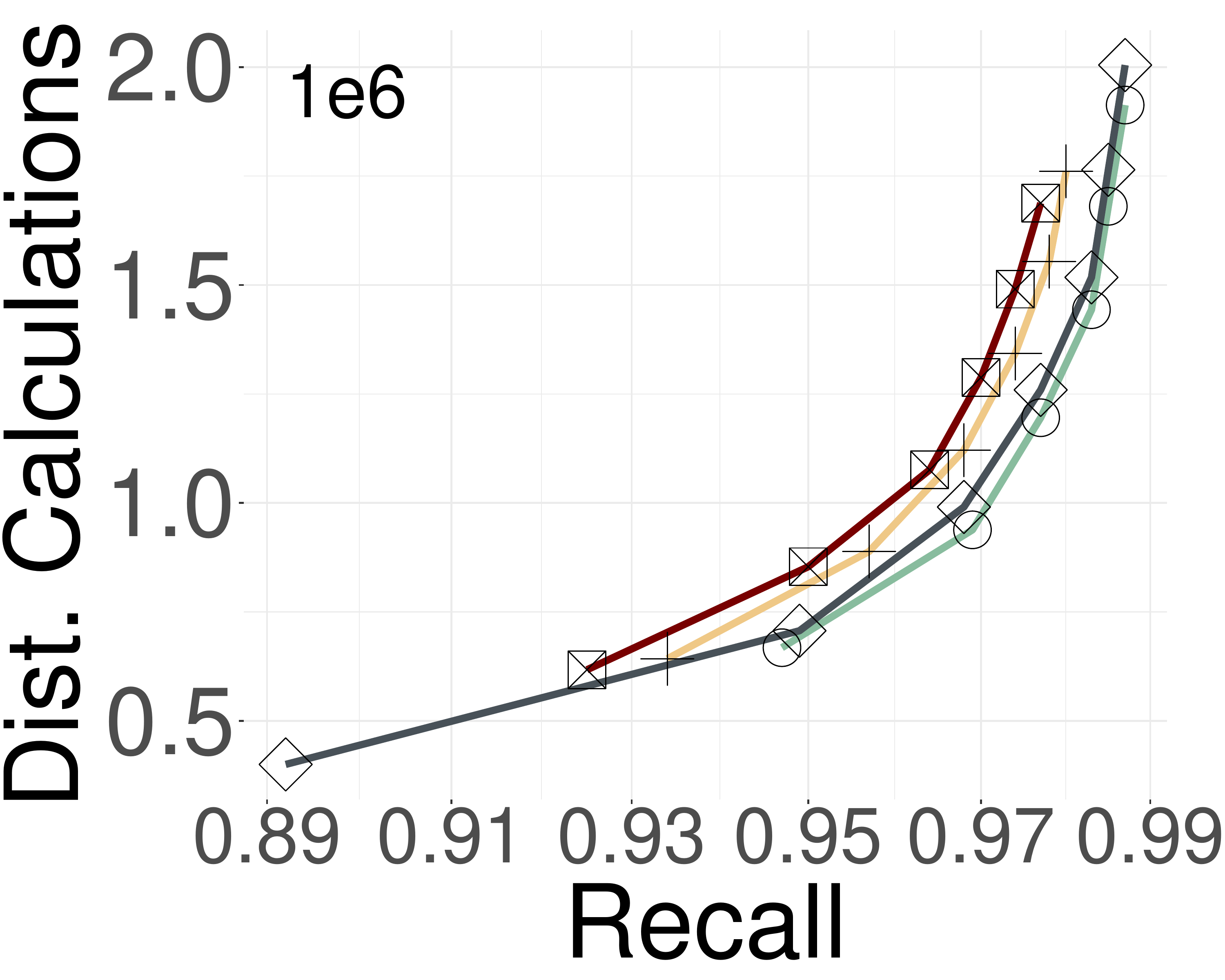}
		\caption{{DEEP25GB}}
		\label{fig:ND:deep25GB}	
		\end{subfigure}	
		\begin{subfigure}{\sffive}
			\centering
			\captionsetup{justification=centering}	
			\includegraphics[width=\textwidth]{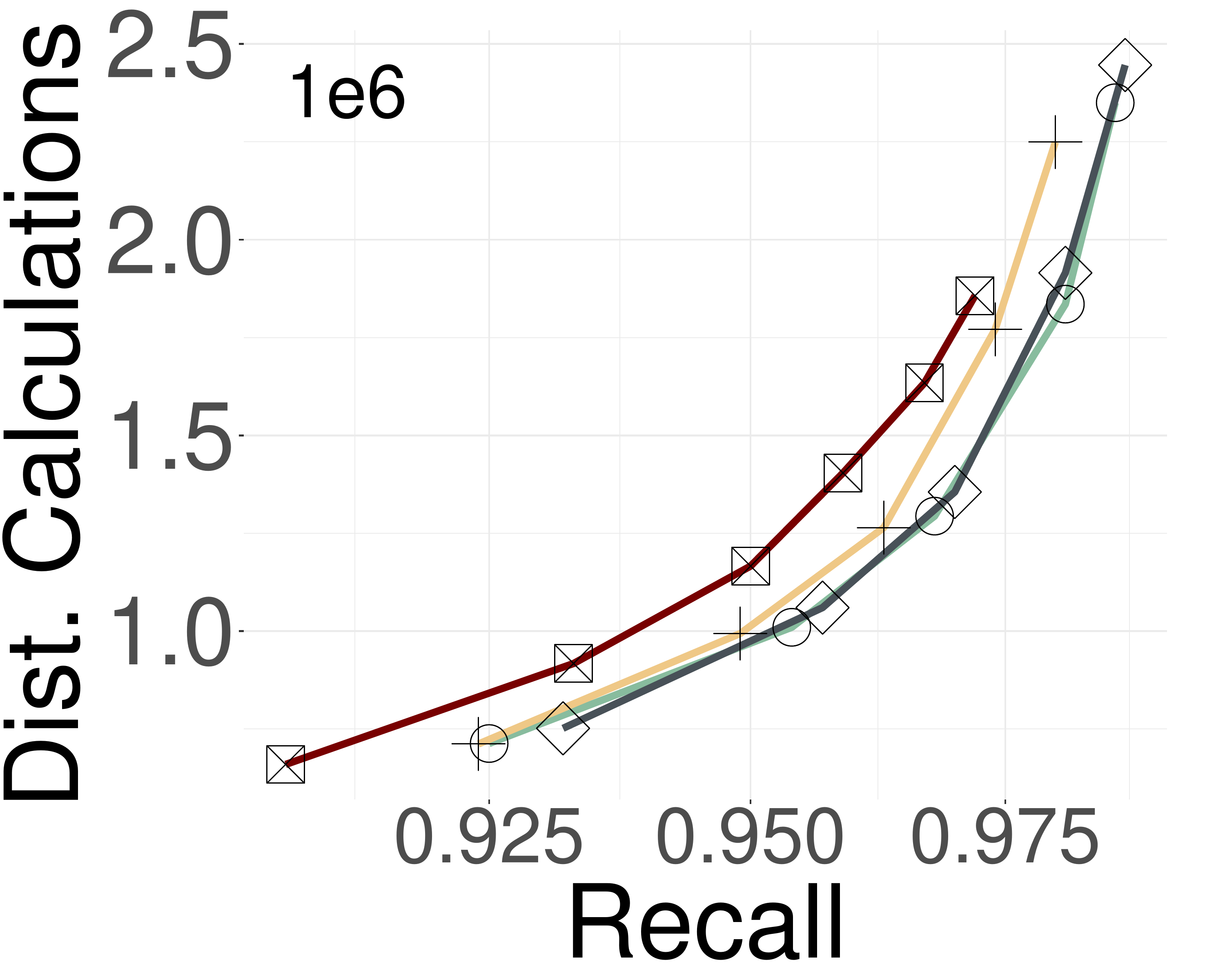}
		\caption{{DEEP100GB}}
		\label{fig:ND:deep100GB}
		\end{subfigure}	
		\begin{subfigure}{\sffive}
			\centering
			\captionsetup{justification=centering}	
			\includegraphics[width=\textwidth]{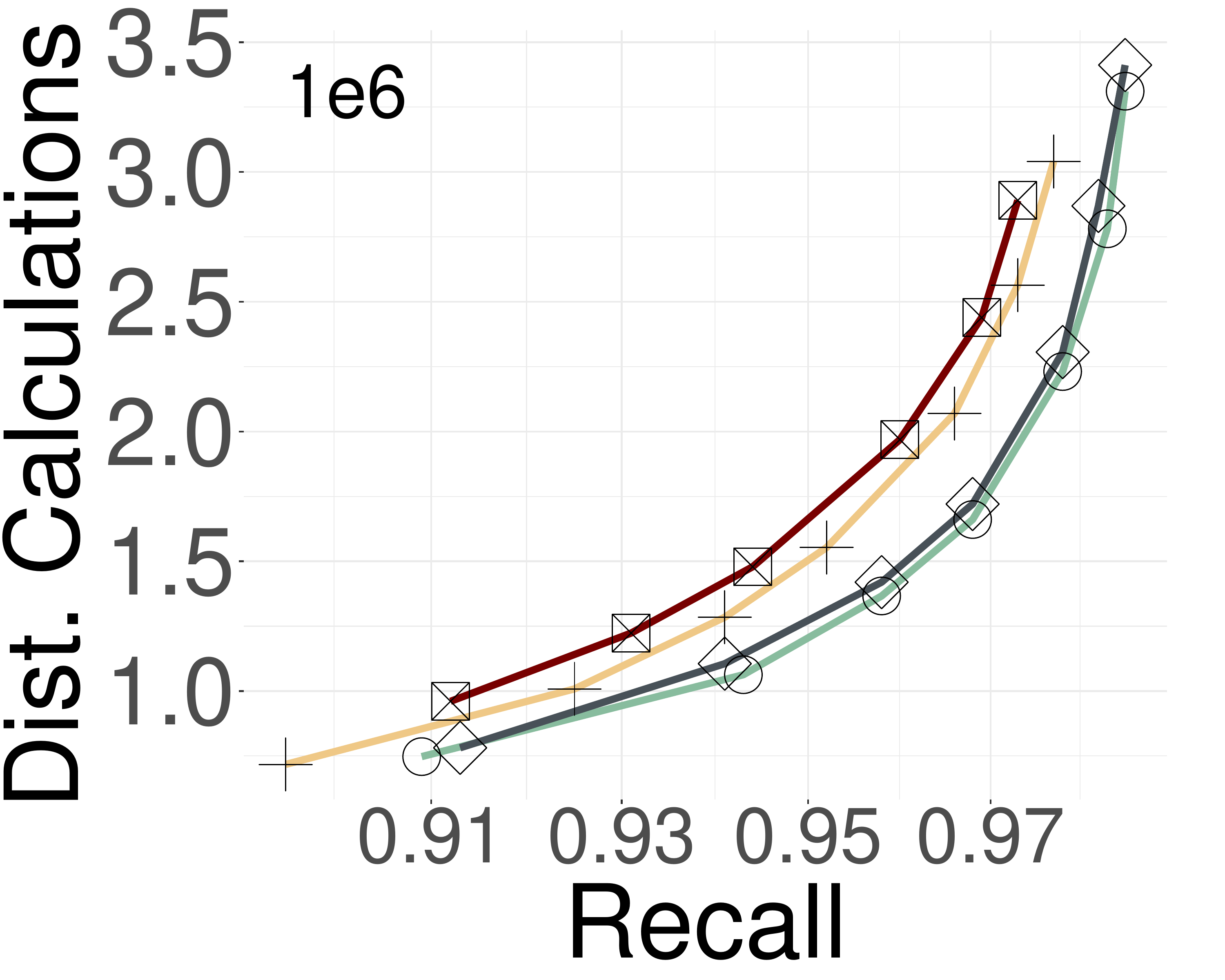}
		\caption{{DEEP1B}}
		\label{fig:ND:deep1b}	
  \end{subfigure}	
  \\
		\begin{subfigure}{\sffive}
			\centering
			\captionsetup{justification=centering}	
			\includegraphics[width=\textwidth]{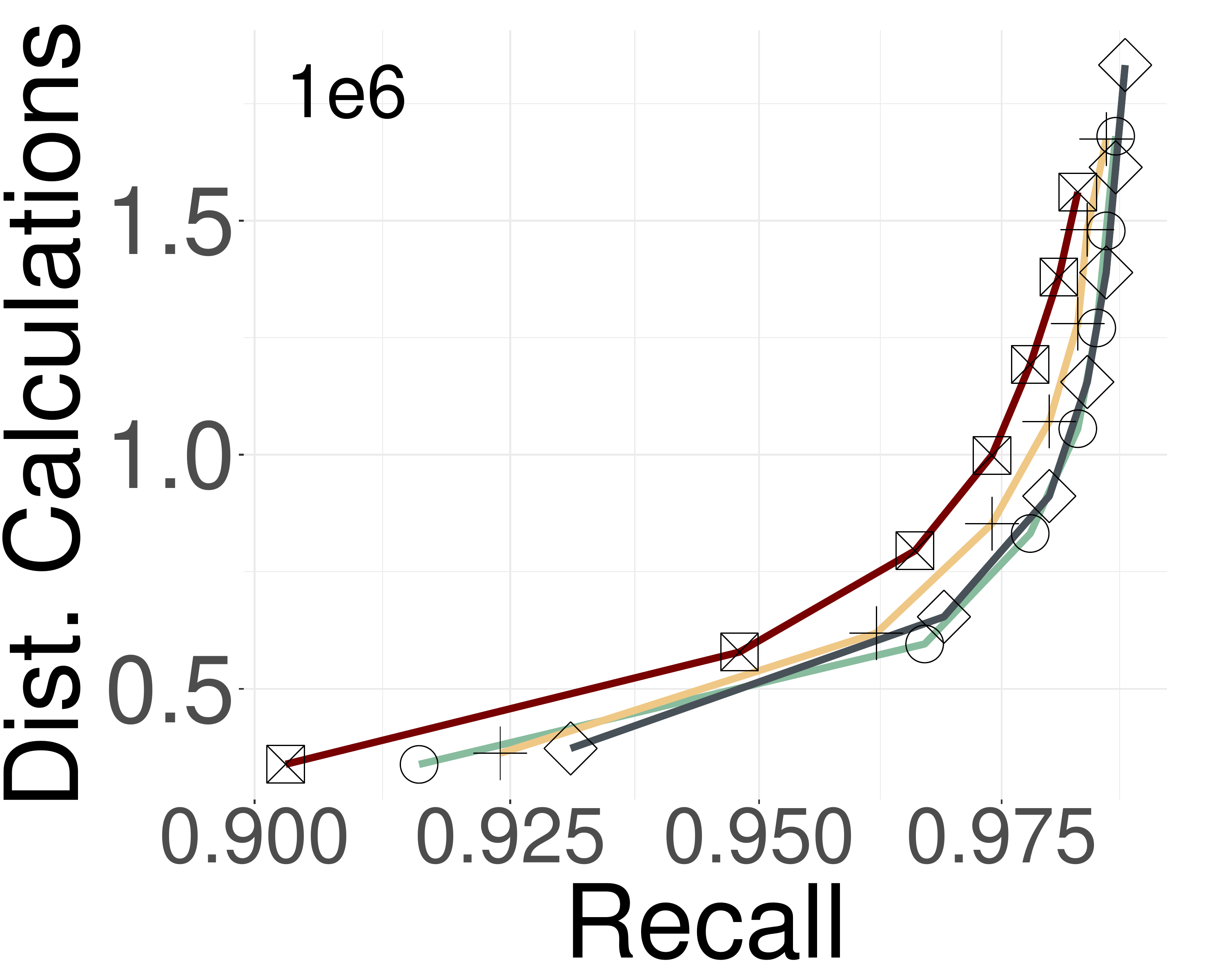}
   \caption{{SIFT25GB}}
		\label{fig:ND:sift25GB}
		\end{subfigure}	
		\begin{subfigure}{\sffive}
			\centering
			\captionsetup{justification=centering}	
			\includegraphics[width=\textwidth]{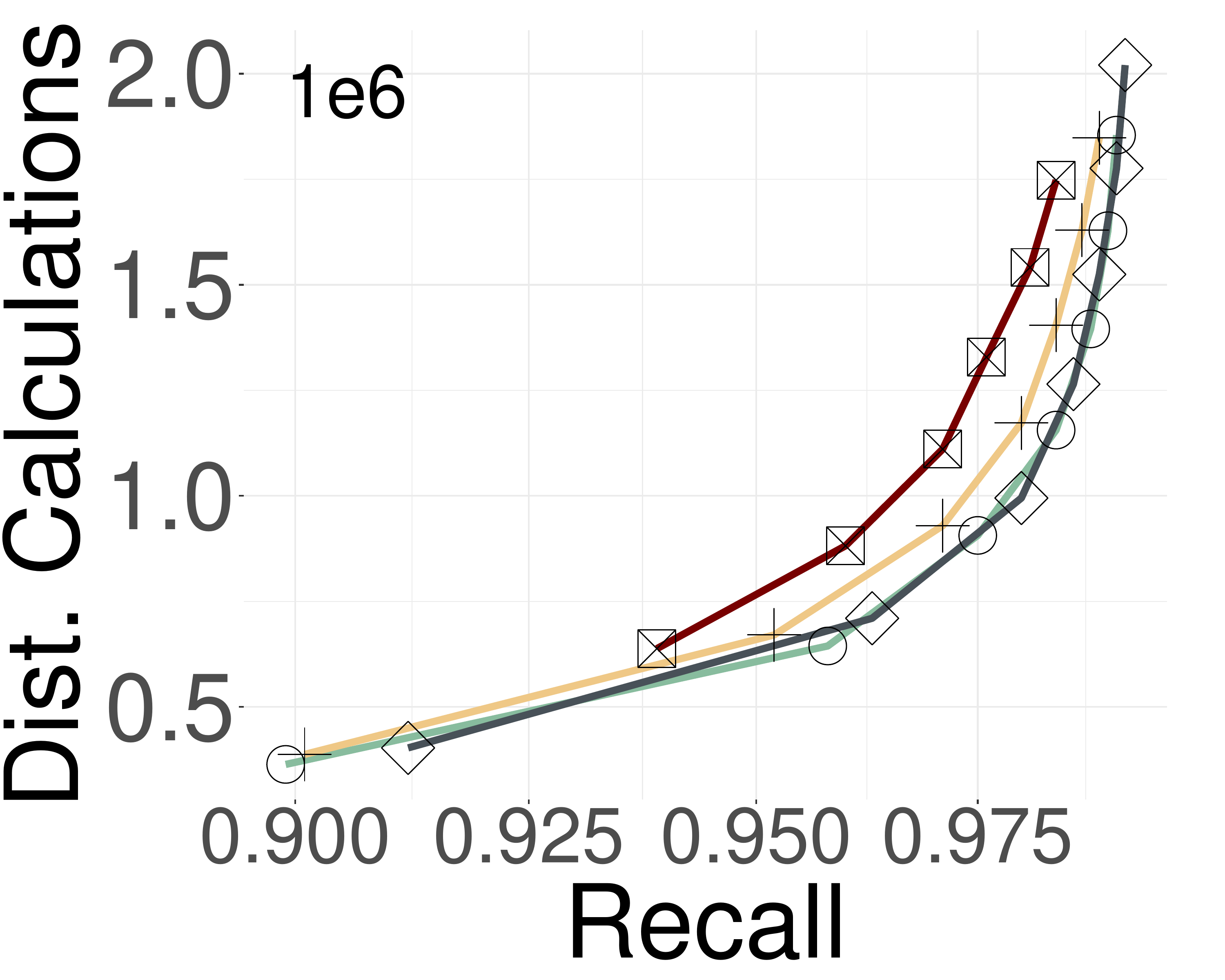}
             \caption{{SIFT100GB}}
		      \label{fig:ND:sift100GB}
		\end{subfigure}	
		\begin{subfigure}{\sffive}
			\centering
			\captionsetup{justification=centering}	
			\includegraphics[width=\textwidth]{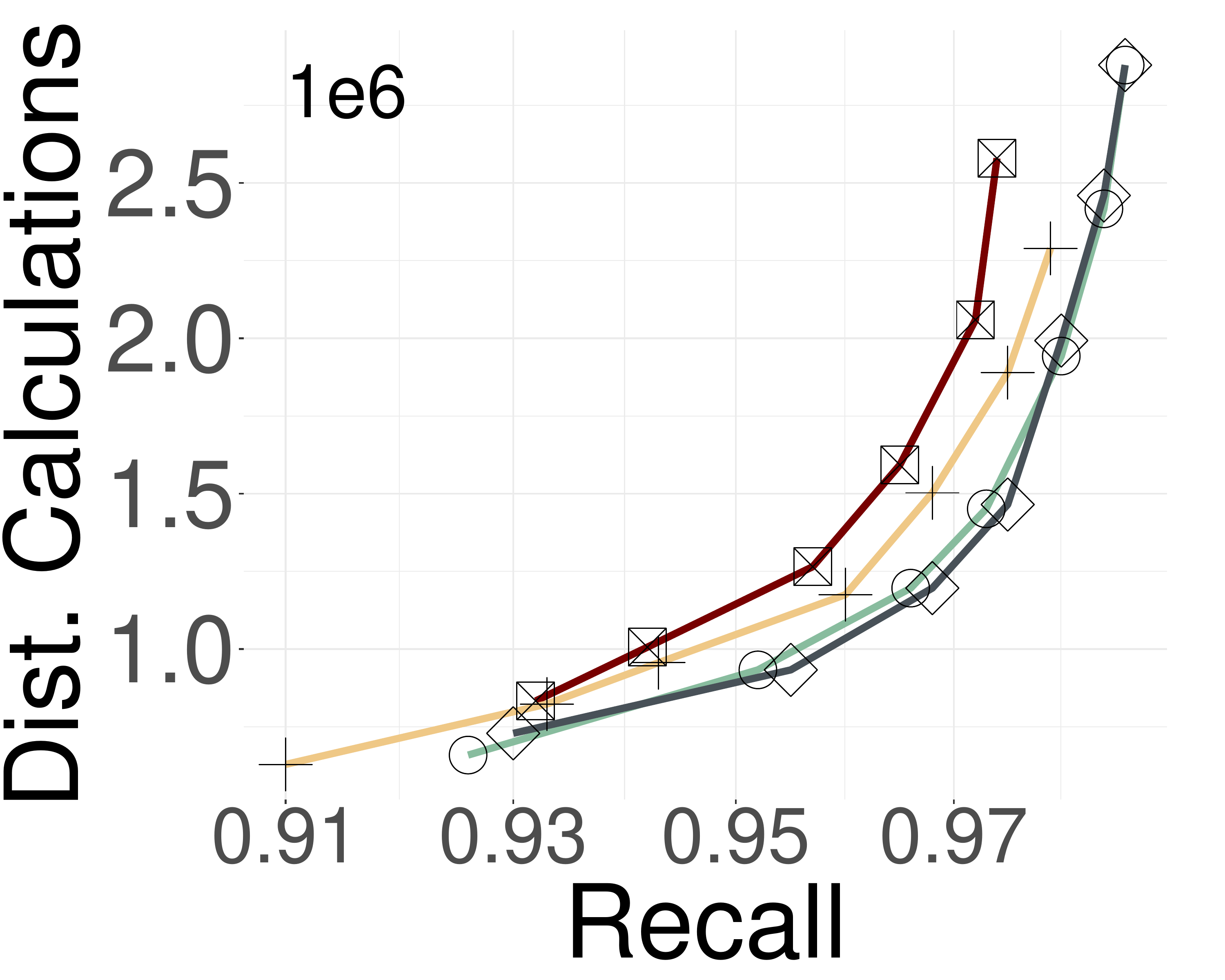}
   \caption{{SIFT1B}}
		\label{fig:ND:sift1b}
		\end{subfigure}	
		\caption{{ND methods performance on real-world datasets}}
	
\label{fig:ND:search:real}
 \end{figure}

As noticed, both RND and MOND lead to the best performance, while RRND can be adjusted through \(\alpha\) to prune similarly to RND using \(\alpha = 1\). Nevertheless, controlling edge density through \(\alpha\) and increasing edge density may deliver better performance in other settings, such as disk-based graph ANN. In particular, Vamana has been adopted in DiskANN~\cite{vamana} as the base graph to reduce the graph search diameter, thereby reducing the number of disk I/Os, which are typically more expensive than computing extra distances per hop. We also note that increasing edge density for in-memory search on hard datasets is promising. Our experiments in the supplementary materials~\cite{url/GASS} on power-law random datasets show that RRND and MOND can deliver better performance on challenging datasets.

In Table~\ref{tab:pruning_ratio}, we report the pruning ratios of the three neighborhood ND methods on the Deep and Sift 25GB datasets. 
 The pruning ratio quantifies the percentage reduction in the size of the candidate neighbor list before and after the diversification step. Higher pruning ratios indicate more aggressive pruning, which directly affects the graph size and memory usage. RND achieves the highest pruning ratios, MOND provides moderate pruning, and RRND exhibits the least pruning. 
 As a result, RND leads to smaller graph sizes and reduced memory requirements, while RRND creates larger graphs with higher memory usage.

\begin{table}[tb]
\centering
{\normalsize
\begin{tabular}{@{}lccc@{}}
\toprule
         & \textbf{RND} & \textbf{MOND} & \textbf{RRND} \\ 
\midrule
\textbf{Deep} & 20\%        & 2\%          & 0.6\%         \\ 
\textbf{Sift} & 25\%        & 4\%          & 0.7\%         \\ 
\bottomrule
\end{tabular}
} 
\caption{Pruning ratios of ND methods on Deep and Sift datasets.}
\label{tab:pruning_ratio}
\end{table}

\subsection{Seed Selection}
State-of-the-art graph-based vector search methods differ in their graph construction strategies but almost universally rely on beam search (Algorithm~\ref{alg:beamsearch}) for query answering. Beam search efficiently retrieves high-quality results on well-connected graphs. A critical factor affecting search efficiency is the choice of initial nodes, called \emph{seeds}. Better seeds reduce the number of visited nodes and speed up the search. Typically, one or more entry nodes warm the beam search priority queue; when multiple seeds are used, the search begins with the seed closest to the query, keeping the others in the queue.

In these experiments, we focus on the four most common SS strategies for the beam search algorithm: SN~\cite{hnsw,elpis}, MD~\cite{nsg,vamana}, KS~\cite{kgraph,nsw11,dpg,vamana,nssg}, and KD~\cite{efanna,SPTAG4,hcnng} (KM and LSH were excluded because they are not among the commonly used seed selection strategies in graph-based methods). We consider the baseline method SF which has not been used in the literature before. 
In these experiments, we focus on the four most common SS strategies for the beam search algorithm: SN~\cite{hnsw,elpis}, MD~\cite{nsg,vamana}, KS~\cite{kgraph,nsw11,dpg,vamana,nssg}, and KD~\cite{efanna,SPTAG4,hcnng} (KM and LSH were excluded because they are not among the commonly used seed selection strategies in graph-based methods). We consider the baseline method SF which has not been used in the literature before. 
These strategies are compared using the same insertion-based graph structure that exploits RND pruning since this is the best baseline from the results in Section~\ref{subsec:experiments-ND}.  
We run 100 queries for each strategy on the Deep and Sift datasets with sizes 25GB, 100GB, and 1B. We extrapolate the results to 1M queries and report the number of distance calculations to achieve a 0.99 accuracy in Figure~\ref{fig:ss:search}. We observe that SN and KS are the most efficient strategies across all scenarios, while SF and MD are the least efficient overall. The KD strategy is competitive on 25GB and 100GB Deep and Sift datasets but its performance deteriorates on billion-scale datasets. 
KS outperforms SN on dataset sizes 25GB and 100GB; however, this trend reverses with the 1B size dataset. 
The difference in distance calculations between SN and KS on the 25GB and 1B datasets is $\sim$1M and $\sim$10M, respectively. 
As the dataset size increases, it becomes imperative to sample more nodes (beyond the beam width utilized during search in KS) to obtain a representative sample of the dataset, thereby enhancing the likelihood of initiating the search closer to the graph region where the query resides (SN adapts its size logarithmically with the growth of the dataset, leading to a better representation of the dataset). Figure~\ref{fig:ss:search} also illustrates that both MD and SF are among the least performing strategies, with MD performing better than SF on Deep and vice-versa on Sift. This indicates neither MD nor SF are effective and robust seed selection strategies.

We now study the effect of SS strategies on indexing performance. We focus on the two best strategies KS and SN and study their effect on the same baseline based on II and RND~\cite{nsw11,dpg,hnsw,nsg,nssg,vamana,elpis,SPTAG4}. This is because these methods are the most impacted by the SS strategy used, since they perform a beam search, which includes a seed selection step, at the insertion of each node.
We build an index using each strategy on Deep1M and Deep25GB and measure distance calculations. 
We calculate the distance overhead of SN compared to KS, and we estimate the number of additional 100-NN queries that the KS-based graph can answer, with 0.99 recall, before the SN-based graph completes its construction.  
We observe (Table~\ref{tab:ss:idx}) that the SN-based graph requires 182 million and 22.3 billion more distance calculations than the KS-based graph on Deep1M and Deep25GB respectively. Furthermore, the KS-based graph can answer ~45K and 1.17 million queries on Deep1M and Deep25GB respectively before the SN-based graph finishes construction.

\newcommand{\sfsix}{0.27\columnwidth}
\begin{figure}[tb]
	\captionsetup{justification=centering}
	\centering	
 		\begin{subfigure}{0.015\columnwidth}
			\centering
			\captionsetup{justification=centering}	
			\includegraphics[width=\textwidth]{../img-png/Experiments/EP/dc.png}
   \vspace{0.14in}
		\end{subfigure}	
		\begin{subfigure}{\sfsix}
			\centering
			\captionsetup{justification=centering}	
			\includegraphics[width=\textwidth]{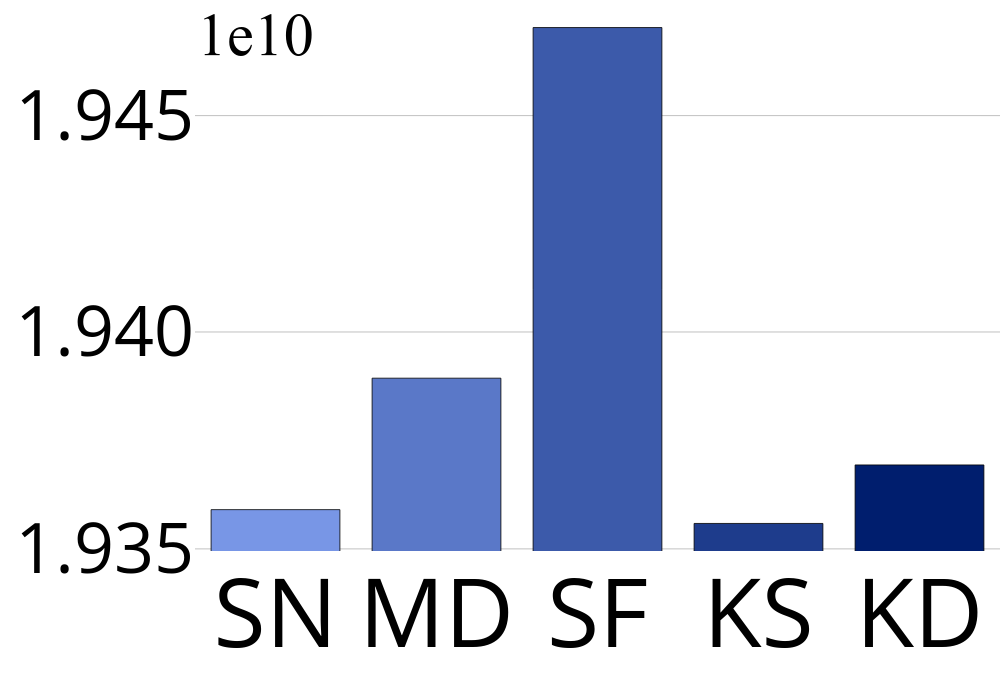}
		\caption{{Deep25GB}}
		\label{fig:ss:deep25gb}
		\end{subfigure}	
		\begin{subfigure}{\sfsix}
			\centering
			\captionsetup{justification=centering}	
			\includegraphics[width=\textwidth]{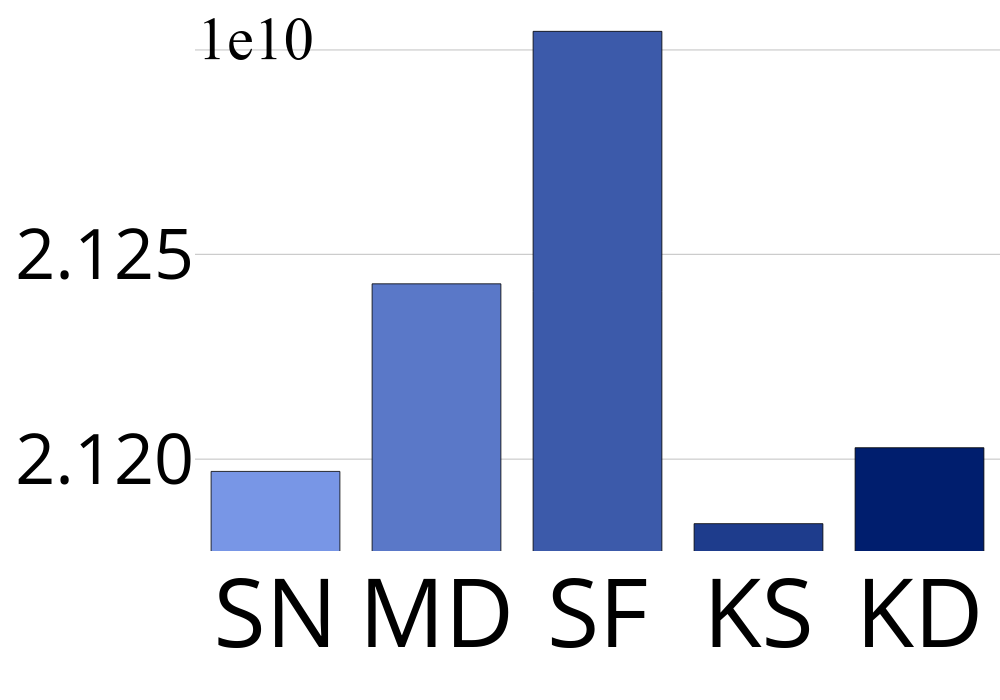}
		\caption{{Deep100GB}}
		\label{fig:ss:deep100gb}
		\end{subfigure}	
		\begin{subfigure}{\sfsix}
			\centering
			\captionsetup{justification=centering}	
			\includegraphics[width=\textwidth]{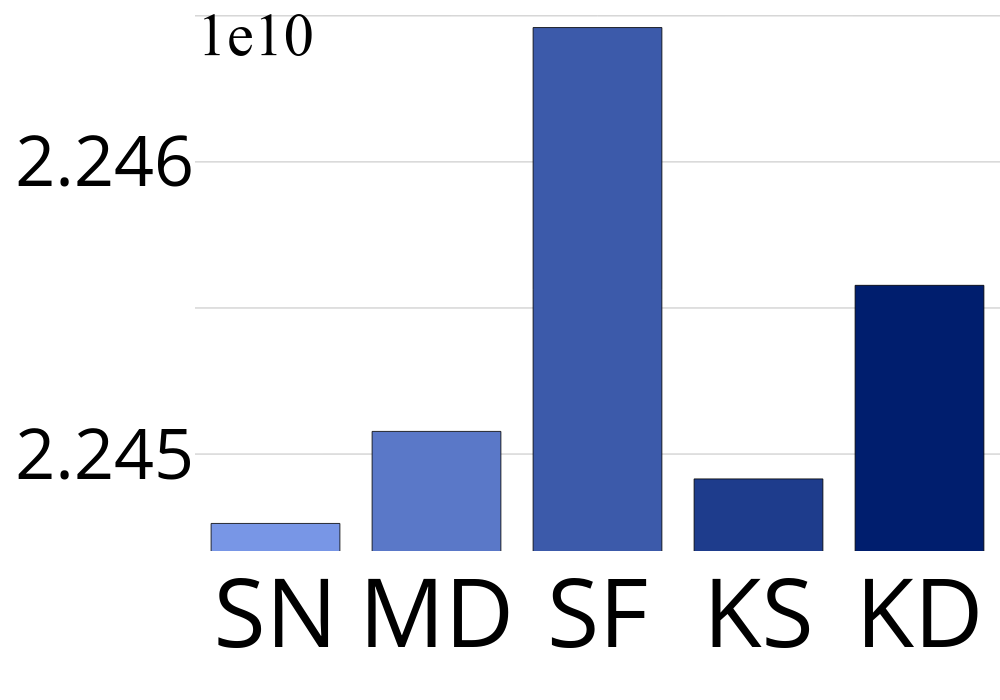}
		\caption{{Deep1B}}
		\label{fig:ss:deep1b}
		\end{subfigure}	
        
 		\begin{subfigure}{0.015\columnwidth}
			\centering
			\captionsetup{justification=centering}	
			\includegraphics[width=\textwidth]{../img-png/Experiments/EP/dc.png}
   \vspace{0.14in}
		\end{subfigure}	
		\begin{subfigure}{\sfsix}
			\centering
			\captionsetup{justification=centering}	
			\includegraphics[width=\textwidth]{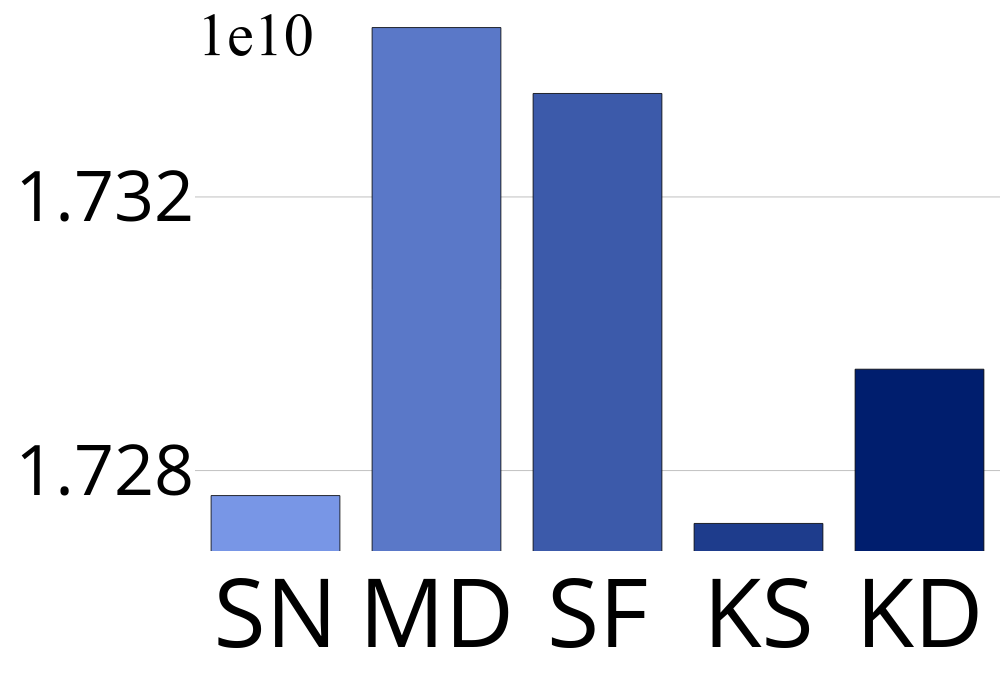}
		\caption{{Sift25GB}}
		\label{fig:ss:sift25gbß}
		\end{subfigure}	
		\begin{subfigure}{\sfsix}
			\centering
			\captionsetup{justification=centering}	
			\includegraphics[width=\textwidth]{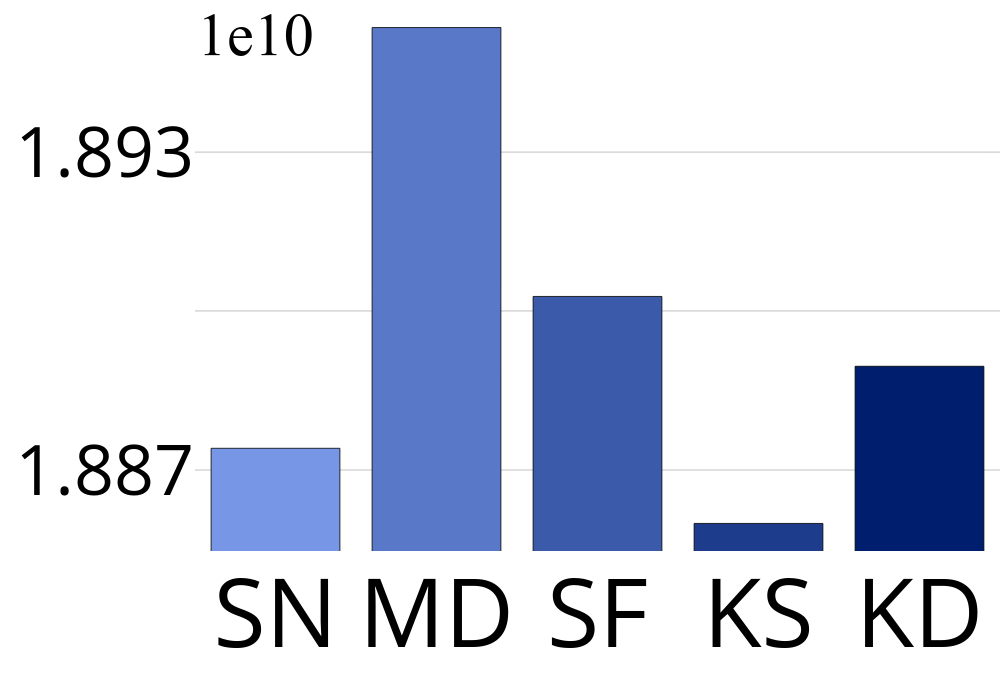}
		\caption{{Sift100GB}}
		\label{fig:ss:sift100gb}
		\end{subfigure}	
		\begin{subfigure}{\sfsix}
			\centering
			\captionsetup{justification=centering}	
			\includegraphics[width=\textwidth]{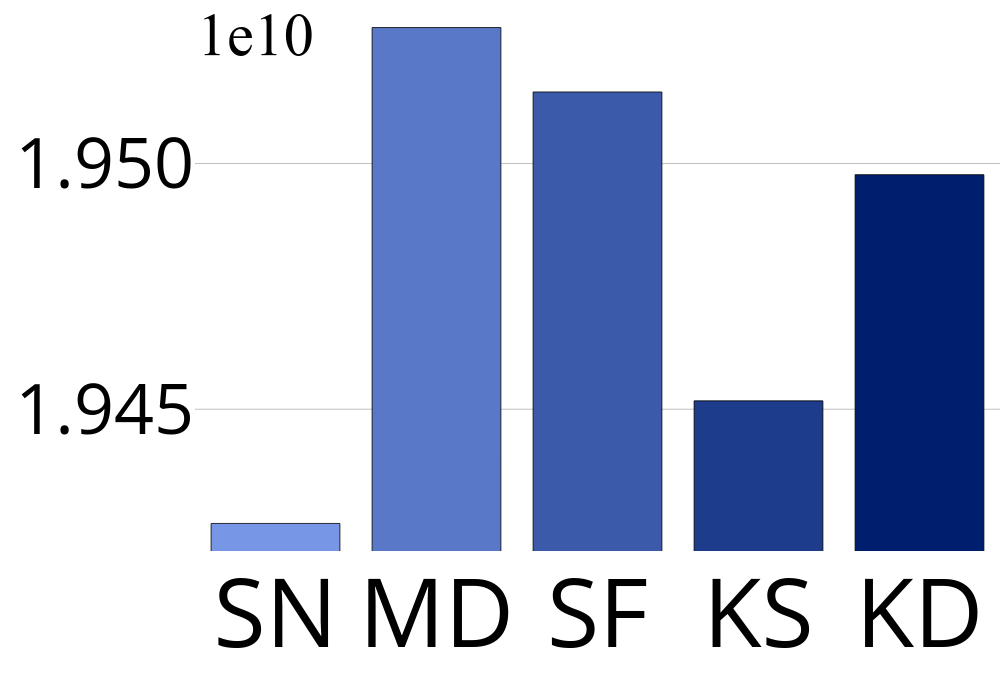}
		\caption{{Sift1B}}
		\label{fig:ss:sift1b}
		\end{subfigure}	
\caption{The impact of SS Methods on query answering}
\label{fig:ss:search}
 \end{figure}

\begin{table}[tb]
\centering
\begin{tabular}{@{}lcc@{}}
\toprule
& \textbf{Deep1M} & \textbf{Deep25GB} \\ 
\midrule
\textbf{Dist. Calculations (SN)} & 4.3 billion & 1.49 trillion \\
\textbf{Dist. Calculations (KS)} & 4.1 billion & 1.46 trillion \\
\midrule
\textbf{Overhead (SN vs. KS)} & 182 million & 22.3 billion \\
\textbf{Additional Queries} & 44,959 & 1,165,870 \\
\bottomrule
\end{tabular}
\caption{The impact of SS methods on Indexing Performance}
\label{tab:ss:idx}
\end{table}

\subsection{Indexing Performance}
We evaluate twelve state-of-the-art vector search methods, varying dataset sizes and reporting total indexing time and memory footprint. For brevity, we present results only for the Deep dataset as trends are consistent across other datasets. Full results are in~\cite{url/GASS}.  We use subsets of Deep ranging from 1 million to 1 billion vectors (equivalent to ~350GB). Indexes are built to allow a 0.99 recall efficiently. Initial experiments on 1 million vectors include all methods. Methods that could not scale to larger datasets are excluded from subsequent experiments. Specifically, HCNNG, SPTAG-BKT, NGT, and SPTAG-KDT take over 24 hours to build indexes on 25GB datasets and exceed 48 hours on 100GB datasets. 
KGraph, DPG, EFANNA, and LSHAPG delivered unsatisfactory results on 25GB, so they were not included in larger datasets. Furthermore, KGraph and EFANNA require over 300GB and 1.4TB of RAM for 25GB and 100GB datasets, respectively. As DPG, NSG, and SSG rely on KGraph and EFANNA, they were also excluded from larger datasets.

\begin{figure*}[!htb]
	\captionsetup{justification=centering}
	\centering	
	\begin{minipage}{\textwidth}
		\begin{subfigure}{\textwidth}
			\centering
			\captionsetup{justification=centering}	
			\includegraphics[width=\textwidth]{../img-png/Experiments/legendall.png}
		\end{subfigure}
	\end{minipage}
	\begin{minipage}{0.19\textwidth}				
			\centering
		\begin{subfigure}{\textwidth}
			\captionsetup{justification=centering}	
			\includegraphics[width=\textwidth]{../img-png/Experiments/Idx_footprint_datasets/idx_time_deep_n.png}
		\end{subfigure}	
		\caption{\karima{Indexing Time }}
		\label{fig:idx:time}
	\end{minipage}	
	\begin{minipage}{0.19\textwidth}				
		\begin{subfigure}{\textwidth}
			\centering
			\captionsetup{justification=centering}	
			\includegraphics[width=\textwidth]{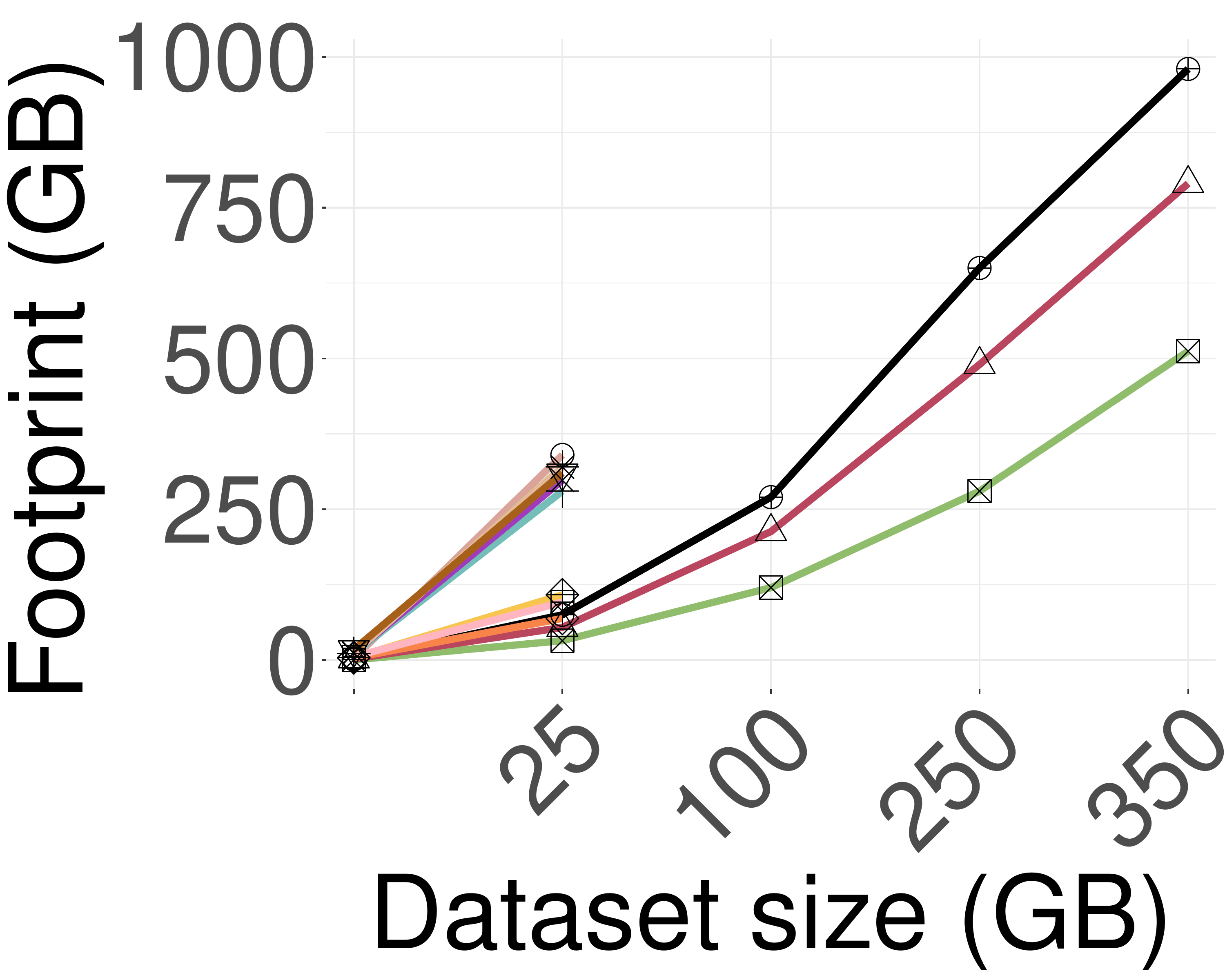}
		\end{subfigure}	
		\caption{\karima{Indexing Memory Footprint}}
		\label{fig:idx:footprint:memory}
	\end{minipage}		
	\begin{minipage}{0.19\textwidth}				
		\begin{subfigure}{\textwidth}
			\centering
			\captionsetup{justification=centering}	
			\includegraphics[width=\textwidth]{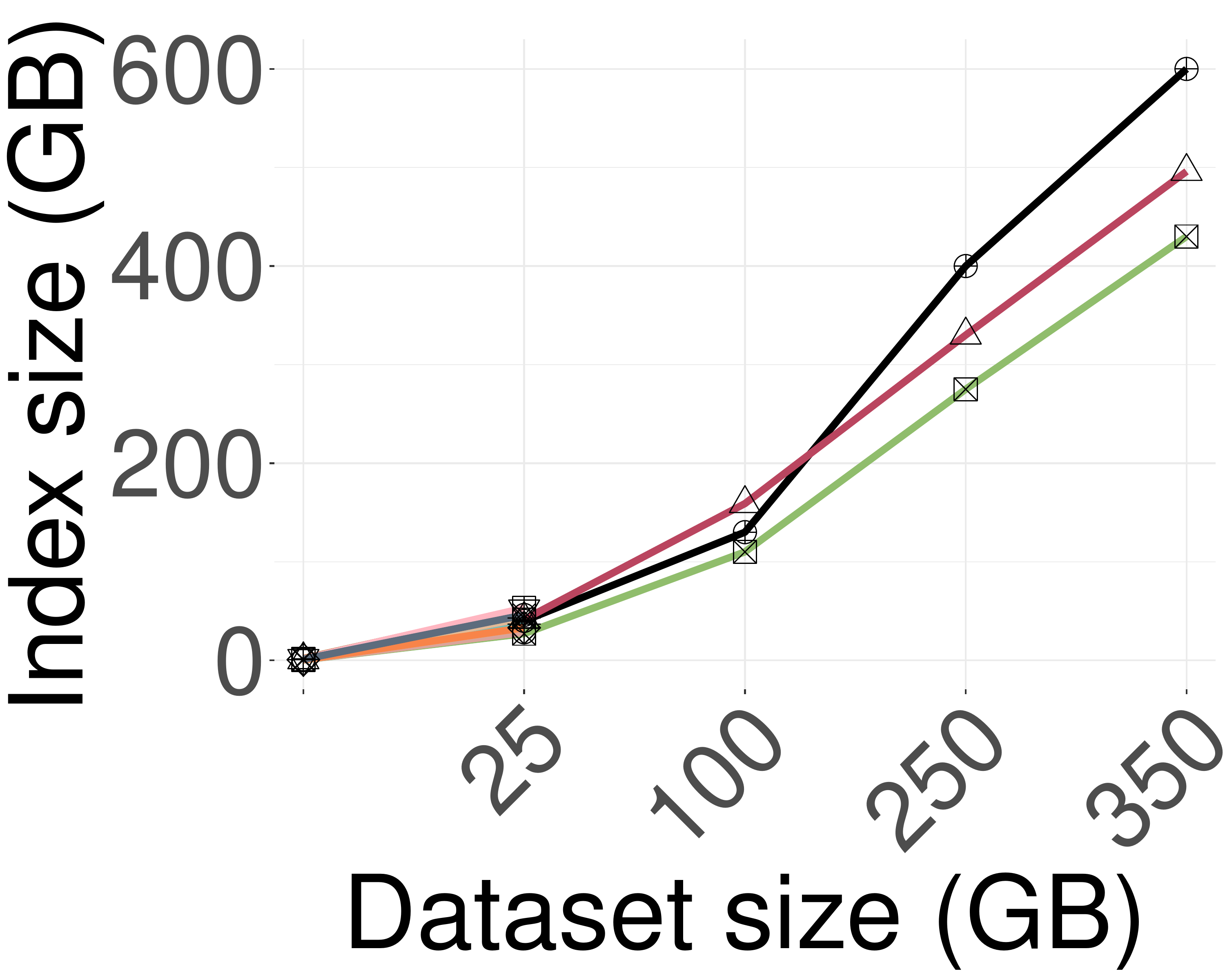}
		\end{subfigure}	
		\caption{\karima{Indexing Disk Footprint}}
		\label{fig:idx:footprint:disk}
	\end{minipage}		
	\begin{minipage}{0.19\textwidth}				
		\begin{subfigure}{\textwidth}
			\centering
			\captionsetup{justification=centering}	
			\includegraphics[width=\textwidth]{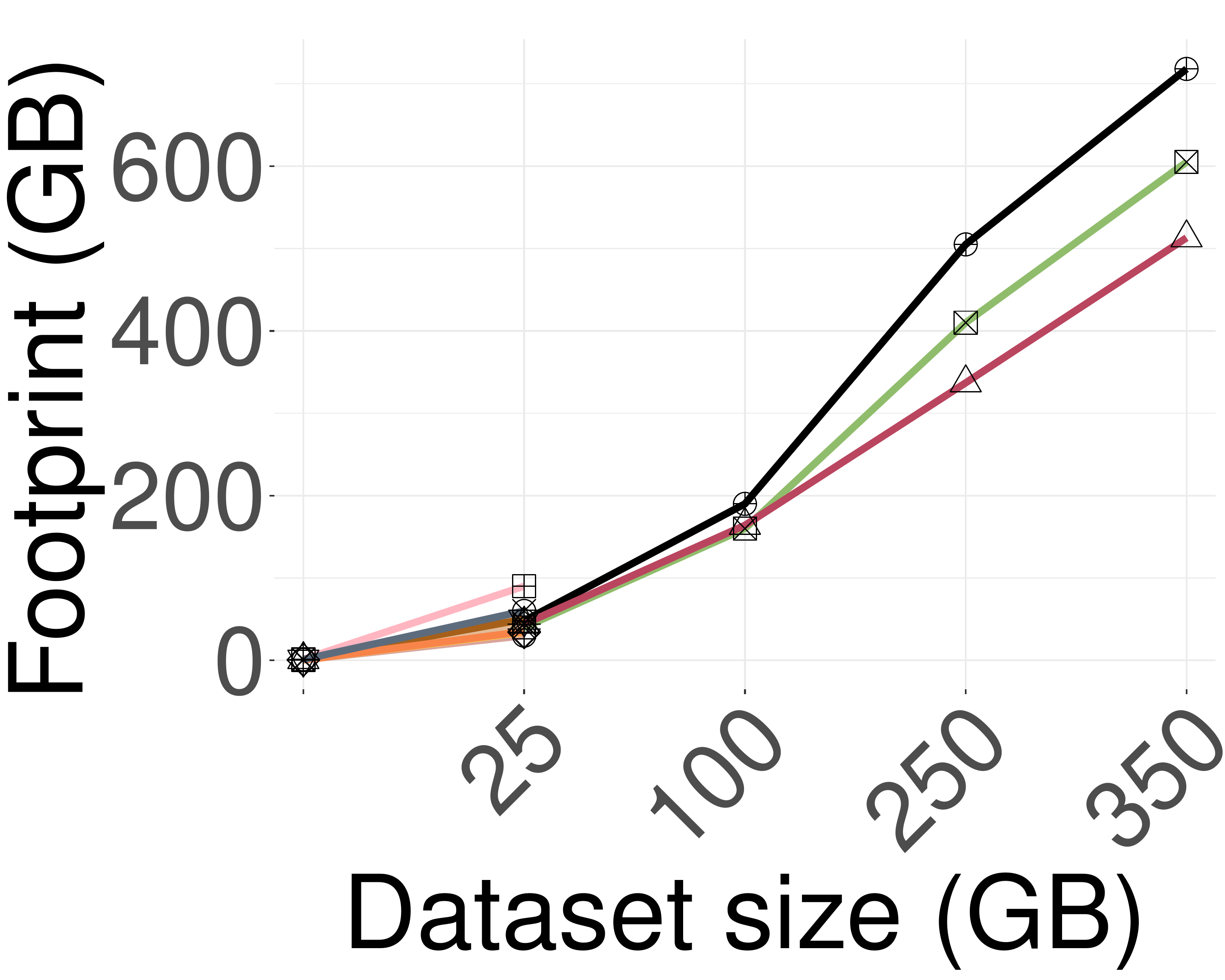}
		\end{subfigure}	
		\caption{\karima{Query Memory Footprint}}
		\label{fig:query:footprint:memory}
	\end{minipage}	
	\begin{minipage}{0.19\textwidth}
		\centering
		\begin{subfigure}{\textwidth}
			\includegraphics[width=\textwidth]{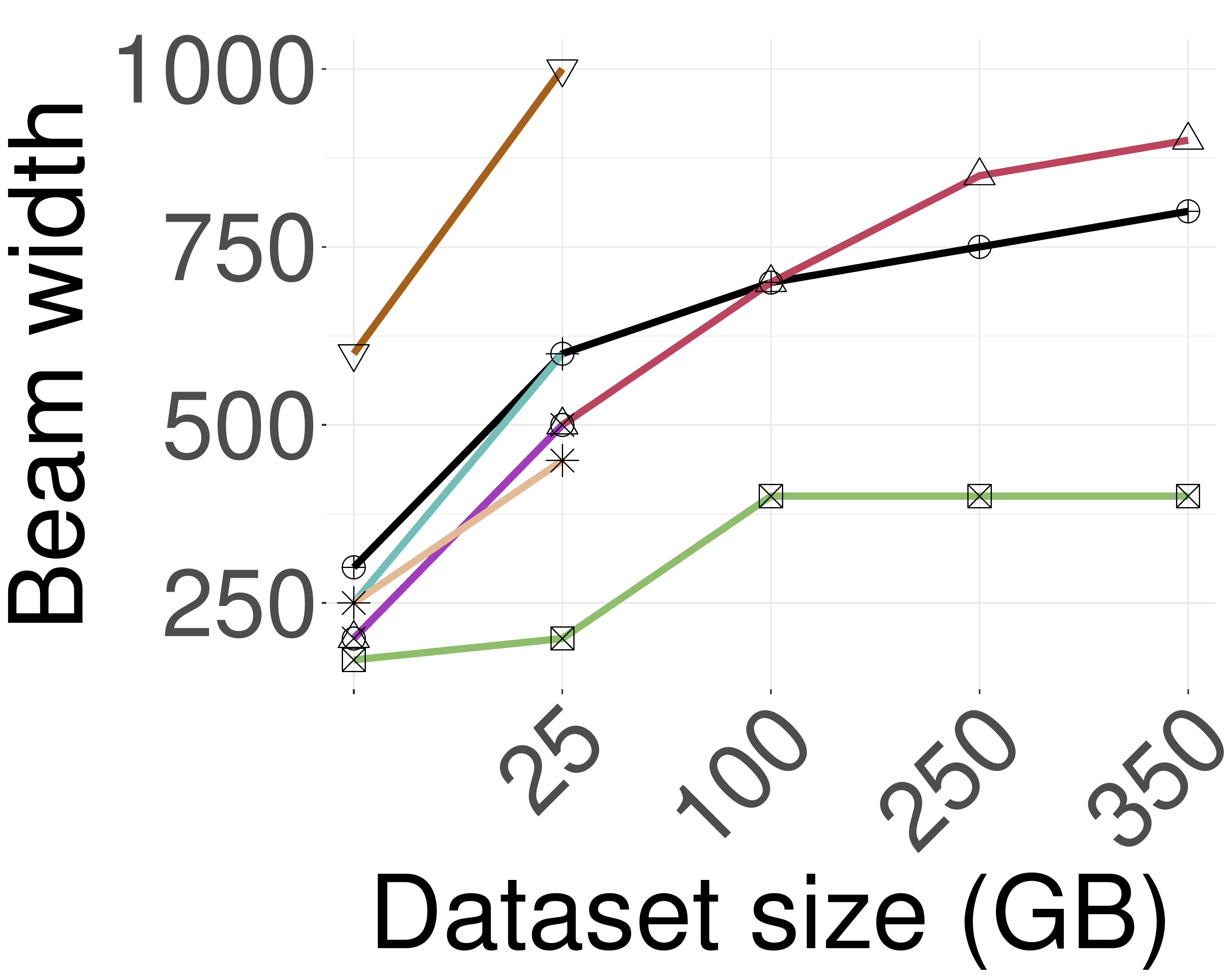}
		\end{subfigure} 
		\caption{\karima{Query Beam Width}}
		\label{fig:query:beam-width}
	\end{minipage}
\end{figure*}

\noindent{\bf Indexing Time.} Figure~\ref{fig:idx:time} demonstrates that II-based approaches have the lowest indexing time across dataset sizes. In particular, the II and DC-based approach ELPIS, is 2.7x faster than HNSW and 4x faster than NSG for both 1M and 25GB dataset sizes, \karima{while HNSW is 1.4x faster than NGT}. Note that NSG's indexing time includes both the construction of its base graph, EFANNA, which is time-intensive, and the refinement with NSG. SPTAG-BKT and SPTAG-KDT exhibit high indexing times, requiring over 25 hours to index the Deep25GB dataset-24 times more than ELPIS, the fastest method. This inefficiency in SPTAG arises from its design, which involves constructing multiple TP Trees and graphs, becoming increasingly costly with larger datasets. On datasets with 100GB and 1B vectors ($\approx$350GB), only HNSW, ELPIS, and Vamana scale with acceptable indexing time, with ELPIS being 2 and 2.7 times faster than HNSW and Vamana, respectively.

\noindent\textbf{Indexing Footprint.}
Figure~\ref{fig:idx:footprint:memory} reports the memory footprint for each index, including the raw data. To perform the evaluation, we record the peak virtual memory usage during construction. SPTAG-BKT and SPTAG-KDT demonstrate efficient memory utilization (1M and 25GB) despite having the highest indexing time. 
For larger datasets, ELPIS has the lowest indexing memory footprint, occupying up to 40\% less memory than HNSW and 30\% less than Vamana during indexing. This is because ELPIS needs a smaller maximum out-degree and beam width compared to its competitors.
HNSW has a higher indexing memory footprint due to its use of a graph layout optimized for direct access to node edges through a large contiguous block allocation~\cite{url/hnsw}. This layout offers a time advantage over adjacency lists by reducing memory indirections and cache misses. However, it can result in quadratic memory growth when using a large maximum out-degree on large-scale datasets.
In Figure \ref{fig:idx:footprint:disk}, we compare the size of method indices, including the raw data. The figure shows that certain methods, such as EFANNA, HCNNG, KGraph, and consequently NSG, SSG, and DPG (which use one of these base graphs), exhibit a significantly larger memory footprint relative to their final index size. For instance, HCNNG consumes substantial memory during indexing, requiring over 200GB for Deep25GB (Fig. \ref{fig:idx:footprint:memory}) due to merging multiple MST from numerous samples generated during hierarchical clusterings. In contrast, its final index size is less than 50GB (Fig. \ref{fig:idx:footprint:disk}).

\subsection{Search Performance}
We now evaluate the search performance of the different methods. All methods were included in the 1M experiments. Some methods were excluded in 25GB plots (KGraph, DPG, SPTAG-KDT, HCNNG, and EFANNA) for the sake of clarity, as their search was significantly slower than the best baselines. Full results are in~\cite{url/GASS}.
Other methods were omitted from the 100GB and 1B dataset sizes due to various limitations. The indexes for SPTAG and NGT could not be built on the larger datasets within 48 hours. EFANNA was excluded due to its high footprint, and likewise for methods based on it such as NSG and SSG.
Finally, KGraph, DPG, and LSHAPG were excluded due to unsatisfactory results on 1M and 25GB.

\noindent\textbf{Query Memory Footprint and Beam Width.}
Figure~\ref{fig:query:footprint:memory} indicates that Vamana, followed by ELPIS, have the lowest memory footprint during search. Even though ELPIS has a smaller index size, it adopts a contiguous memory storing during search just like HNSW, which increases the index footprint when loaded into memory. Besides, Figure~\ref{fig:query:beam-width} shows that Elpis requires the smallest beam width to reach similar query accuracy. Having a very high beam width indicates that the beam search requires to visit a wider area to and making more distance calculations to retrieves the NN answers.

\begin{figure*}[h]
  \centering      
        \includegraphics[width=\linewidth]{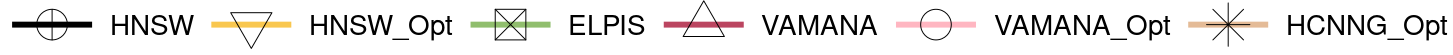}
 \begin{minipage}{0.73\linewidth}
 \centering
    \begin{minipage}{\textwidth}
        \centering
        \captionsetup{justification=centering}
        \begin{subfigure}{0.32\textwidth}
            \centering
            \captionsetup{justification=centering}
            \includegraphics[width=\textwidth]{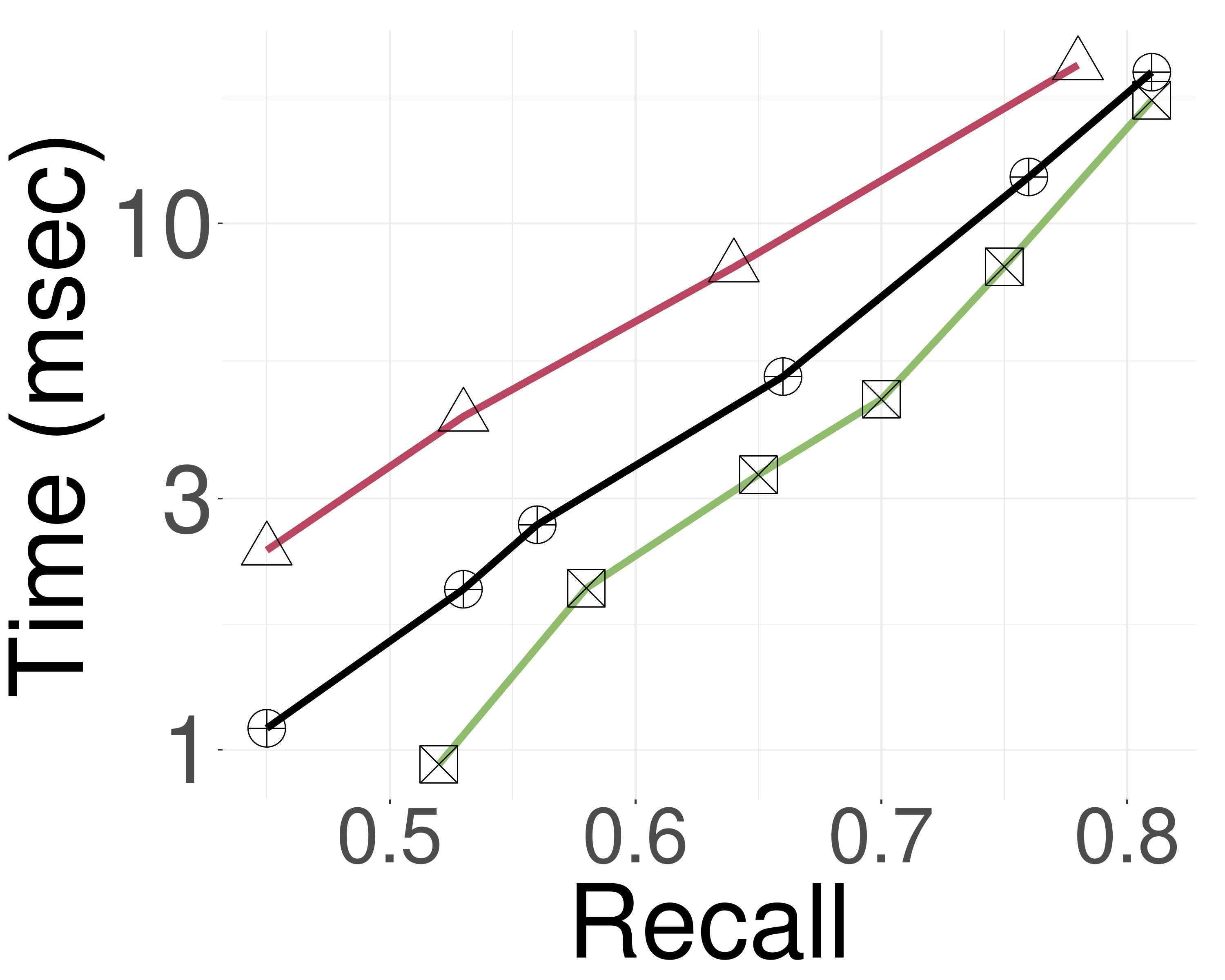}
            \caption{\karima{Text2Image}}
            \label{fig:query:performance:1B:t2i:10NN}
        \end{subfigure}
        \begin{subfigure}{0.305\textwidth}
            \centering
            \captionsetup{justification=centering}
            \includegraphics[width=\textwidth]{../img-png/Experiments/search/1B/deep_10nn.png}
            \caption{\karima{Deep}}
            \label{fig:query:performance:1B:deep:10NN}
        \end{subfigure}
        \begin{subfigure}{0.305\textwidth}
            \centering
            \captionsetup{justification=centering}
            \includegraphics[width=\textwidth]{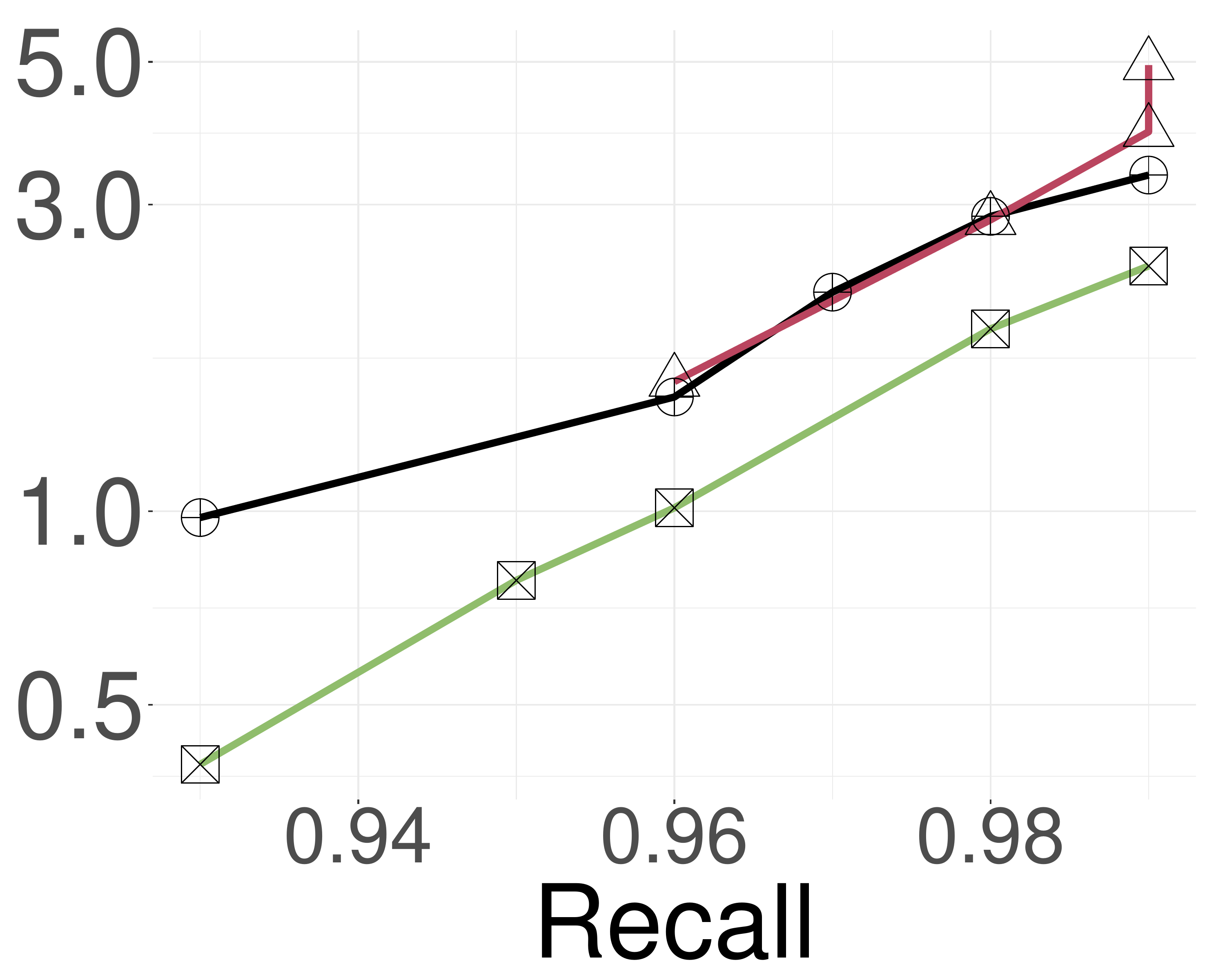}
            \caption{\karima{Sift}}
            \label{fig:query:performance:1B:sift:10NN}
        \end{subfigure}
    \end{minipage}
    \caption{\karima{1B Datasets}}
    \label{fig:query:performance:1B}
    \end{minipage}
    \begin{minipage}{0.26\linewidth}
        \centering
        \captionsetup{justification=centering}
        \includegraphics[width=\textwidth]{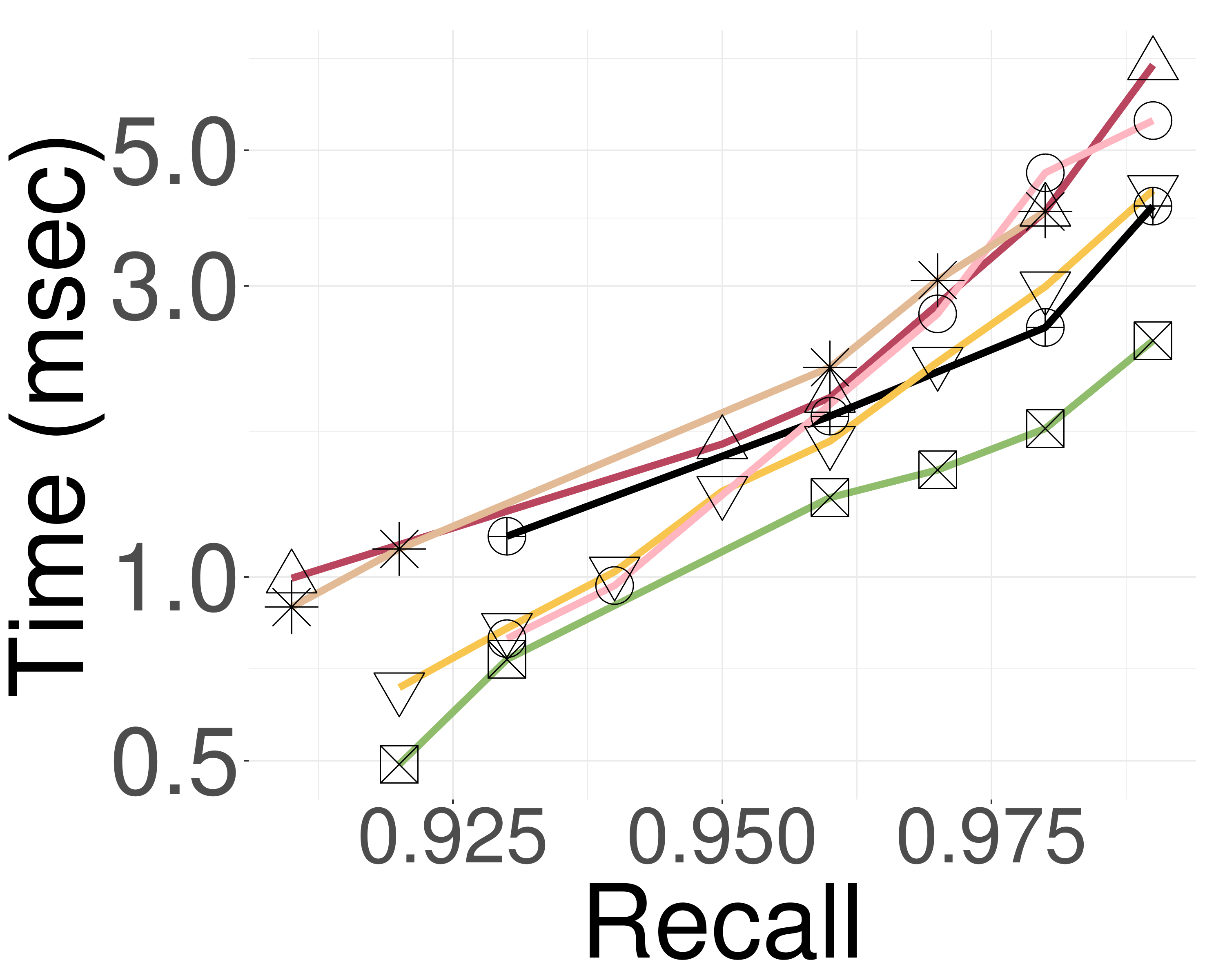}
        \caption{\karima{Optimized Implementations (Deep1B)}}
        \label{fig:optimized_impl}
    \end{minipage}
\end{figure*}

\noindent\textbf{Real Datasets.} 
On datasets with 1M vectors (Figure~\ref{fig:query:performance:1M}), ELPIS and NSG/SSG perform best on Sift1M, achieving the highest performance for 0.99 and lower recall, respectively. For Seismic1M, HCNNG and ELPIS share the top rank. NGT, SSG, and NSG excel on Deep1M, while HCNNG leads on SALD1M at the highest recall, followed by SPTAG and NSG at lower recall levels. In ImageNet1M, NSG/SSG and HNSW rank as the top performers. 
Across most scenarios, SSG and NSG show similar performance. However, LSHAPG demonstrates limitations, requiring more computation to achieve high accuracy. Its probabilistic rooting prunes promising neighbors, requiring a larger beam width and tighter $L$-bsf lower-bound distance during search.
When moving to 25GB datasets (Figure~\ref{fig:query:performance:25GB}), SSG, NSG, NGT and HCNNG experience a drop in performance, and ELPIS takes the lead with the best overall performance together with SPTAG-BKT for SALD25GB .
It is worth noting that none of the methods achieved an accuracy over 0.8 on the Seismic dataset, leading us to report results for these lower recall values. 
The significant indexing footprint of NSG prevented us from extending its evaluation to larger datasets, as constructing the EFANNA graph (which NSG depends on) requires more memory than the available 1.4TB. 
For hard query workloads in Figure~\ref{fig:search:query:performance:25GB:hard} we compare the best-performing methods from the two most performing graph paradigms, ND-based and DC-based methods, including HNSW, NSG, ELPIS and SPTAG-BKT. 
SPTAG-BKT achieves the overall best performance for 1\% noise query set, as we increase the noise up to 10\%, SPTAG-BKT's performance deteriorates, which we can relate to SPTAG BKT structures failing to identify good seed points. At the same time, the other competitors gain an advantage, with ELPIS taking the lead. 
When analyzing very large datasets of 1 billion vectors, Figure~\ref{fig:query:performance:1B} shows the superiority of ELPIS which is up to an order of magnitude faster at achieving 0.95 accuracy, thanks to its design that supports multi-threading for single query answering
This trend is consistent across subsets ranging from 100GB (Figure~\ref{fig:query:performance:100GB}) to 250GB (detailed results are reported in~\cite{url/GASS}).

\noindent{\bf Data Distributions.} We assess top performers representing different paradigms (EFANNA, Vamana, SSG, HNSW, ELPIS, and SPTAG-BKT) on challenging datasets (Fig. ~\ref{fig:datacomp}). Results (Figs. ~\ref{fig:query:performance:25GB:rand:pow1:10NN} and ~\ref{fig:query:performance:25GB:rand:pow50:10NN}) indicate that ELPIS consistently achieves high accuracy across skewness levels (0 to 50), outperforming other methods. As skewness increases, search becomes easier so most graph-based approaches improve but ELPIS maintains its superiority.

\newcommand{\soneM}{0.15}
\begin{figure*}[t]
    \centering
            \captionsetup{justification=centering}	
            \includegraphics[width=\textwidth]{../img-png/Experiments/legendall.png}
            
    \begin{minipage}{\textwidth}
        \centering
        \captionsetup{justification=centering}
        \captionsetup[subfigure]{justification=centering}
        \begin{subfigure}{0.015\textwidth}
            \raisebox{1cm}{\includegraphics[width=\textwidth]{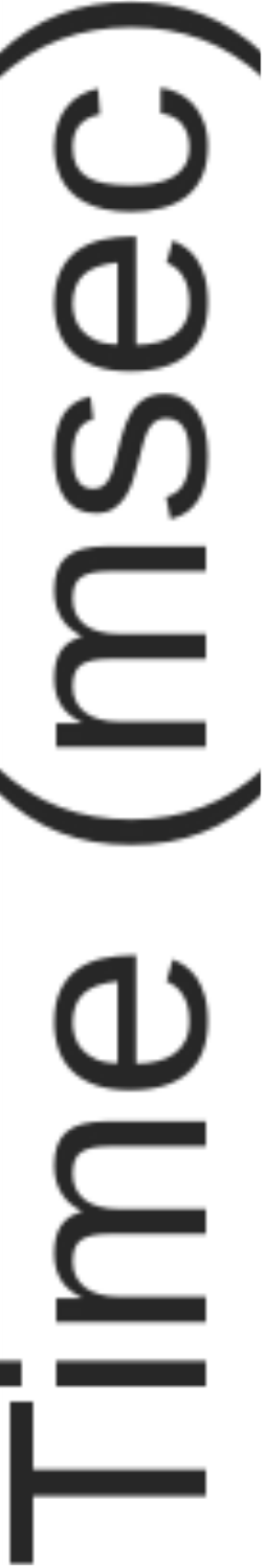}} 
        \end{subfigure}
        \begin{subfigure}{\soneM\textwidth}
            \includegraphics[width=\textwidth]{../img-png/Experiments/search/1M/deep_10nn.png}
            \caption{\karima{\textbf{Deep}}} 
            \label{fig:query:performance:1M:deep:10NN}
        \end{subfigure}
        \begin{subfigure}{\soneM\textwidth}
            \includegraphics[width=\textwidth]{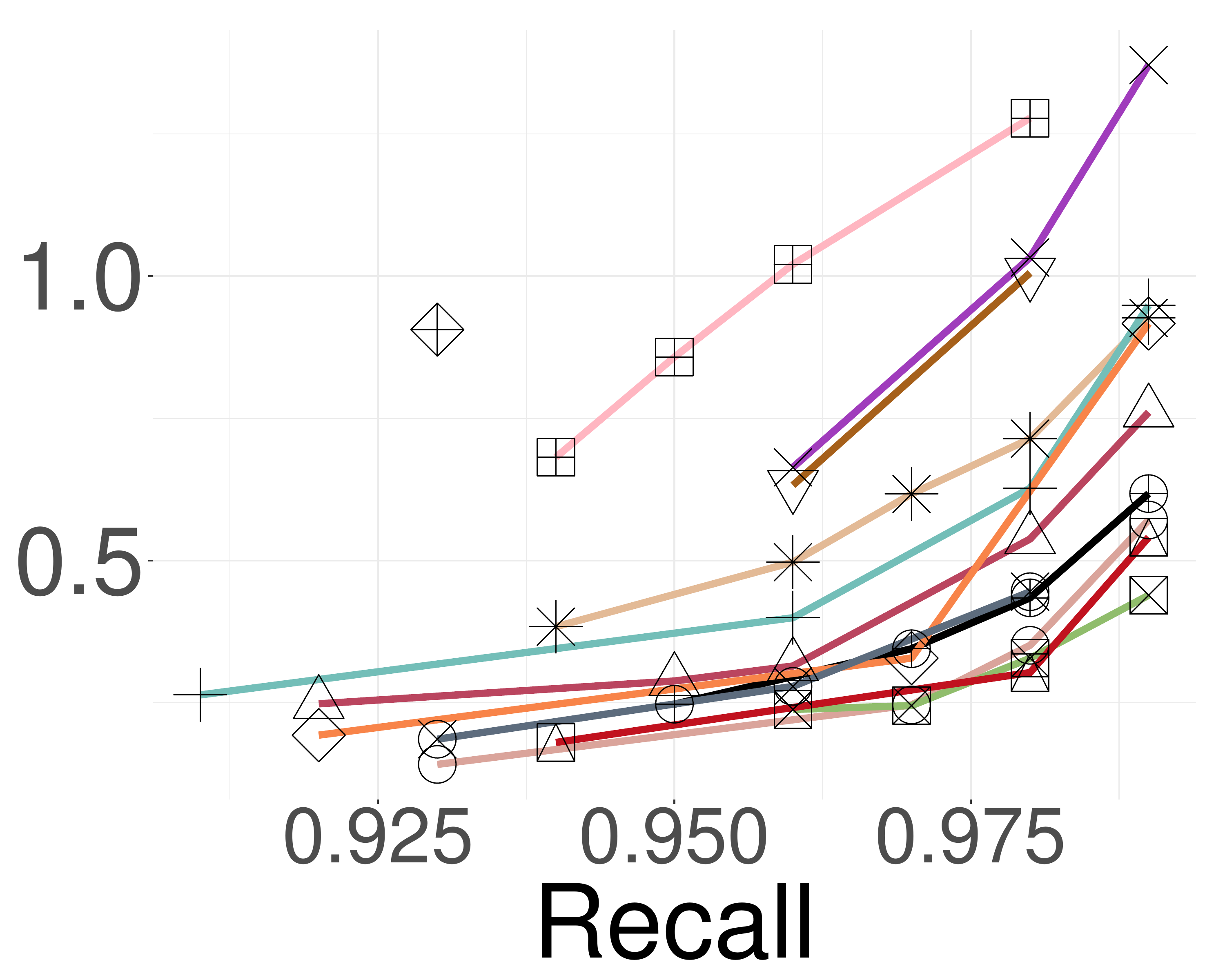}
            \caption{\karima{\textbf{Sift}}} 
        \label{fig:query:performance:1M:sift:10NN}
        \end{subfigure}
        \begin{subfigure}{\soneM\textwidth}
            \includegraphics[width=\textwidth]{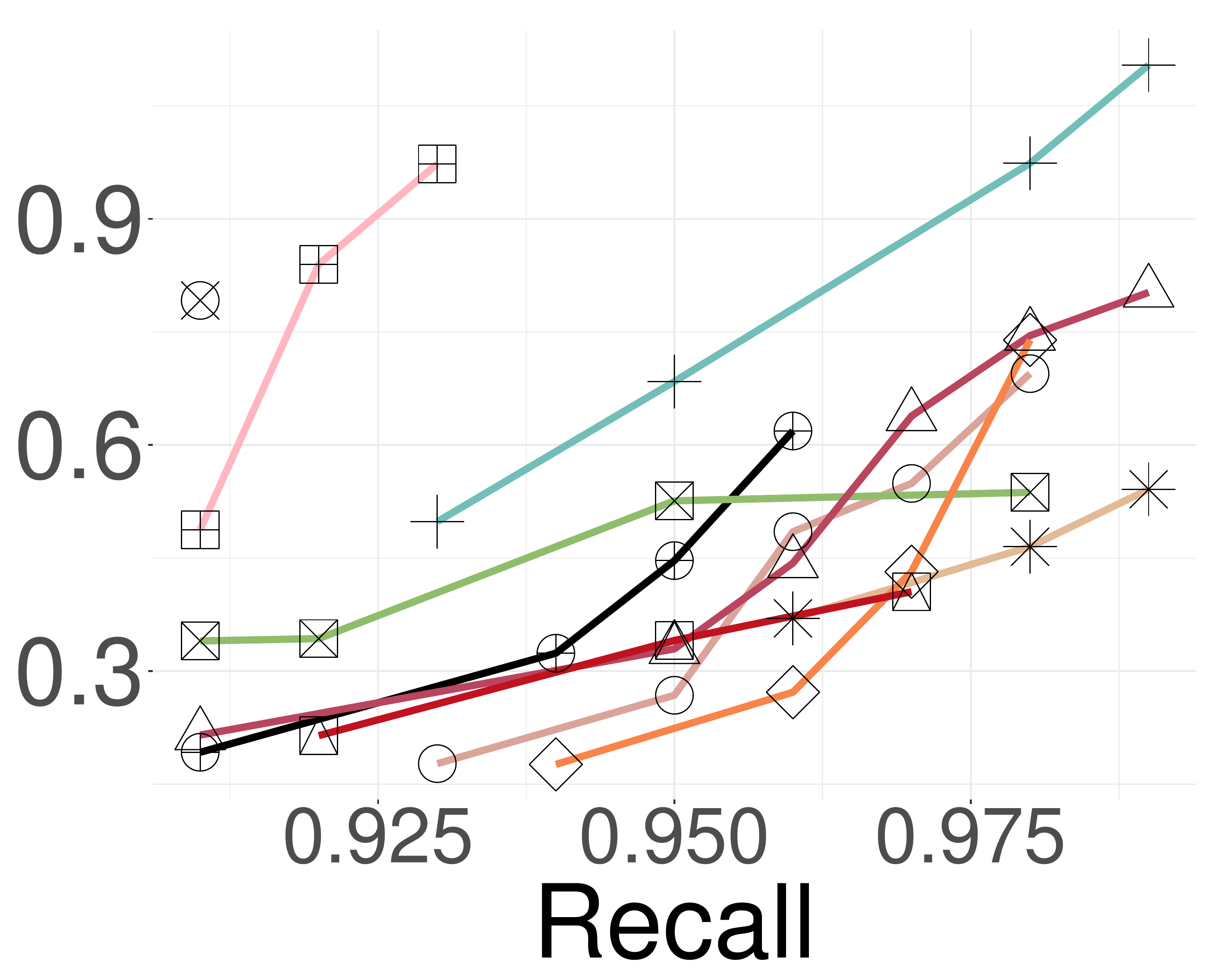}
            \caption{\karima{\textbf{SALD}}} 
            \label{fig:query:performance:1M:sald:10NN}
        \end{subfigure}
        \begin{subfigure}{\soneM\textwidth}
            \includegraphics[width=\textwidth]{../img-png/Experiments/search/1M/seismic_10nn.png}
            \caption{\karima{\textbf{Seismic}}} 
            \label{fig:query:performance:1M:seismic:10NN}
        \end{subfigure}
        \begin{subfigure}{\soneM\textwidth}
            \includegraphics[width=\textwidth]{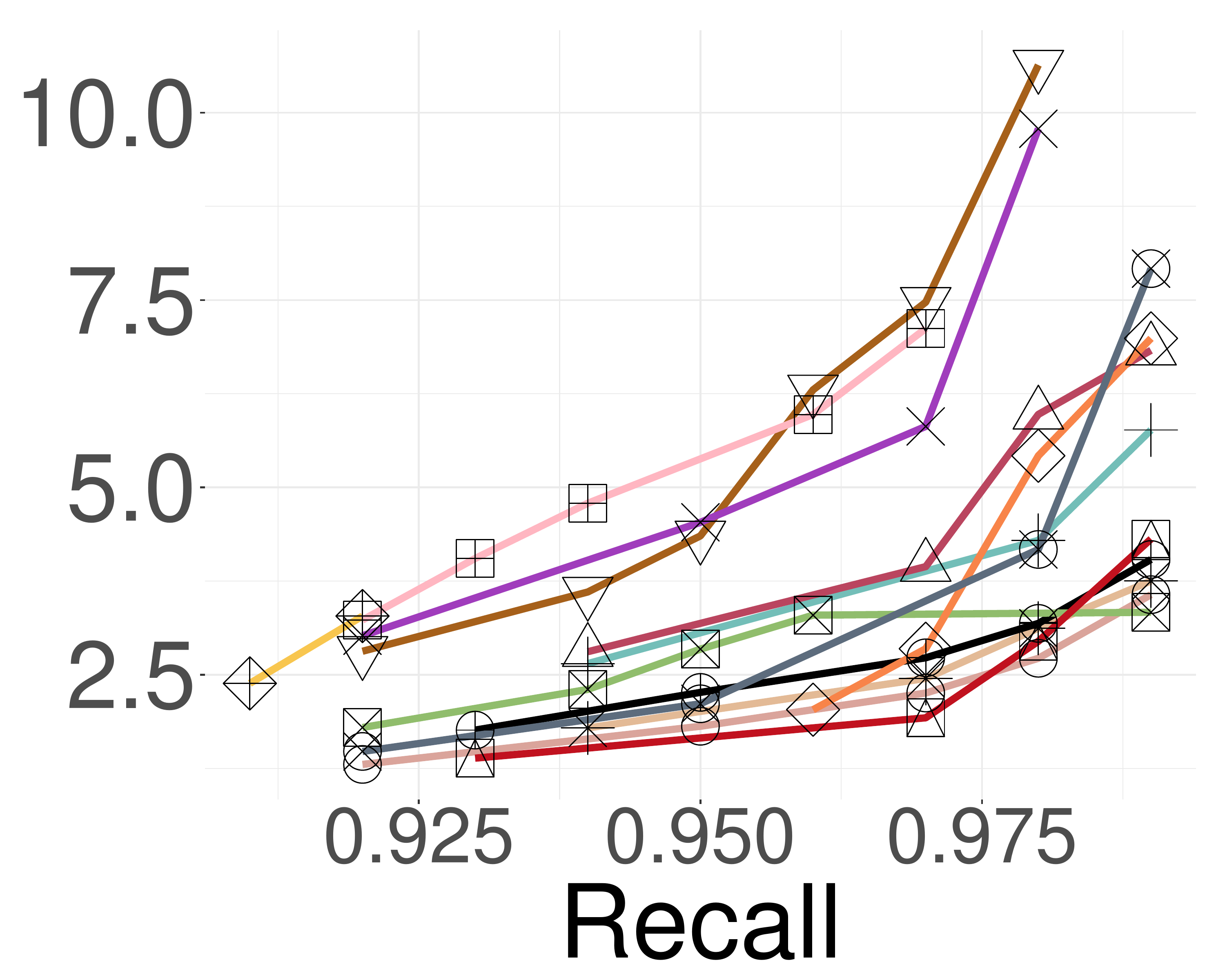}
            \caption{\karima{\textbf{Gist}}} 
            \label{fig:query:performance:1M:gist:10NN}
        \end{subfigure}
        \begin{subfigure}{\soneM\textwidth}
            \includegraphics[width=\textwidth]{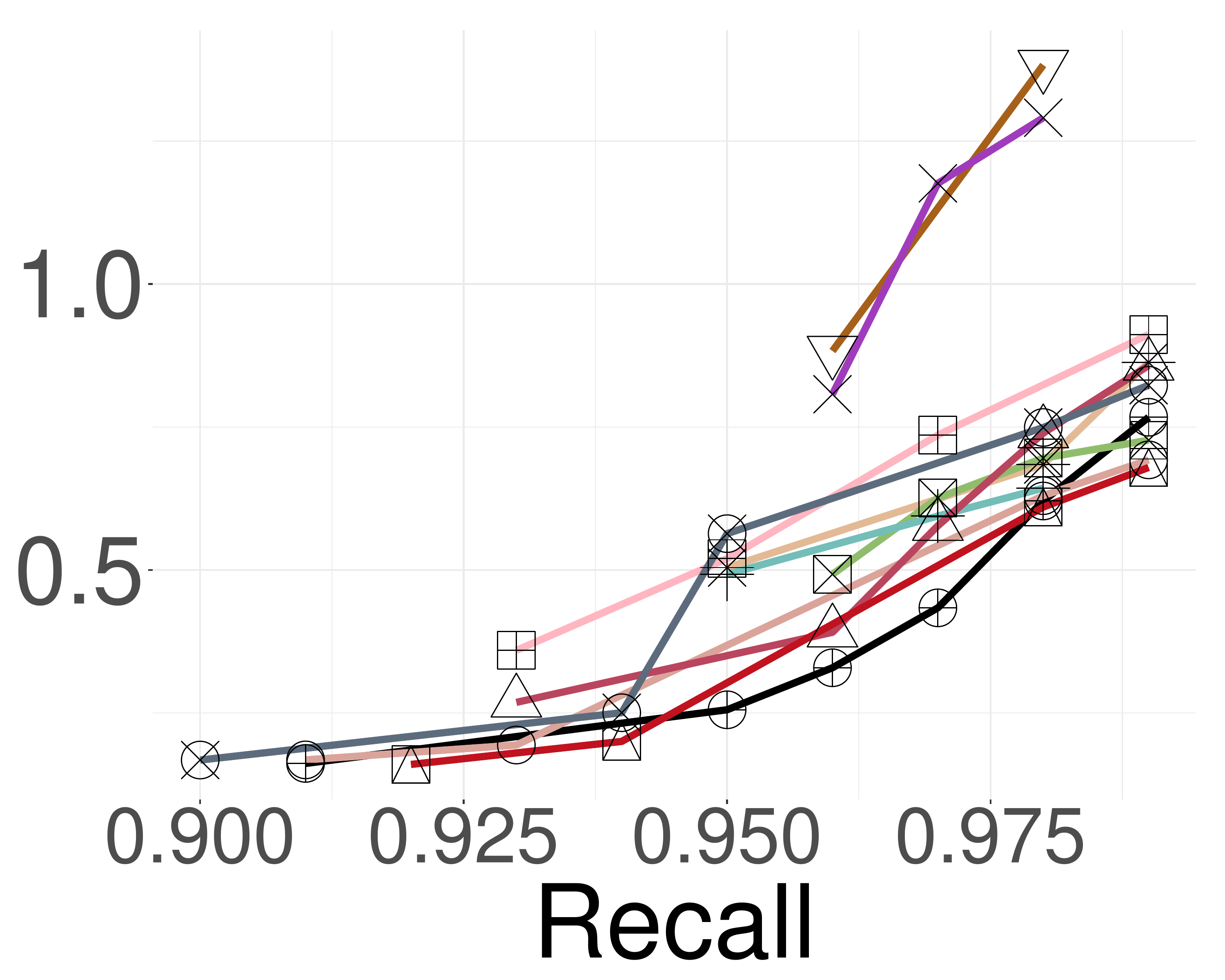}
            \caption{\karima{\textbf{Imagenet}}} 
            \label{fig:query:performance:1M:imagenet:10NN}
        \end{subfigure}
        
        \caption{\karima{Query performance on 1M vectors}}
        \label{fig:query:performance:1M}
    \end{minipage}
\end{figure*}

\newcommand{\ylabelwidth}{0.03\textwidth} 
\newcommand{\subfigwidthA}{0.31\textwidth}
\newcommand{\subfigwidthB}{1\textwidth}
\begin{figure*}[t]
  \centering
  \captionsetup{justification=centering}
  \captionsetup[subfigure]{font=bf,justification=centering}  
  
  \begin{minipage}{\textwidth}
    \includegraphics[width=\textwidth]{Experiments/legendall}
  \end{minipage}
  \begin{minipage}{0.60\textwidth}
    \begin{subfigure}{\ylabelwidth}
      \raisebox{1cm}{\includegraphics[width=\linewidth]{Experiments/time}}
    \end{subfigure}
    \begin{subfigure}{\subfigwidthA}
      \includegraphics[width=\linewidth]{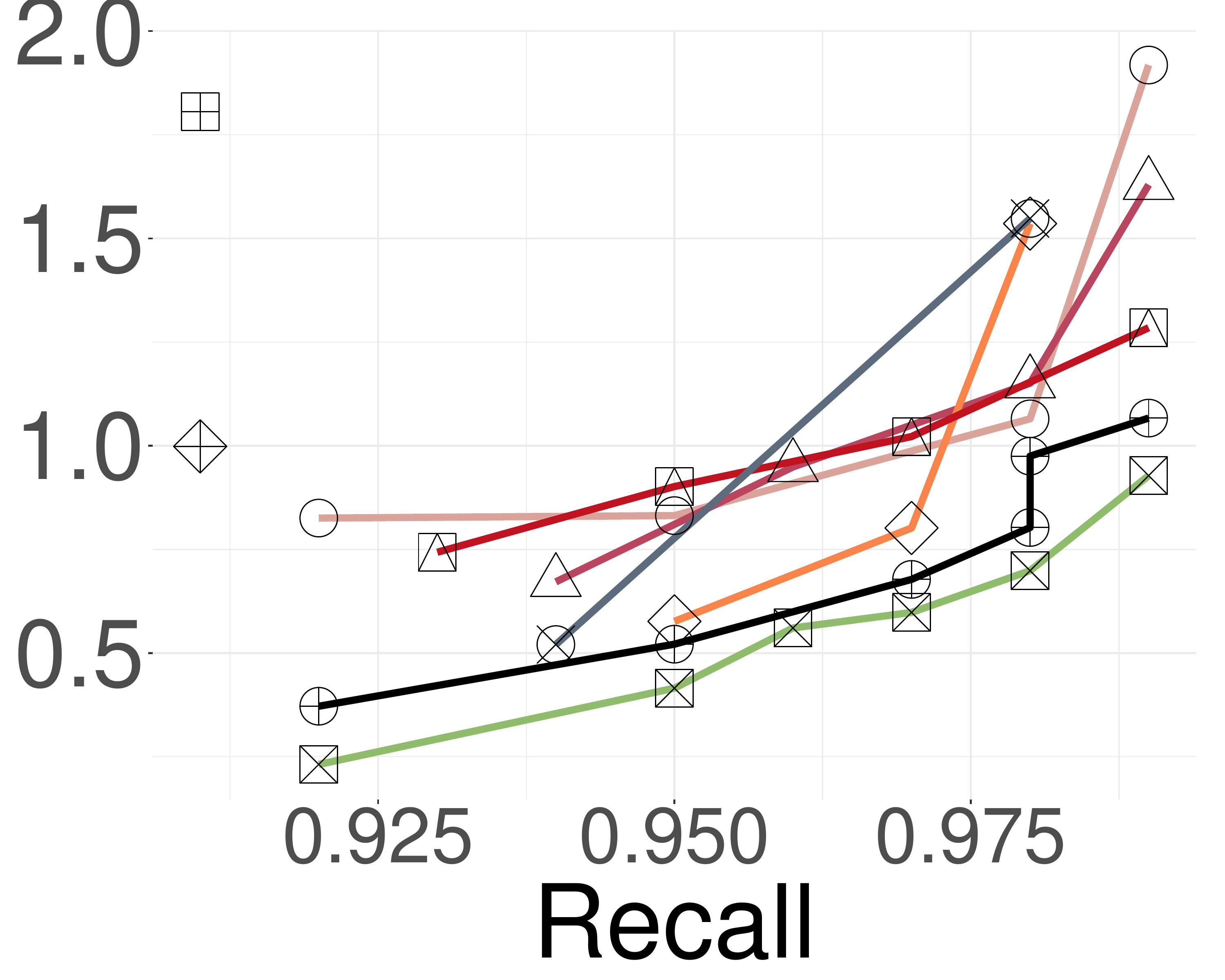}
      \caption{\karima{Deep}}
      \label{fig:query:performance:25GB:deep:10NN}
    \end{subfigure}
    \begin{subfigure}{\subfigwidthA}
      \includegraphics[width=\linewidth]{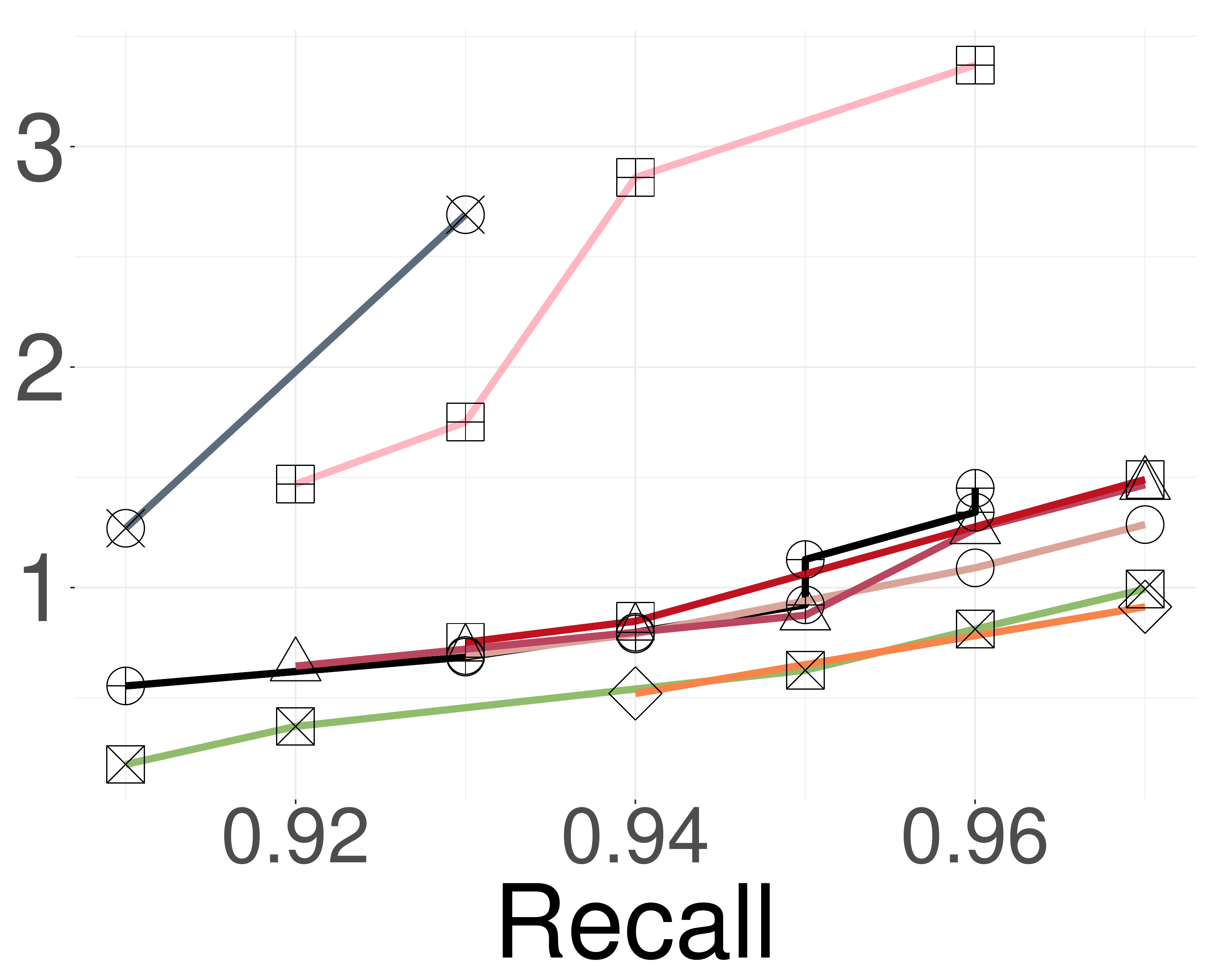}
      \caption{\karima{SALD}}
      \label{fig:query:performance:25GB:sald:10NN}
    \end{subfigure}
    \begin{subfigure}{\subfigwidthA}
      \includegraphics[width=\linewidth]{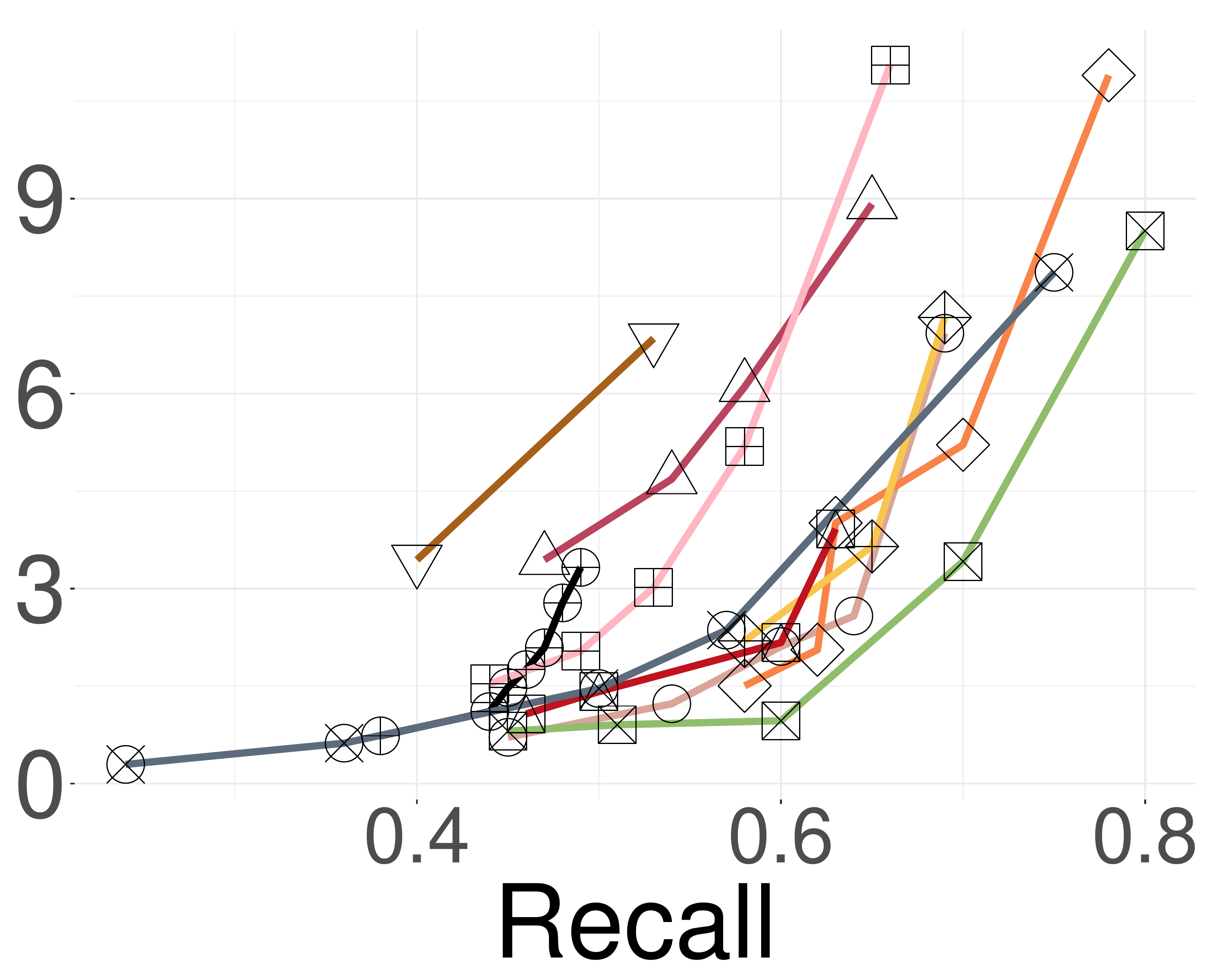}
      \caption{\karima{Seismic}}
      \label{fig:query:performance:25GB:seismic:10NN}
    \end{subfigure}

    \begin{subfigure}{\ylabelwidth}
      \raisebox{1cm}{\includegraphics[width=\linewidth]{Experiments/time}}
    \end{subfigure}
    \begin{subfigure}{\subfigwidthA}
      \includegraphics[width=\linewidth]{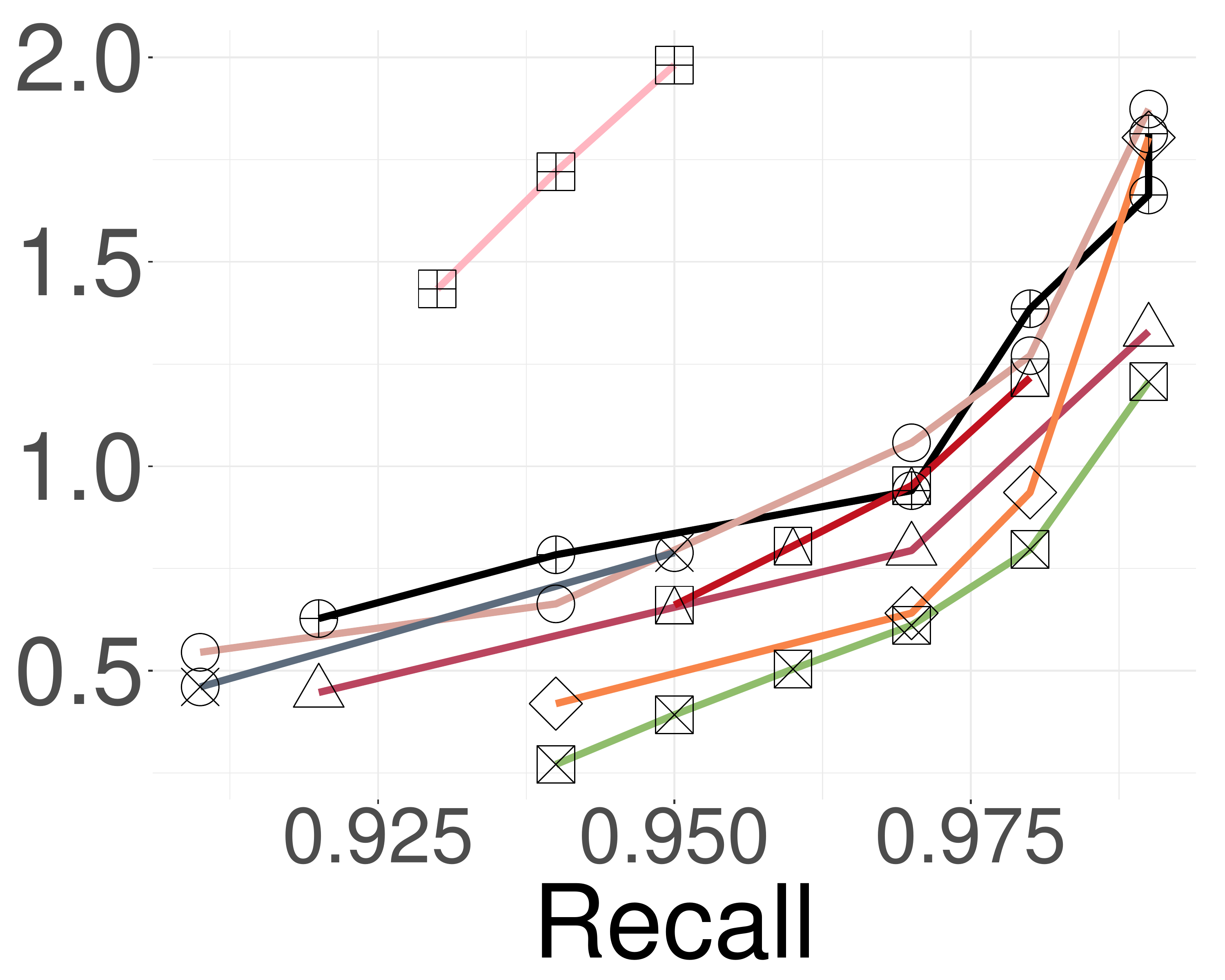}
      \caption{\karima{Sift}}
      \label{fig:query:performance:25GB:sift:10NN}
    \end{subfigure}
    \begin{subfigure}{\subfigwidthA}
      \includegraphics[width=\linewidth]{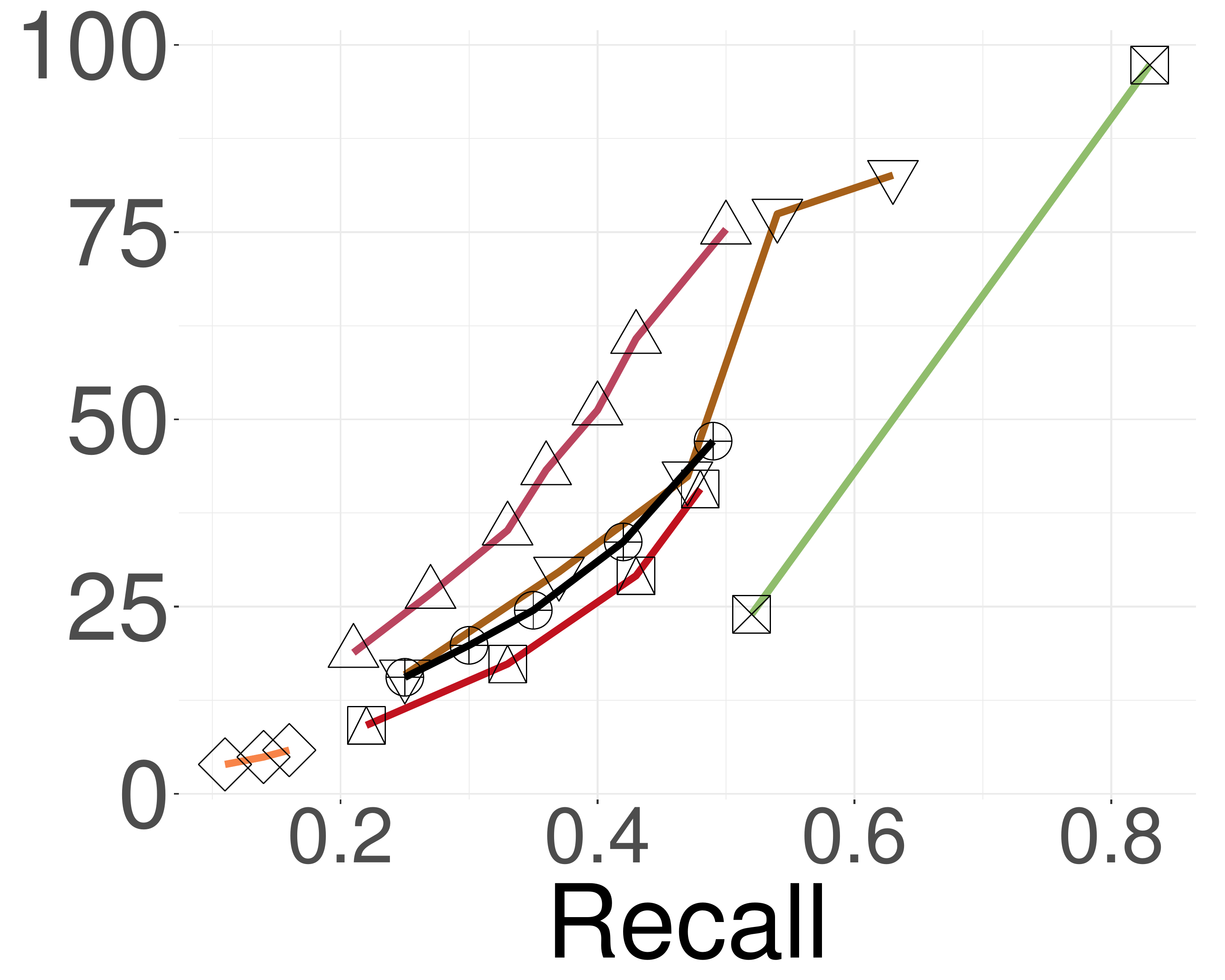}
      \caption{\karima{RandPow0}}
      \label{fig:query:performance:25GB:rand:pow1:10NN}
    \end{subfigure}
    \begin{subfigure}{\subfigwidthA}
      \includegraphics[width=\linewidth]{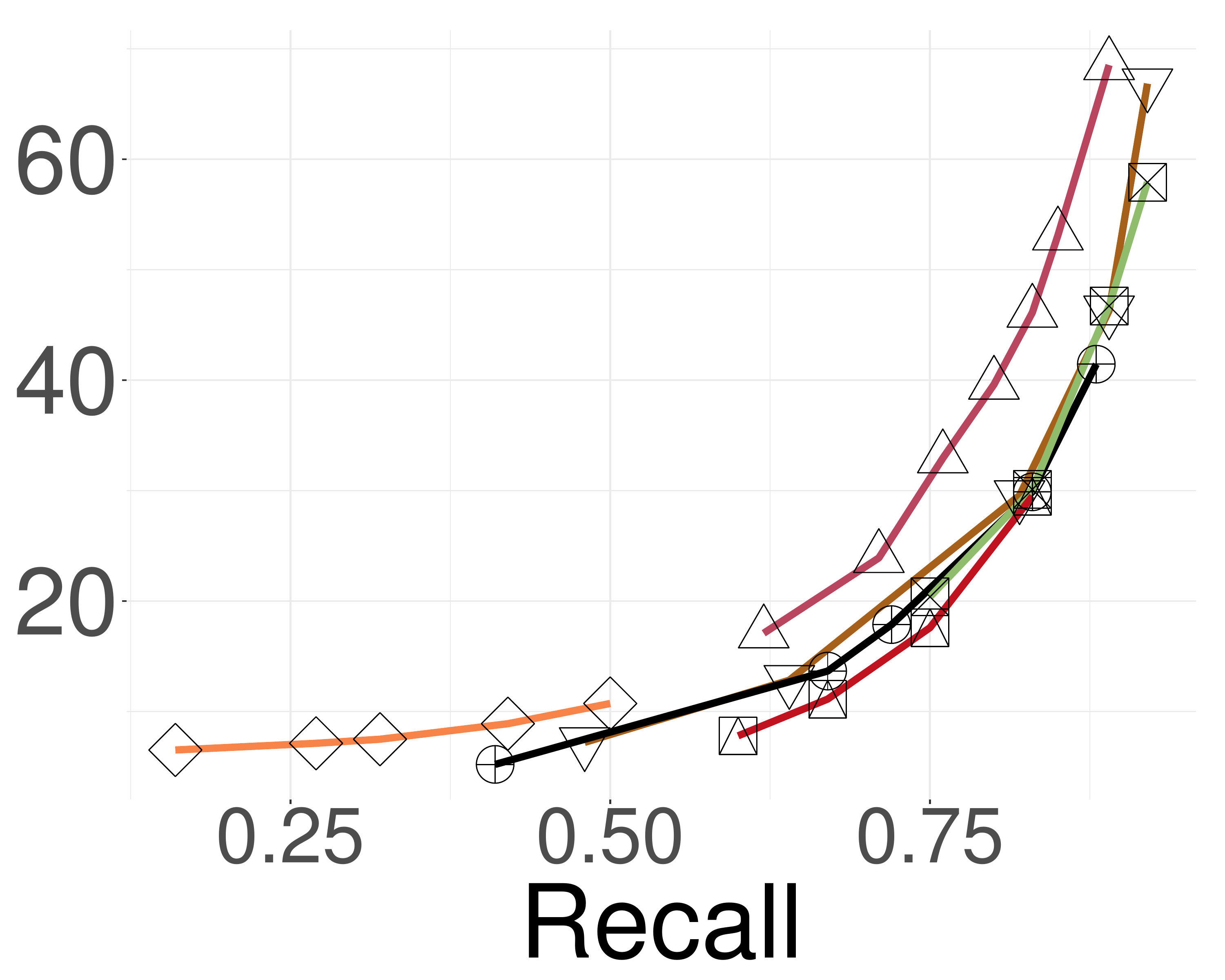}
      \caption{\karima{RandPow50}}
      \label{fig:query:performance:25GB:rand:pow50:10NN}
    \end{subfigure}
    \caption{\karima{25GB datasets}}
    \label{fig:query:performance:25GB}
  \end{minipage}
  \begin{minipage}{0.19\textwidth}
    \begin{subfigure}{\subfigwidthB}
      \includegraphics[width=\linewidth]{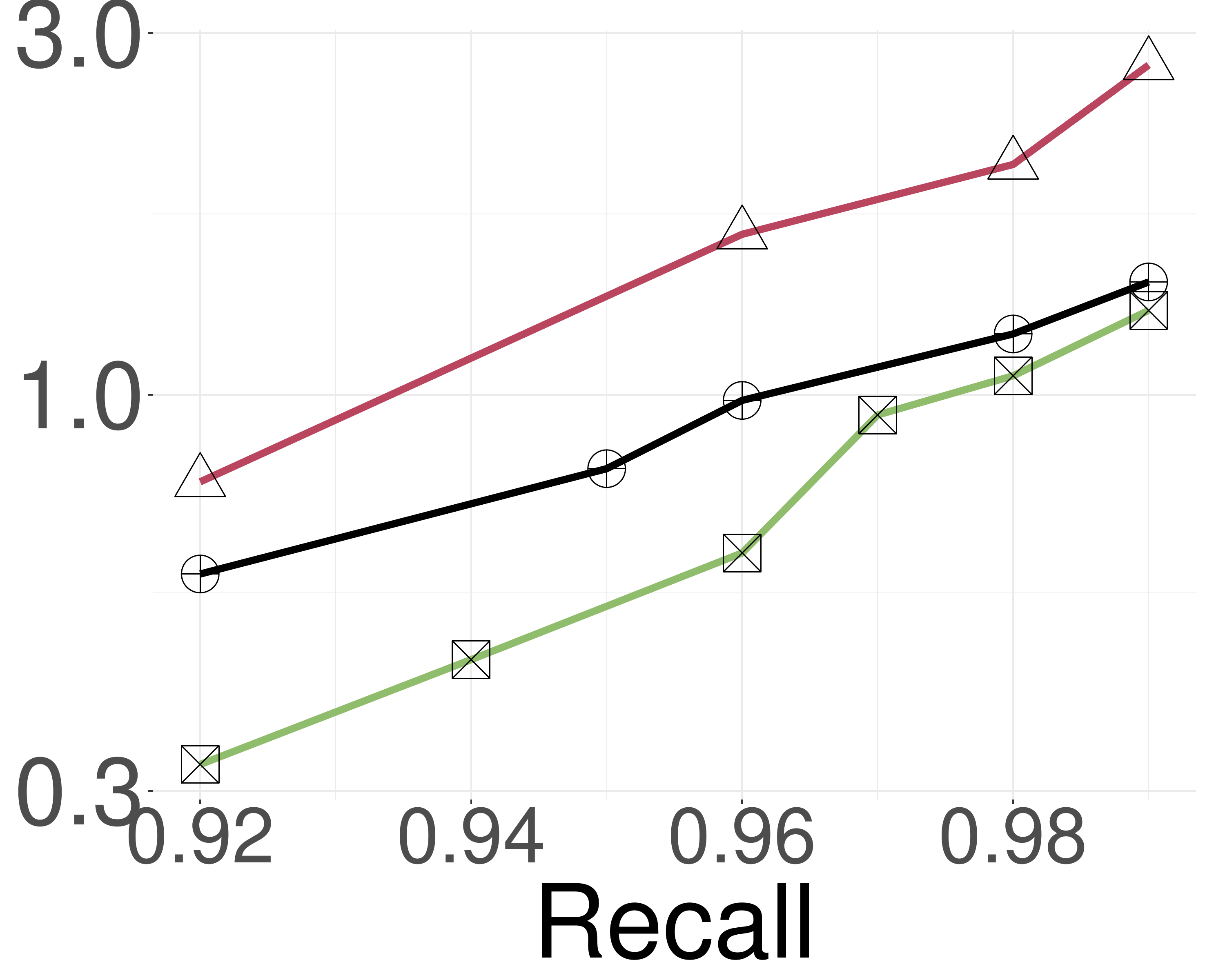}
      \caption{\textbf{Deep}}
      \label{fig:query:performance:100GB:deep:10NN}
    \end{subfigure}
    \begin{subfigure}{\subfigwidthB}
      \includegraphics[width=\linewidth]{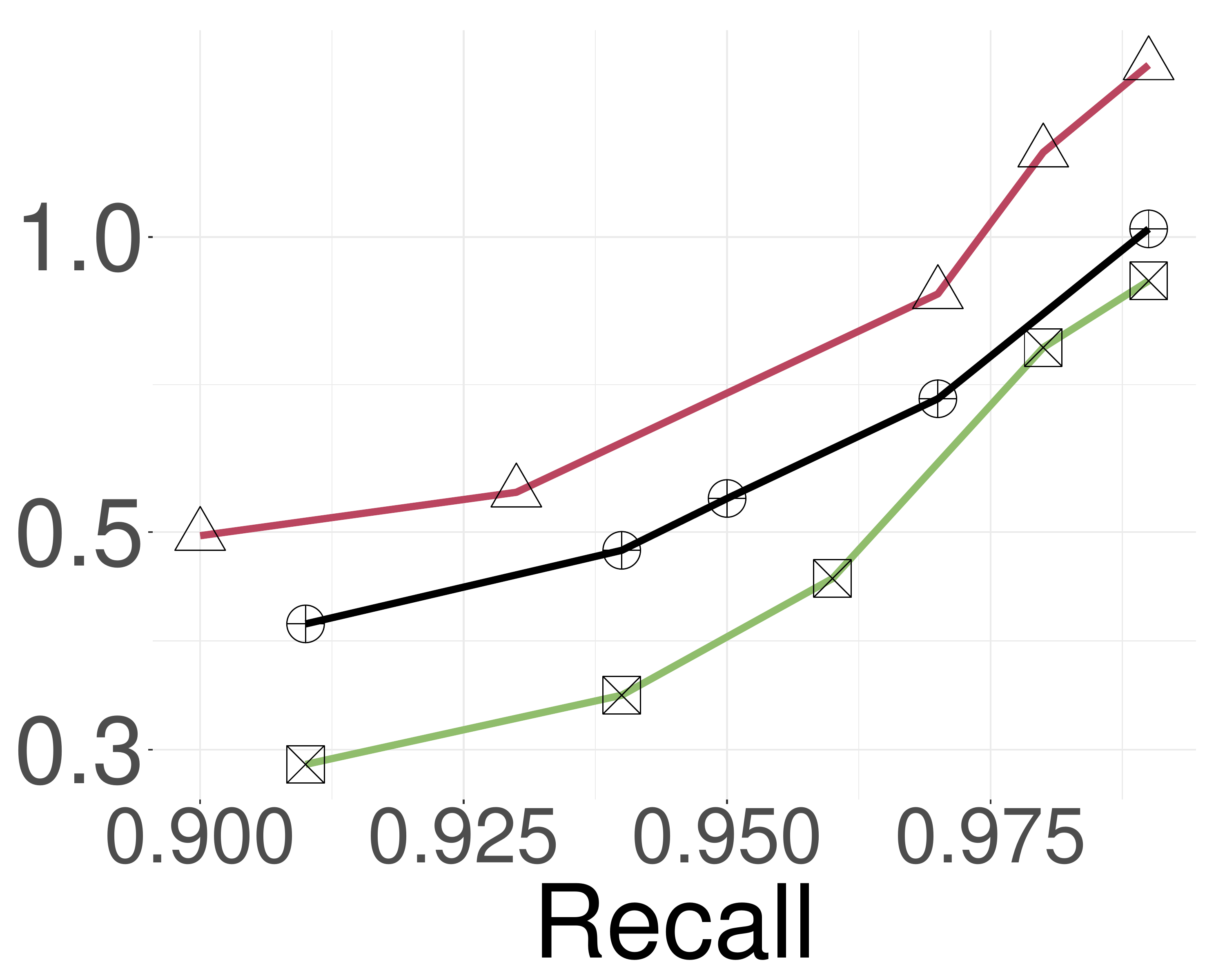}
      \caption{\textbf{Sift}}
      \label{fig:query:performance:100GB:sift:10NN}
    \end{subfigure}
    \caption{100GB datasets}
    \label{fig:query:performance:100GB}
  \end{minipage}
  \begin{minipage}{0.19\textwidth}
    \begin{subfigure}{\subfigwidthB}
      \includegraphics[width=\linewidth]{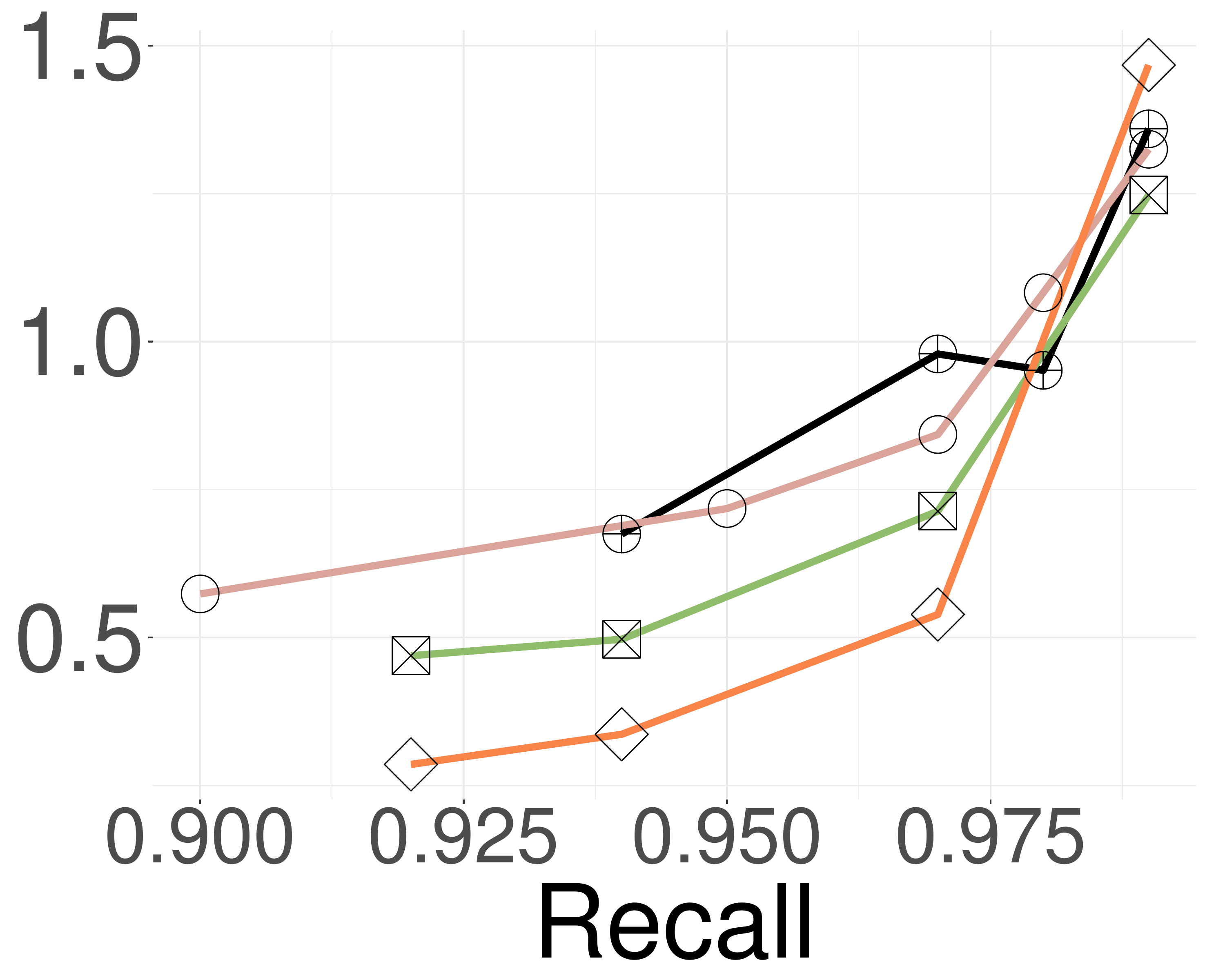}
      \caption{\karima{1\% noise}}
      \label{fig:search:query:performance:25GB:hard:1p}
    \end{subfigure}
    \begin{subfigure}{\subfigwidthB}
      \includegraphics[width=\linewidth]{Experiments/search/25/deep10p_10nn}
      \caption{\karima{10\% noise}}
      \label{fig:search:query:performance:25GB:hard:10p}
    \end{subfigure}
    \caption{\karima{Varying workloads}}
    \label{fig:search:query:performance:25GB:hard}
  \end{minipage}
\end{figure*}

\karima{\noindent{\bf Implementation Impact.} 
We evaluate the performance of original implementations of the best performing methods on 1B experiments, i.e., Vamana, HNSW, and ELPIS against optimized methods from the ParlayANN library~\cite{parlayann} (Vamana\_Opt, HNSW\_Opt, and HCNNG\_Opt) on Deep1B.  
Figure~\ref{fig:optimized_impl} indicates that Vamana\_Opt and HNSW\_Opt are faster for recall below 0.97 compared to their original counterparts, due to more efficient data structures~\cite{parlayann, parlayanncode}. However, at higher recall, this advantage diminishes as distance computations dominate; 
HCNNG\_Opt is competitive with Vamana and HNSW, while ELPIS maintains a performance lead. 
}

\end{document}